\documentclass[12pt]{article}
\usepackage{amsmath}
\usepackage[colorlinks=true,linkcolor=blue, citecolor=blue]{hyperref}%
\usepackage{amsfonts}
\usepackage{amssymb}
\usepackage{amsthm}
\usepackage{graphicx}
\usepackage[ruled,vlined]{algorithm2e}
\usepackage{hyperref}
\usepackage[all]{hypcap}
\usepackage{subfig}
\usepackage{graphicx}
\usepackage[countmax]{subfloat}
\usepackage{setspace}
\usepackage{subfig}
\usepackage{gensymb}
\usepackage{color}
\usepackage{bm}
\usepackage{authblk}
\usepackage{fullpage}
\usepackage{url}
\usepackage[authoryear]{natbib}
\usepackage{etoolbox}
\usepackage{amsthm}
\usepackage{multirow}
\usepackage[]{algorithm2e}
\usepackage{hyperref}
\usepackage{mathrsfs}
\newtheorem*{theorem1*}{Problem Statement}
\newtheorem*{proposition1*}{Proposition 1}
\newtheorem*{proposition2*}{Proposition 2}
\newtheorem*{proposition3*}{Proposition 3}

\theoremstyle{remark}

\makeatletter

\newif\ifabbreviation
\pretocmd{\thebibliography}{\abbreviationfalse}{}{}
\AtBeginDocument{\abbreviationtrue}

\newtheorem*{proposition1_1*}{Proposition 1}
\newtheorem*{proposition1_2*}{Proposition 2}
\newtheorem*{proposition1_3*}{Proposition 3}
\newtheorem*{proposition1_4*}{Proposition 4}
\newtheorem*{proposition1_5*}{Proposition 5 (Kalman filter in the spectral space)}
\newtheorem*{proposition1_6*}{Proposition 6}
\newtheorem*{proposition1_7*}{Proposition 7}

\newtheoremstyle{exampstyle}
{4pt} % Space above
{1pt} % Space below
{} % Body font
{} % Indent amount
{\bfseries} % Theorem head font
{.} % Punctuation after theorem head
{.5em} % Space after theorem head
{} % Theorem head spec (can be left empty, meaning `normal')

\theoremstyle{exampstyle} 
\theoremstyle{exampstyle} 
\theoremstyle{exampstyle} 
\theoremstyle{exampstyle} 
\theoremstyle{exampstyle} 
\theoremstyle{exampstyle}

\usepackage{bm}
\usepackage{soul}
\usepackage{wrapfig}
\usepackage[]{algorithm2e}

\usepackage{titlesec}

\titlespacing\section{0pt}{12pt plus 4pt minus 2pt}{0pt plus 2pt minus 2pt}
\titlespacing\subsection{0pt}{12pt plus 4pt minus 2pt}{0pt plus 2pt minus 2pt}
\titlespacing\subsubsection{0pt}{12pt plus 4pt minus 2pt}{0pt plus 2pt minus 2pt}

\begin{document}

%\pagestyle{empty}
%\begin{center}

\title{Inverse Models for Estimating the Initial Condition of Spatio-Temporal Advection-Diffusion Processes}
%\title{Statistical Inverse Modeling for Estimating Smooth Initial Conditions of Advection-Diffusion Processes}
%\author{}
\author[1]{Xiao Liu} %\thanks{A.A@university.edu}}
\author[2]{Kyongmin Yeo} %\thanks{B.B@university.edu}}
\affil[1]{Department of Industrial Engineering\\ University of Arkansas}
\affil[2]{IBM T. J. Watson Research Center}

%\author{Blinded Version} %\thanks{A.A@university.edu}}
%\affil{(*only major changes are highlighted in \textcolor{blue}{blue})}

\date{ }

\maketitle

\vspace{0.5cm}
%
%Research Team \footnote{\noindent IBM Research}
%and
%

%\today

%\end{center}
\singlespacing
\begin{abstract}
Inverse problems involve making inference about unknown parameters of a physical process using observational data. This paper investigates an important class of inverse problems---the estimation of the initial condition of a spatio-temporal advection-diffusion process using spatially sparse data streams. Three spatial sampling schemes are considered, including irregular, non-uniform and shifted uniform sampling. The irregular sampling scheme is the general scenario, while computationally efficient solutions are available in the spectral domain for non-uniform and shifted uniform sampling. For each sampling scheme, the inverse problem is formulated as a regularized convex optimization problem that minimizes the distance between forward model outputs and observations. The optimization problem is solved by the Alternating Direction Method of Multipliers algorithm, which also handles the situation when a linear inequality constraint (e.g., non-negativity) is imposed on the model output. Numerical examples are presented, code is made available on GitHub, and discussions are provided to generate some useful insights of the proposed inverse modeling approaches.

\end{abstract}

\noindent\textbf{Key words:} {\em Inverse Models, Spatio-Temporal Processes, Advection-Diffusion Processes, Alternating Direction Method of Multipliers}

\clearpage
%\onehalfspacing
\doublespacing
\section{Introduction} \label{sec:one}

\subsection{Motivating Examples} \label{sec:motivation}
Inverse problems involve making inference about unknown parameters of a physical process using observational data, and are widely found in scientific and engineering applications. For example, in urban air quality and environmental monitoring,  inverse problems aim at quickly pinpointing the sources of instantaneous emissions of gaseous pollutants that cause public health concerns \citep{Eckhardt2008, Martinez2014, Hwang2019}, or detecting  fugitive emissions  due to accidental releases from industrial operations \citep{Hosseini2016, Klein2016}. In healthcare applications, inverse models have been employed to obtain heart-surface potentials from body-surface measurements, known as the inverse ECG problem \citep{Yao2021}. 
In Seismology, inverse problems aim at getting information about the structure of the forces acting in the earthquake's focus from seismic waves at Earth's surface \citep{Apostol2019}. Inverse modeling has also found its applications in detecting the impact location of the missing Malaysian Airlines MH370, using the drift of marine debris \citep{Miron2019} or acoustic-gravity waves \citep{Kadri2019}. %More applications can be found in \cite{Issartel2007, }. 
%The recent expansion of domestic natural gas production from shale resources has improved the economic, security, and environmental outlook of future energy portfolio. However, at least 2\% of this gas resource is wasted through leaks of methane at production sites. Data-driven inverse modeling approaches have been used to quickly locate emission sources using sensor monitoring data streams \citep{Klein2016}. 

\begin{figure}[h!]  
	\begin{center}
		\includegraphics[width=0.65\textwidth]{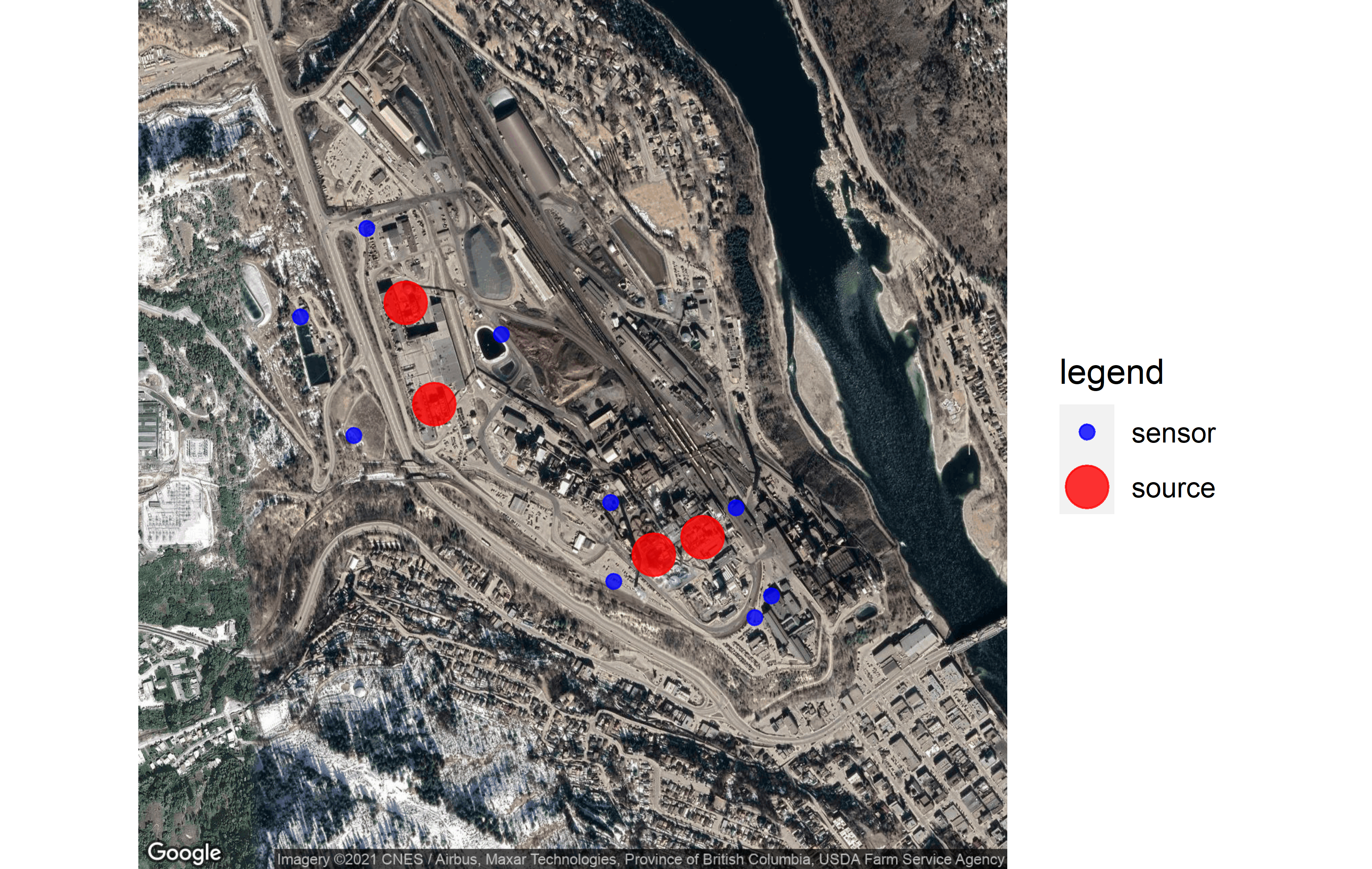}
		\centering
		\vspace{-6pt}
		\caption{A spatial area that contains four potential ZnSO$_\text{4}$ emission sources (big red circles) and nine receptors for ground-level measurements (small blue circles)}
		\label{fig:map}
	\end{center}
\end{figure}

This paper investigates an important class of statistical inverse problems---the estimation of the initial condition of a spatio-temporal advection-diffusion process using spatially sparse data streams. Consider the detection of accidental releases of fugitive emissions from industrial operations \citep{Hosseini2016}. Figure \ref{fig:map} shows a $2\times 2$ $\text{km}^2$ spatial area that includes a large lead-zinc smelter located in Trail, British Columbia, Canada. The four large red circles indicate the potential emission sources of Zinc Sulphate (ZnSO$_\text{4}$), while the small blue circles indicate the locations of nine receptors (i.e., sensors) deployed to detect accidental ZnSO$_\text{4}$ leak. The transport of ZnSO$_\text{4}$  is governed by an advection-diffusion equation in the form of a Partial Differential Equation (PDE). In case of accidental ZnSO$_\text{4}$ releases, sensor monitoring data are used to estimate probable emission locations. This inverse problem requires a statistical model that (i) establish the explicit and interpretable link between observations, emission sources, and process parameters (e.g. wind, diffusivity and decay) by integrating the underlying advection-diffusion physics, (ii) incorporate sensing data streams to estimate the initial condition when ZnSO$_\text{4}$ is released, and (iii) handle data arising from different sensor network layouts, such as irregular, uniform, non-uniform, nested, etc. 

%Such a physical process is often described by a Partial Differential Equation (PDE):
%\vspace{-8pt}
%\begin{equation} \label{eq:SPDE1_case}
%c(t, \bm{s}) + \bm{\vec{v}}^T\triangledown c(t, \bm{s}) - \triangledown \cdot [\bm{D} \triangledown  c(t, \bm{s})] = \Phi(t, \bm{s})
%\vspace{-6pt}
%\end{equation}	
%where $c(t, \bm{s})$ is the contamination concentration [mg $\text{m}^{-3}$], $\bm{\vec{v}}$ is the wind vector [m $\text{s}^{-1}$], $\bm{D}$ is the diffusivity [$\text{m}^{2}$ $\text{s}^{-1}$], and $\Phi(t, \bm{s})$ is an instantaneous emission source term [mg $\text{m}^{-3}$ $\text{s}^{-1}$]. 

%\begin{equation} \label{eq:source_case}
%\Phi(t, \bm{s}) = \delta(t-0) \sum_{j=1}^{4} \phi_j(\bm{s}), \quad\quad J\geq 0
%\end{equation}	
%The advection-diffusion process (\ref{eq:SPDE1_case}) is a special case of the general form (\ref{eq:SPDE1}) in Section \ref{sec:general}.
%The PDE (\ref{eq:SPDE1_case}), as well as its variations, serves as the governing equation behind a large class of physical phenomena where particles, energy or other physical quantities are transferred inside a system. 

\vspace{-40pt}
\subsection{Statistical Inverse Models and Literature Review}
%Statistical inverse problems, utilizing spatio-temporal sensor data streams, hinge on the statistical models employed to describe the underlying physical processes. A popular statistical spatio-temporal modeling approach approximates spatio-temporal processes by random fields (e.g., Gaussian Random Field), and are primarily driven by the correlation among observations  in space and time \citep{Cressie2011, Banarjee2012, Heaton2019, Yao2017, Yan2018, Ezzat2019, Fang2019, Prakash2021}. Such approaches have gained tremendous success in the literature. 

%(\textbf{spatio-temporal models}) 
An inverse model typically involves formulating an optimization problem that minimizes the distance between forward model output and observations \citep{Constantinescu2019}. 
Consider a physical process governed by an equation $F(\xi,\bm{\Theta}) = 0$ with $\xi$ and $\bm{\Theta}$ respectively being the state and parameter (unknown). Because the state of the process $\xi$ must depend on the parameter $\bm{\Theta}$ following the governing equation, we may define a mapping from $\bm{\Theta}$ to $\xi$, i.e., $\mathcal{F}(\bm{\Theta}) = \xi$, known as the parameter-to-observable map. Once the observations $\bm{Y}$ of the process are available, %Given the observations $\bm{Y}$ of the physical process, we may consider a statistical model $\bm{Y}=\xi+\epsilon^{\text{obs}}$ where $\xi$ satisfies the governing equation $F(\xi,\bm{\Theta}^{\text{true}}) = 0$ with $\bm{\Theta}^{\text{true}}$ being the true value of $\bm{\Theta}$, and $\epsilon^{\text{obs}}$ captures the observation error or model uncertainty. However, $\bm{\Theta}^{\text{true}}$ is often unknown and the system state $\xi$ is only obtained by solving the governing equation for some given value of $\bm{\Theta}$. Such a mapping from $\bm{\Theta}$ to $\xi$ is known as the parameter-to-observable map, $\mathcal{F}(\bm{\Theta}) = \xi$. Let  $\mathcal{L}(\mathcal{F}(\bm{\Theta}), \bm{Y})$ be a loss function, 
an inverse problem can be conceptually formulated as $\hat{\bm{\Theta}} = \text{argmin}_{\bm{\Theta}}\mathcal{L}(\mathcal{F}(\bm{\Theta}), \bm{Y})$ where $\mathcal{L}$ is some pre-defined loss function. 
%\vspace{-20pt}
%\begin{equation} \label{eq:inverse0}
%\hat{\bm{\Theta}} = \text{argmin}_{\bm{\Theta}}\mathcal{L}(\mathcal{F}(\bm{\Theta}), \bm{Y}), \quad\quad \text{s.t.   }  F(\xi,\bm{\Theta}) = 0.
%\end{equation}
%\vspace{-26pt}
%where $\mathcal{L}(\mathcal{F}(\bm{\Theta}), \bm{Y})$ is a loss function. 
%It is seen that an inverse model hinges on the chosen forward map $\mathcal{F}$, and the choices of the forward models give rise to different formulations of inverse problems \citep{Kaipio2005, Constantinescu2019}. 
For example, 
%\cite{Rajaona2015} proposed an adaptive Bayesian inference algorithm for estimating the parameters of a hazardous atmospheric release processes. 
\cite{Hwang2019} proposed a Bayesian inverse model to estimate the two-dimensional source functions by exploiting the adjoint advection-diffusion operator. The authors used the finite difference method to solve both the forward and backward physics models, and constructed the likelihood function for the emission rate given observations.
%also see \cite{Keats2007}. 
\cite{Oates2019} proposed an inverse model to estimate time-dependent parameters in an electrical potential model for industrial hydrocyclone equipment. Bayesian methods were employed to incorporate statistical models for the errors in the numerical solution of the physical equation.
\cite{Yeo2019} proposed a spectral method for source detection of advection-diffusion processes. The authors used the Gaussian radial basis functions to approximate a smooth emission function over space, and the spectral coefficients are modeled by generalized polynomial chaos. 

Note that, the physics model $F$ is typically solved by converting the PDE to a large system of Ordinary Differential Equations (ODE) given a finite difference discretization of the physical domain. When the dimension of the discretization is high, it is often computationally expensive to obtain the forward model output by directly solving the governing equation $F$. Statistical surrogate modeling is thus used to construct the parameter-to-observable map $\mathcal{F}$  \citep{Mak2018,  Qian2019, Gul2018}. For example, 
%following two main approaches: projection-based and data-driven-based approaches. For projection-based surrogate models, 
%\cite{Mak2018} proposed the surrogate model of large eddy simulations for design evaluation and physics extraction leveraging the idea of Proper Orthogonal Decomposition. \cite{Swischuk2019} and \cite{Qian2019} investigated physics-informed approaches for learning low-dimensional reduced-order models for large-scale dynamical systems governed by PDE. \cite{Gul2018} proposed an in-situ emulator that accommodates user-specified levels of the qualitative factors for uncertainty quantification of computer simulations. 
Gaussian Processes (GP) have been extensively investigated for constructing statistical surrogate models \citep{Hung2015, Deng2017, Gramacy2020, Zhang2021, Sauer2021}. For advection-diffusion processes, in particular, 
%For example, \cite{Hung2015} investigated the construction of a GP kriging model for functional data. 
%and developed a Gibbs sampling-based expectation maximization algorithm that enables the Kronecker product-based approach. 
%\cite{Deng2017} proposed an additive GP model for computer experiments with an additive correlation structure for qualitative factors and a multiplicative correlation structure for quantitative factors.  
\cite{Sigrist2015} obtained a GP by solving a PDE with an advection-diffusion operator that does not vary in space and time, and \cite{Liu2020} extended this approach by considering spatially-varying advection-diffusion.
%\cite{Zhang2021} investigated the distance-distributed design for GP surrogate models, and \cite{Sauer2021} proposed the deep GP surrogate models for handling abrupt regime changes in training data. 
%Reviews on building surrogate models for engineering applications are available in \cite{Asher2015} and \cite{Gramacy2020}. 
In recent years, physics-informed machine learning is rapidly emerging for data-driven discovery of governing physics and state/parameter/operator inference which are physically meaningful. For example, 
%\cite{Raissi2017} proposed the Hidden Physics Model for learning PDEs from noisy measurement data by assigning GP priors to the latent solutions of nonlinear PDEs. %also see \cite{Raissi2018} and \cite{Raissi2018b}. 
\cite{Raissi2019} proposed a deep learning framework for solving both forward and inverse problems for nonlinear partial differential equations. 
%Discovery of governing physical equations from measurement data through machine learning has also been investigated for nonlinear dynamical systems \citep{Brunton2016, Schaeffer2017, Lin2018}. Sparsity-promoting techniques are used to determine the fewest terms in dynamic governing equations, and obtain models that are parsimonious with a balance between model complexity and descriptive ability. In \cite{Lusch2018}, deep learning is used to discover the representations of Koopman eigenfunctions from data, which enable the embedding of dynamics on a low-dimensional manifold. 
\cite{Kang2021} proposed an active learning approach to estimate the unknown differential equations. An adaptive design criterion combining the D-optimality and the maximin space-filling criterion is used to reduce the experimental data size, where the D-optimality accounts for the unknown solution of the differential equations and its derivatives.

%(\textbf{inverse models}) 
%Finally, based on the spatio-temporal models or surrogate models, deterministic inverse models have been widely employed \citep{Issartel2007, Eckhardt2008, Martinez2014}. 
%This approach, adopted from atmospheric data assimilation, involves formulating a PDE-constrained convex optimization problem that minimizes the distance  between forward simulations and observations \citep{Constantinescu2019}. 
%At the high level, consider a PDE, $F(\xi,\bm{\Theta}; \Phi) = 0$, that governs a physical process with state $\xi$, a set of parameters $\bm{\Theta}$ and forcing $\Phi$. The process can be observed by sensors in space and time, and the observed data are given by $\bm{Y}=\mathcal{F}(\bm{\Theta}^{\text{true}})+\epsilon^{\text{obs}}$, where $\mathcal{F}(\bm{\Theta}^{\text{true}}) = \xi$ is the parameter-to-observable map, and $\xi$ solves $F(\xi,\bm{\Theta}; \Phi) = 0$. Hence, let  $\mathcal{L}(\mathcal{F}(\bm{\Theta}), \bm{Y})$ be a properly defined loss function, an inverse problem is formulated as $\bm{\Theta}^* = \text{argmin}_{\bm{\Theta}}\mathcal{L}$, s.t. $F(\xi,\bm{\Theta}; \Phi) = 0$. To handle the uncertainty associated with the underlying processes and noisy observations, statistical approaches have been employed. 

\color{black}
\vspace{-12pt}
\subsection{Problem Statement, Contributions and Overview} \label{sec:problem}
In this paper, we investigate a statistical inverse model that aims to estimate the initial condition (over the entire spatial domain) of an advection-diffusion process from spatially sparse sensor measurements. The problem can be formally stated as follows:

%This paper investigates the inverse modeling for spatio-temporal advection-diffusion processes. Let $\xi(t, \bm{s})$ denote a physical advection-diffusion process. As is the case in many scientific and engineering applications, the process $\xi(t, \bm{s})$ can only be observed at discrete times and discrete spatial locations. Then, this paper focuses on the following problem: 
\vspace{-4pt}
\begin{theorem1*}
	Let $\xi(t, \bm{s})$ be an advection-diffusion process monitored at $M$ spatial locations for $L$ discrete time periods, this paper is concerned with an inverse problem that estimates $\xi(0, \bm{s})$ over the entire spatial domain utilizing spatially sparse sensor data streams. %In the spectral domain, the paper is concerned with the estimation of the coefficients, $\eta(\bm{k})$ for $\bm{k}\in\mathcal{K}$, that determine $\xi(0, \bm{s})$ through (\ref{eq:source_Fouier}). 
	%estimates the coefficients $\eta(\bm{k})$ for $\bm{k}\in\mathcal{K}$, which gives the instantaneous source $\Phi(0,\bm{s})$ at time $0$ through the transform (\ref{eq:source_Fouier}). %The optimally select a number of $M$ spatial locations over the spatial domain such that $Q(0,\bm{s})$ can be accurately estimated. 	
	%Suppose that multiple impulses occur at time $t_0$ within the spatial domain $[0,1]^2$, i.e., $\delta_{t,t_0}Q_{t_0}(\bm{s})>0$ for some $s\in [0,1]^2$, the goal is to optimally select a number of $M$ spatial locations over the spatial domain such that all Fourier coefficients $\bm{\alpha}_{t_0} = (\alpha_{t_0}(\bm{k}_1), \alpha_{t_0}(\bm{k}_2), ..., \alpha_{t_0}(\bm{k}_N))^T$ at time $t_0$ can be uniquely estimated within $L\geq 1$ time periods, $t_0=t_1<t_2<\cdots<t_L$. Without loss of generality, we let $t_0=0$.
\end{theorem1*}

\vspace{-16pt}
\begin{figure}[h!]  
	\begin{center}
		\includegraphics[width=1\textwidth]{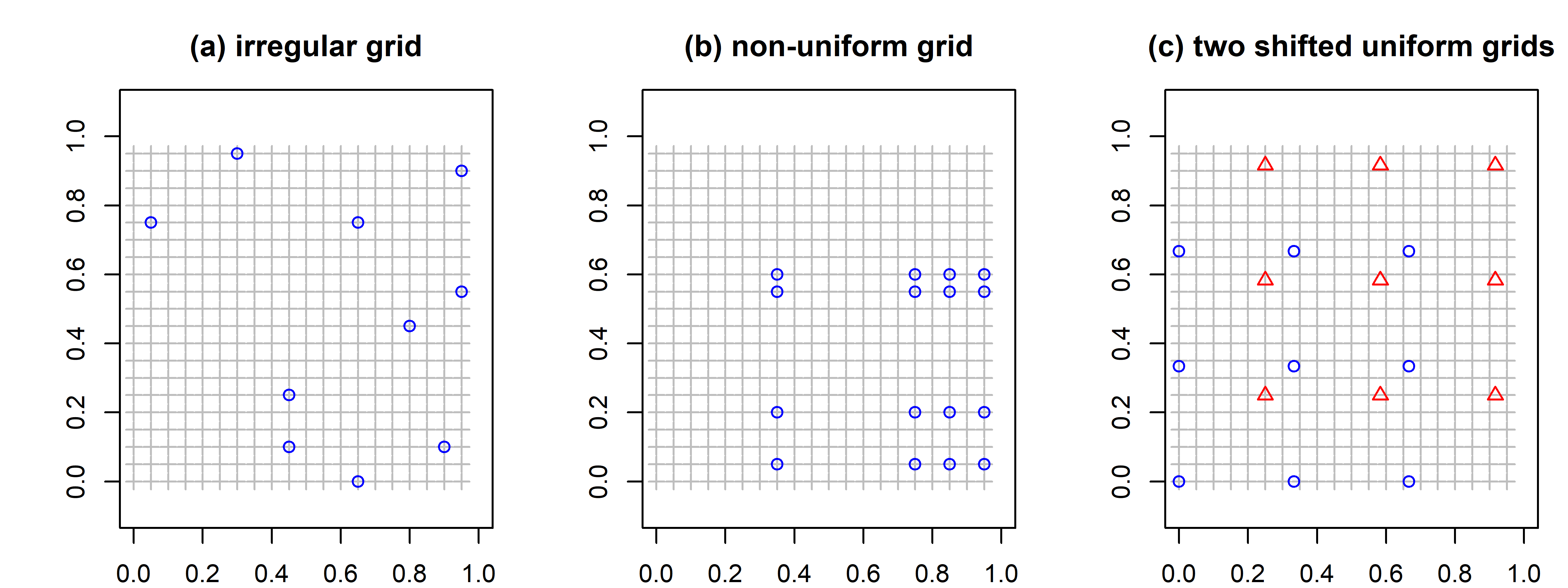}
		\centering
		\vspace{-20pt}
		\caption{Illustrations of the three spatial sampling schemes considered in this paper: (a) irregular grid, (b) non-uniform grid, and (c) shifted uniform grids (nested).}
		\label{fig:grid1}
	\end{center}
\end{figure}
%The three sampling schemes are respectively defined in Sections \ref{sec:general} and \ref{sec:special}.

\vspace{-26pt}
In particular, three important spatial sampling schemes (i.e., network layout) are considered: \textit{irregular}, \textit{non-uniform}, and \textit{shifted uniform} sampling.  Note that,  (i) the irregular sampling (Figure \ref{fig:grid1}a) is the general scenario that includes the non-uniform, shifted uniform, and uniform sampling as its special cases; (ii) the two special cases, i.e., non-uniform and shifted uniform sampling (Figures \ref{fig:grid1}b and \ref{fig:grid1}c), are also investigated because computationally efficient solutions are available in the spectral domain for the two special schemes. In practice, non-uniform sampling is often used to minimize acquisition time, sensor installation cost and power consumption, and is particularly useful for monitoring low-activity signals \citep{Venkataramani2001, Beyrouthy2015}. Shifted uniform sampling  (also known as the nested array or difference co-array) involves two nested uniform sensing networks, and significantly increases the degrees of freedom of linear arrays. By nesting two or more uniform linear arrays, shifted uniform sampling can provide $\mathcal{O}(M^2)$ degrees of freedom using only $M$ physical sensors, and thus mitigate the issue of spectral aliasing in spectral analysis \citep{Pal2010, Qin2021}.

%\vspace{-24pt}
Contributions of this paper are summarized as follows: (i) This paper proposes the first inverse model based on a forward spatio-temporal model for advection-diffusion processes proposed in \cite{Liu2020}. This forward model, which provides the parameters-to-observables map $\mathcal{F}$ for our inverse model, decomposes a physical spatio-temporal process by the linear combination of spatial bases and a multivariate random process of spectral coefficients. The temporal dynamics of spectral coefficients is determined by the advection-diffusion equation so as to integrate the governing physics into statistical models; see Section \ref{sec:preliminaries}. 
In this paper, following the idea of spectrum decomposition, the estimation of $\xi(0, \bm{s})$ over the entire spatial domain can be performed by estimating the spectral coefficients at time zero that determine $\xi(0, \bm{s})$. 
(ii) Because estimating the spectral coefficients at time zero requires sufficient observations over space and time given a sensor network layout, Section \ref{sec:property} performs theoretical investigations and obtains sufficient and necessary conditions for the spectral coefficients at time zero to be uniquely estimated. When such conditions are not met, spectral coefficients cannot be uniquely determined and this is known as spectral aliasing in signal processing. 
(iii) In Section \ref{sec:problem}, we further argue that it is not always possible to uniquely estimate all spectral coefficients at time zero. The sensor network layout and the number of observations in space and time are often subject to practical constraints. Hence, we often require the spectral coefficients to be estimated under the scenarios where neither the sufficient nor necessary conditions are met. It is also noted that, the spectral coefficients may rapidly decay at high-frequency modes if  $\xi(0, \bm{s})$ is smooth. To cope with this issue, Section \ref{sec:problem} presents a regularized inverse problem that estimates the spectral coefficients at time zero given any irregular sensor network layout. The regularization induces both the sparsity in spectral coefficients and the smoothness of neighboring spectral coefficients. Section \ref{sec:special} presents the special results when data are obtained from non-uniform and shifted uniform sampling, under which computationally efficient solutions are available in the spectral domain.
(iv) Finally, Section \ref{sec:ADMM} develops the Alternating Direction Method of Multipliers (ADMM) algorithm for efficiently solving the proposed regularized inverse problem. We then extend the proposed inverse model and the ADMM algorithm to handle non-negativity constraint on $\xi(0, \bm{s})$, i.e., $\xi(0, \bm{s})\geq 0$. 
Section \ref{sec:numerical} provides comprehensive numerical investigations. 
%The first example is based on simulated data which aims to provide useful insights about the proposed inverse model, while the second example revisits the motivating example in Section \ref{sec:motivation}. 
Sensitivity analysis is performed to demonstrate the robustness of the proposed method.

%\section{The Model} \label{sec:physics}
%\vspace{8pt}
%The paper is organized as follows. Section 2 presents the main results under the irregular sampling scheme (i.e., the general case). Section 3 further investigates the two special cases when data are sampled at non-uniform or shifted uniform sampling grids. Section 4 provides the technical details on solving the inverse problems using the Alternating Direction Method of Multipliers (ADMM). Numerical investigations are performed in Section 5 to generate some critical insights of the proposed approaches, and a case study is also included to illustrate the application of the proposed method in the context of a real application. Section 6 concludes the paper and highlights some future research directions.   

\vspace{-16pt}
\section{Inverse Modeling under General Irregular Sampling} \label{sec:general}

%We are concerned with the estimation of the coefficients, $\eta(\bm{k})$ for $\bm{k}\in\mathcal{K}$, when data are collected from an irregular grid with $M$ sensors and for $L$ discrete time periods.

\vspace{-10pt}
\subsection{Preliminaries} \label{sec:preliminaries}
%Section \ref{sec:general} firstly present the inverse problem under the general irregular sampling scheme as shown in Figure \ref{fig:grid1}a. 
\vspace{-6pt}
Consider a physical spatio-temporal advection-diffusion process $\xi(t, \bm{s})$ given by a PDE:
%\vspace{-14pt}
 %and let $Y(t,\bm{s})$ be the observation of $\xi(t, \bm{s})$ at time $t$ and location $\bm{s}$:
%This paper focuses on the invese modeling for spatio-temporal data arising from a generic form of advection-diffusion processes:
%\vspace{-12pt}
%\begin{equation} \label{eq:Y}
%\begin{split}
%Y(t,\bm{s}) = \xi(t, \bm{s}) + \epsilon^{\text{obs}}, \quad\quad\quad \bm{s}\in\mathbb{S}, \quad t \geq 0
%\mathcal{A} \xi(t, \bm{s}) = \phi(t, \bm{s}), \quad\quad\quad \bm{s}\in\mathbb{S}, \quad t \geq 0
%\end{split}
%\end{equation}	
%\noindent where $\epsilon^{\text{obs}}$ represents the observation error.
\vspace{-16pt}
\begin{equation} \label{eq:SPDE1}
\mathcal{A} \xi(t, \bm{s}) = \phi(t, \bm{s}), \quad\quad \bm{s}\in\mathbb{S},t\geq0 
\end{equation}	

\vspace{-18pt}
\noindent where $\mathbb{S}$ is the spatial domain, $\phi(t, \bm{s})$ is the source term, %$\varepsilon_t(\bm{s})$ is a spatio-temporal error process, 
and the advection-diffusion operator $\mathcal{A}$ is given by $\mathcal{A}\xi(t, \bm{s}) = \dot{\xi}(t, \bm{s}) + \bm{\vec{v}}^T\triangledown \xi(t, \bm{s}) - \triangledown \cdot [\bm{D} \triangledown  \xi(t, \bm{s})] + \zeta \xi(t, \bm{s})$	
%\vspace{-14pt}
%\begin{equation} \label{eq:A}
%\mathcal{A}\xi(t, \bm{s}) = \dot{\xi}(t, \bm{s}) + \bm{\vec{v}}^T\triangledown \xi(t, \bm{s}) - \triangledown \cdot [\bm{D} \triangledown  \xi(t, \bm{s})] + \zeta \xi(t, \bm{s})
%\end{equation}	
with $\bm{\vec{v}}$, $\bm{D}$, $\zeta$, $\triangledown$ and $\triangledown\cdot$ respectively being the velocity field, diffusion tensor, decay, gradient and divergence operator. The PDE (\ref{eq:SPDE1}) serves as the governing equation behind an extremely large class of physical phenomena where particles and energy are transferred inside a system. 
%In this Partial Differential Equation (PDE),
%for example, species concentration for mass transfer, temperature for heat transfer, pollutant concentrations for air quality, radar reflectivity for dynamic weather systems, etc.  

%The process $\xi(t, \bm{s})$ is monitored by spatially-sparse sensors at discrete times. 
In this paper, the process $\xi(t, \bm{s})$ can only be observed by spatially distributed sensors at discrete times. Hence, the inverse problem is concerned with estimating the initial condition $\xi(0, \bm{s})$ over the entire spatial domain.
%Given the observations of $\xi(t, \bm{s})$ from $M$ spatial locations for $L$ discrete time periods, we are interested in estimating $\xi(0, \bm{s})$ over the entire spatial domain. In particular, 
%about spatially-sparse instantaneous sources,  $\Phi(t, \bm{s})$, of the following form:
%\begin{equation} \label{eq:source}
%\Phi(t, \bm{s}) = \delta(t-0) \sum_{j=1}^{J} \phi_j(\bm{s}), \quad\quad J\geq 0
%\end{equation}	
%where $\delta$ is the Dirac function and $J$ represents the number of sources in space.
%In particular, let 
Following \cite{Liu2020}, $\xi(0, \bm{s})$ is assumed to be spanned by a finite number of orthogonal spatial Fourier basis functions, 
\vspace{-16pt}
\begin{equation} \label{eq:source_Fouier}
\xi(0, \bm{s}) = \sum_{\bm{k}\in\mathcal{K}} \eta(\bm{k}) f_{\bm{k}}(\bm{s})
\end{equation}

\vspace{-12pt}
\noindent where $\bm{k}=(k_1, k_2)^T \in \mathcal{K}$ is the wavenumber, $f_{\bm{k}}(\bm{s})=e^{\imath 2\pi \bm{s}^T \bm{k}}$ is the Fourier basis function, $\eta(\bm{k})$ is the coefficient that determines the weight of each Fourier mode, and 
%\begin{equation} \label{eq:tensor_k}
%\mathcal{K} = (-\frac{N_1}{2}+1, -\frac{N_1}{2}+2, \cdots, \frac{N_1}{2})\otimes (-\frac{N_2}{2}+1, -\frac{N_2}{2}+2, \cdots, \frac{N_2}{2}). 
%\end{equation}
\vspace{-12pt}
\begin{equation} \label{eq:tensor_k}
\mathcal{K} = \left\{(k_1, k_2)^T; k_1 = -\frac{N_1}{2}+1, -\frac{N_1}{2}+2, \cdots, \frac{N_1}{2}, k_2 = -\frac{N_2}{2}+1, -\frac{N_2}{2}+2, \cdots, \frac{N_2}{2}\right\}. 
\end{equation}

\vspace{-12pt}
Note that, the equality in (\ref{eq:source_Fouier}) holds when the initial condition is band-limited with the high-frequency parts of its Fourier expansion decaying rapidly to exactly zero. 
%if the process is discretized on a $N_1 \times N_2$ mesh grid, or, $\xi(0, \bm{s})$ is smooth and the high-frequency parts of its Fourier expansion are truncated. 
%Let the initial condition of the PDE (\ref{eq:SPDE1}) be given by
%\begin{equation} \label{eq:initial}
%\xi(0, \bm{s}) = \Phi(0, \bm{s})
%\end{equation}	
%and further assume that $\Phi(0, \bm{s})$, $\bm{s} \in \mathbb{S}$, is spanned by a finite number of spatial Fourier functions, 
%\begin{equation} \label{eq:source_Fouier}
%\Phi(0, \bm{s}) = \sum_{\bm{k}\in\mathcal{K}} \eta(\bm{k}) f_{\bm{k}}(\bm{s})
%\end{equation}
%where $f_{\bm{k}}(\bm{s})=e^{\imath 2\pi \bm{s}^T \bm{k}}$ is the Fourier basis function, $\bm{k}$ is the wavenumber, $\eta(\bm{k})$ is the Fourier coefficient at $\bm{k}=(k_1, k_2)^T \in \mathcal{K}$, and 
%%\begin{equation} \label{eq:tensor_k}
%%\mathcal{K} = (-\frac{N_1}{2}+1, -\frac{N_1}{2}+2, \cdots, \frac{N_1}{2})\otimes (-\frac{N_2}{2}+1, -\frac{N_2}{2}+2, \cdots, \frac{N_2}{2}). 
%%\end{equation}
%\begin{equation} \label{eq:tensor_k}
%\mathcal{K} = \left\{(k_1, k_2)^T; k_1 = -\frac{N_1}{2}+1, -\frac{N_1}{2}+2, \cdots, \frac{N_1}{2}, k_2 = -\frac{N_2}{2}+1, -\frac{N_2}{2}+2, \cdots, \frac{N_2}{2}\right\}. 
%\end{equation}
Based on (\ref{eq:source_Fouier}), it has been shown that the process $\xi(t,\bm{s})$ remains in $\mathbb{S}$ for $t\geq 0$ and also admits a spectral representation: $\xi(t, \bm{s}) = \sum_{\bm{k}\in\mathcal{K}} \alpha(t, \bm{k}) f_{\bm{k}}(\bm{s})$, %\vspace{-16pt}
%\begin{equation} \label{eq:represent}
%\xi(t, \bm{s}) = \sum_{\bm{k}\in\mathcal{K}} \alpha(t, \bm{k}) f_{\bm{k}}(\bm{s})
%\end{equation}
%\vspace{-12pt}
where $\alpha(t, \bm{k})$ is the Fourier coefficient evolving over time, and $\alpha(0, \bm{k})=\eta(\bm{k})$ \citep{Sigrist2015, Liu2020}. 
%Without loss of generality, $N_1$ and $N_2$ are assumed to be even integers throughout the paper, and $N = N_1 \times N_2$.

%$\tau_1< \tau_2 < \cdots$.
%and on a rectangular spatial mesh system $\mathbb{S}$ defined by a tensor product of two one-dimensional collocation sets,
%\begin{equation} \label{eq:tensor_s}
%\mathbb{S} = (0,\frac{1}{N_1},\frac{2}{N_1},\cdots,\frac{N_1-1}{N_1}) \otimes (0,\frac{1}{N_2},\frac{2}{N_2},\cdots,\frac{N_2-1}{N_2}), 
%\end{equation} 

%Here, $L$ can be viewed as the delay for source (impulse) detection in the inverse problem. It is not hard to see that, there must exist an important trade-off between $M$ and $L$: a smaller $L$ requires a higher $M$, and vice versa. Mathematically, for all Fourier coefficients $\bm{\alpha}_{t_0}$ to be uniquely determined, the minimum number of spatial sampling locations is bounded from below by a function of $L$:
%\begin{equation} \label{eq:fstar}
%M \geq f^*(L)
%\end{equation}
%where $f^*$ is a convex function which is monotone decreasing in $L$. The relationship (\ref{eq:fstar}) is established in Section *.*. 

%Without loss of generality, $t_0=0$.

Next, consider a sensor network with $M$ sensors at spatial locations $\bm{s}_1,\bm{s}_2,...,\bm{s}_M$. Let a column vector $\bm{Y}(l) = (Y(l,\bm{s}_1),Y(l,\bm{s}_2),...,Y(l,\bm{s}_M))^T$ contain the observations arising from the advection-diffusion process (\ref{eq:SPDE1}) at time $l$ ($l=1,2,...,L$),  
%and on a rectangular spatial mesh system $\mathbb{S}$ defined by a tensor product of two one-dimensional collocation sets,
%\begin{equation} \label{eq:tensor_s}
%\mathbb{S} = (0,\frac{1}{N_1},\frac{2}{N_1},\cdots,\frac{N_1-1}{N_1}) \otimes (0,\frac{1}{N_2},\frac{2}{N_2},\cdots,\frac{N_2-1}{N_2}). 
%\end{equation} 
and let a $M\times L$ matrix $\bm{Y}$ be a collection of the observations from the $L$ time periods: $\bm{Y} = [\bm{Y}(1), \bm{Y}(2), \cdots,  \bm{Y}(L)]$. 
%\vspace{-12pt}
%\begin{equation} 
%\bm{Y} = \begin{bmatrix}
%| & | &  & |\\ 
%\bm{Y}(1) & \bm{Y}(2) & \cdots & \bm{Y}(L)\\ 
%| & | &  & |
%\end{bmatrix}
%\end{equation}
%\begin{equation} 
%\bm{Y} = [\bm{Y}(1), \bm{Y}(2), \cdots,  \bm{Y}(L)]
%\end{equation}
%\vspace{-12pt}
Then, a spatio-temporal model based on the PDE (\ref{eq:SPDE1}) is proposed in \cite{Liu2020}:
\vspace{-18pt}
\begin{equation} \label{eq:Y}
\bm{Y} = \bm{F}\bm{E}\bm{G} + \bm{V}.
\end{equation}

\vspace{-28pt}
\noindent Here, 

\vspace{-6pt}
$\bullet$ $\bm{F}$ is an $M\times N$ matrix of the Fourier basis functions ($N=N_1\times N_2$), $\bm{F} = (
	\bm{f}_{\bm{k}_1} , \bm{f}_{\bm{k}_2} , \cdots , \bm{f}_{\bm{k}_N})$
%	\begin{equation} \label{eq:F}
%	\bm{F} = \begin{bmatrix}
%	| & | &  & |\\ 
%	\bm{f}_{\bm{k}_1} & \bm{f}_{\bm{k}_2} & \cdots & \bm{f}_{\bm{k}_N}\\ 
%	| & | &  & |
%	\end{bmatrix}
%	\end{equation}
%\vspace{-12pt}
%	\begin{equation} \label{eq:F}
%	\bm{F} = (
%	\bm{f}_{\bm{k}_1} , \bm{f}_{\bm{k}_2} , \cdots , \bm{f}_{\bm{k}_N})
%	\end{equation}
%	
%	\vspace{-12pt}
%	\noindent 
	where $\bm{f}_{\bm{k}_i} = (e^{\imath 2\pi \bm{s}_1^T \bm{k}_i}, e^{\imath 2\pi \bm{s}_2^T \bm{k}_i}, \cdots, e^{\imath 2\pi \bm{s}_M^T \bm{k}_i})^T$ and $\imath$ is the imaginary unit. 
%	\begin{equation} \label{eq:F}
%	\bm{F} = 
%	\begin{pmatrix} e^{\imath 2\pi \bm{s}_1^T \bm{k}_1} & e^{\imath 2\pi \bm{s}_1^T \bm{k}_2} & \cdots & e^{\imath 2\pi \bm{s}_1^T \bm{k}_N}\\ 
%	e^{\imath 2\pi \bm{s}_2^T \bm{k}_1} & e^{\imath 2\pi \bm{s}_2^T \bm{k}_2} & \cdots & e^{\imath 2\pi \bm{s}_2^T \bm{k}_N} \\
%	\vdots & \vdots & \ddots & \vdots \\
%	e^{\imath 2\pi \bm{s}_M^T \bm{k}_1} & e^{\imath 2\pi \bm{s}_M^T \bm{k}_2} & \cdots & e^{\imath 2\pi \bm{s}_M^T \bm{k}_N}
%	\end{pmatrix}
%	\end{equation}

$\bullet$ $\bm{E}=\mathrm{diag}(\bm{\alpha}(0,\bm{k}))=\mathrm{diag}(\bm{\eta})$ is a $N\times N$ matrix of the spectral coefficients at time $0$, and $\bm{\eta}$ is a vector that contains $\eta(\bm{k})$ for all $\bm{k}\in\mathcal{K}$.

$\bullet$ $\bm{G}$ is a $N\times L$ matrix,  $\bm{G} = (\bm{g}(1) , \bm{g}(2) , \cdots , \bm{g}(L))$,
%	\begin{equation} \label{eq:G}
%	\bm{G} = \begin{bmatrix}
%	| & | &  & |\\ 
%	\bm{g}(1) & \bm{g}(2) & \cdots & \bm{g}(L)\\ 
%	| & | &  & |
%	\end{bmatrix}
%	\end{equation}
%\vspace{-6pt}
%	\begin{equation} \label{eq:G}
%	\bm{G} = 
%	(\bm{g}(1) , \bm{g}(2) , \cdots , \bm{g}(L))
%	\end{equation}
%	
%	\vspace{-12pt}
%	\noindent 
	which captures the temporal evolution of the elements in $\bm{E}$. Here,  $\bm{g}(l)=(g_{1}(l),...,g_{N}(l))^T$ is a column vector where  $g_{j}(l)=e^{\gamma_{j}(l-1)\Delta}$, $\Delta$ is the sampling interval in time, and $\gamma_{j}= -\bm{k}^T_j \bm{D}\bm{k}_j - \zeta - \imath  \bm{\vec{v}}^T \bm{k}_j$ for $j=1,2,\cdots,N$.
	
%	\vspace{-12pt}
%	\begin{equation} \label{eq:gamma}
%	%\begin{split}
%	%\gamma_{j}= &  \int_{\mathbb{S}}\left(-\bm{k}^T_j \bm{D}_{\bm{s}}\bm{k}_j - \zeta_{\bm{s}} - \imath  (\bm{\vec{v}}_{\bm{s}}^T \bm{k}_j - [\triangledown\cdot \bm{D}_{\bm{s}}]\bm{k}_j)    \right)d\bm{s}
%	\gamma_{j}= -\bm{k}^T_j \bm{D}\bm{k}_j - \zeta - \imath  \bm{\vec{v}}^T \bm{k}_j,     
%	\quad\quad j=1,2,...,N.
%	%\end{split}
%	\end{equation}
	%	\item $\bm{W}$ is a $N\times L$ matrix, $\bm{W}=(\bm{w}_1,\bm{w}_2,...,\bm{w}_L)$, where $\bm{w}_i \sim N(0,\bm{\Sigma}_{\bm{w}}^{(i)})$ and 
	%	\begin{equation} \label{eq:Sigma_w}
	%	\bm{\Sigma}_{\bm{w}}^{(i)} = \int_{0}^{(i-1)\Delta} \exp(\mathrm{diag}(\gamma_{j})(\Delta-\tau))\bm{H}\exp^*(\mathrm{diag}(\gamma_{j})(\Delta-\tau))
	%	\end{equation}
	%	where $\bm{H}=\mathrm{diag}(\tilde{h}(\bm{k}_j))$ with $\tilde{h}$ being the spectral density. $\bm{W}$ corresponds to the error process $\varepsilon_t(\bm{s})$ in (\ref{eq:SPDE1}) +
	
	%\vspace{-16pt}
$\bullet$ $\bm{V}=(\bm{v}(1),\bm{v}(2),...,\bm{v}(L))$ is a $M\times L$ matrix that captures the measurement error, and $\bm{v}(l)$ is multivariate Gaussian, $N_M(0,\bm{\Sigma}_{\bm{v}})$, for $l=1,2,...,L$.

%(Here, need to defined $\bm{D}$, and $\bm{x}$)Hence, let $\bm{x}_i$, $i=1,2,...,M$ be a number of $M$ locations selected, 

%Let $\bm{D}$ be a $M\times N$ selection matrix that defines the $M$ selected spatial locations. For any row of $\bm{D}$, only the $j$th element is 1 if the location $\bm{s}_j$ is selected, while all other elements in this row are zero. 
%Let $\bm{\tilde{V}}=\bm{F}\bm{W} + \bm{V}$, we express model (\ref{eq:Y}) as follows:
%\begin{equation} \label{eq:Y_tilde}
%\bm{Y} = \bm{F} \bm{A} \bm{G} + \bm{\tilde{V}}.
%\end{equation}    

Readers may refer to  \cite{Sigrist2015} and \cite{Liu2020} for details of (\ref{eq:Y}).
%while the latter allows the operator $\mathcal{A}$ to vary in space (the non-linear case). 
The model (\ref{eq:Y}) is based on the classical solution of nonlinear dynamical systems using the spectral theory and eigenfunction expansions, and serves as the foundation based on which the inverse models are to be established in this paper.
%\subsection{The Sufficient and Necessary Conditions}
%We first present the inverse problem under the general case by focusing on the irregular sampling scheme shown in Figure \ref{fig:grid1}(a). In other words, we are concerned with the c collected from an irregular grid (with $M$ sensors) for $L$ discrete time periods.

\vspace{-12pt}
\subsection{The Inverse Problem and Its Basic Properties} \label{sec:property}
 In an inverse modeling problem considered in this paper, $\bm{Y}$ is the sensor observation, both $\bm{F}$ and $\bm{G}$ are pre-computed, and the goal is to estimate the coefficient vector $\bm{\eta}$ that determines the initial condition $\xi(0, \bm{s})$. %In statistics, a model is identifiable if it is theoretically possible to learn the true values of underlying parameters from an infinite number of observations. As discussed above, 
Note that, the spectral coefficients $\bm{\eta}$ may not be uniquely determined given
insufficient observations over space and time and particular sensor network layouts (known as spectral aliasing). Proposition 1 firstly establishes necessary conditions for all components in $\bm{\eta}$ to be uniquely estimated. 

\vspace{-6pt}
\begin{proposition1*}
	Given the observations of the process (\ref{eq:SPDE1}) from $M$ sensors and for $L$ discrete time periods, all spectral coefficients in $\bm{\eta}$ can be uniquely estimated from the model (\ref{eq:Y}) if at least one of the following two conditions is met:
	
	Condition A: given the velocity $\bm{\vec{v}}$ and diffusivity $\bm{D}$, there exist no $\bm{k}_{j_1}$ and $\bm{k}_{j_2}$ ($\bm{k}_{j_1},\bm{k}_{j_2} \in \mathcal{K}$ and $j_1 \neq j_2$) such that
	\vspace{-20pt}
	\begin{equation} 
	\bm{\vec{v}}^T(\bm{k}_{j_1}-\bm{k}_{j_2}) = 0 \quad\text{and}\quad \bm{k}_{j_1}^T \bm{D} \bm{k}_{j_1} = \bm{k}_{j_2}^T \bm{D} \bm{k}_{j_2}.
	\end{equation} 
	
	\vspace{-16pt}
	Condition B: There exist at least two sampling locations $\bm{s}$ and $\bm{s}'$ such that neither of the following conditions holds:
	
	\vspace{-48pt}
	\begin{subequations}
    \begin{align}
      2\bm{k}_{j_1}(\bm{s}-\bm{s}') \in \mathbb{Z}^{(\text{odd})},\quad 2\bm{k}_{j_2}(\bm{s}-\bm{s}') \in \mathbb{Z}^{(\text{odd})} \\
      2\bm{k}_{j_1}(\bm{s}-\bm{s}') \in \mathbb{Z}^{(\text{even})},\quad 2\bm{k}_{j_2}(\bm{s}-\bm{s}') \in \mathbb{Z}^{(\text{even})}.
    \end{align}
  \end{subequations}

\end{proposition1*}

\vspace{-18pt}
All proofs are presented in the Supplemental Materials.  As shown by this proposition, whether all spectral coefficients in $\bm{\eta}$ can be uniquely estimated depends on key physical parameters of the underlying process, such as the velocity $\bm{\vec{v}}$ and diffusivity $\bm{D}$ as one might naturally expect. 

Next, we investigate the sufficient condition for all components in $\bm{\eta}$ to be uniquely estimated. In general, the sufficient condition requires either sufficiently large spatial observations (i.e., large $M$), or sufficiently large temporal observations (i.e., large $L$), or both. This is intuitively true and detailed discussions are presented as follows. 

(\textbf{When the spatial observations are large}). If $M\geq N$, the left inverse of $\bm{F}$ exists. Let $\bm{F}_{L}^{-1}$ be the left inverse of $\bm{F}$, (\ref{eq:Y}) can be re-written as $[\bm{F}_{L}^{-1}\bm{Y}]^T = \bm{G}^T \bm{E}^T + [\bm{F}_{L}^{-1}\bm{V}]^T$. 
%\begin{equation} 
%\bm{F}_{L}^{-1}\bm{Y} = \bm{A} \bm{G} + %\bm{F}_{L}^{-1}\bm{\tilde{V}}
%\end{equation} 
%and
%\vspace{-18pt}
%\begin{equation} 
%[\bm{F}_{L}^{-1}\bm{Y}]^T = \bm{G}^T \bm{A}^T + [\bm{F}_{L}^{-1}\bm{V}]^T.
%\end{equation} 
Hence, let $\bm{G}^T_{\cdot,j}$ be the $j$th column of $\bm{G}^T$, $j=1,2,...,N$, we have
\vspace{-16pt}
\begin{equation} \label{eq:Y_left_vec}
\mathrm{vec}([\bm{F}_{L}^{-1}\bm{Y}]^T) =
\mathrm{diag}( \{ \bm{G}^T_{\cdot,j}\}_{j=1}^{N})\bm{\eta} + 
\mathrm{vec}([\bm{F}_{L}^{-1}\bm{\tilde{V}}]^T)
\end{equation}

\vspace{-12pt}
\noindent where $\text{vec}(\cdot)$ denotes matrix vectorization, and $\mathrm{diag}( \{ \bm{G}^T_{\cdot,j}\}_{j=1}^{N})$ is a block diagonal matrix with the column vector $\bm{G}^T_{\cdot,j}$ being its  $j$th block.
The exponential structure of $g_{i,j}$ in $\bm{G}$ guarantees that the $LN \times N$ matrix $\mathrm{diag}( \{ \bm{G}^T_{\cdot,j}\}_{j=1}^{N})$ is full column rank, and all elements in $\bm{\eta}$ can be uniquely determined. Note that, when $M\geq N$, the sampling frequency in space exceeds the Nyquist frequency---the largest bandwidth that can be sampled without aliasing. 

%In fact, for any given number of spatial samples $M$, as long as the number of temporal observations $L$ is not less than the row rank of $\bm{G}$, all component in $\bm{\eta}$ can be uniquely identified. The sufficient condition is formally presented in Proposition 2. 

(\textbf{When the temporal observations are large}). 
A large value of $L$ corresponds to another scenario where the temporal samples are abundant. By examining the expression of $\gamma_j$ in (\ref{eq:Y}), it is possible to find $j$ and $j'$ ($j \neq j'$) such that $\gamma_{j} = \gamma_{j'}$. In other words, it is possible that $\bm{G}$ is row rank deficient with identical rows. 

Let $\{1,2,...,N\} \overset{\mathcal{M}}{\rightarrow}\Psi =\{\Psi_1,\Psi_2,...,\Psi_{\tilde{N}} \}$
%\vspace{-8pt}
%\begin{equation} \label{eq:mapping}
%\{1,2,...,N\} \overset{\mathcal{M}}{\rightarrow}\Psi =\{\Psi_1,\Psi_2,...,\Psi_{\tilde{N}} \}
%\end{equation}
%\vspace{-12pt}
%\noindent 
be a mapping where $\Psi_i$ ($i=1,2,...,\tilde{N}$) is a set such that $\gamma_{j}=\gamma_{j'}$ for $j,j'\in\Psi_i$.
In other words, the mapping $\mathcal{M}$ defines a partition of $\bm{G}$ where each partition contains identical rows, and the row rank of $\bm{G}$ is given by $\tilde{N}$.

%\vspace{8pt}
\begin{proposition2*}
	If the number of temporal samples $L$ is greater than the (row) rank of $\bm{G}$, i.e., $L>\tilde{N}$, the sufficient condition for all components in $\bm{\eta}$ to be uniquely determined is
	\vspace{-16pt}
	\begin{equation} \label{eq:rank}
	\mathrm{rank}(\tilde{\bm{F}}_i) = |\Psi_i|, \quad\quad \forall  i=1,...,\tilde{N}
	\end{equation}
	
	\vspace{-16pt}
	\noindent where $\tilde{\bm{F}}_i = \{f_{m,j}\}_{m=1,...,M, j\in\Psi_i}$ is a $M\times|\Psi_i|$ matrix, and $|\Psi_i|$ represents the cardinality of the set $\Psi_i$ for $i=1,...,\tilde{N}$.
\end{proposition2*}

%for any $j=1,2,...,N$, there exist at least two sampling locations, $\bm{s}_{m}$ and $\bm{s}_{m'}$, such that %$\pi(\bm{s}_{m}-\bm{s}_{m'})^T\bm{k}_j \notin \mathbb{Z}$.
%\begin{equation} 
%\pi(\bm{s}_{m}-\bm{s}_{m'})^T\bm{k}_j %\notin \mathbb{Z}.
%\end{equation} 

\vspace{-12pt}
\subsection{A Regularized Inverse Problem} \label{sec:problem}
Propositions 1 and 2 above establish sufficient and necessary conditions for all components in $\bm{\eta}$ to be uniqued determined from spatially distributed sensor data streams. Despite the theoretical values rooted in the two propositions, real applications may not always require all components in $\bm{\eta}$ to be uniquely estimated. For example, the numbers of spatial/temporal samples as well as the locations where sensors can be deployed are always subject to practical constraints. 
%In addition, $\bm{\eta}$ can be sparse as the spectral coefficients corresponding to the high-frequency modes are often close to zero. 
Hence, we often require $\bm{\eta}$ to be estimated under the scenarios where neither sufficient nor necessary condition is met. 

We first re-write (\ref{eq:Y}) and define $\vec{\bm{\mathscr{Y}}}$,  $\bm{\mathscr{X}}^{(\text{P-I})}$, and $\vec{\bm{\mathscr{V}}} \sim N(\bm{0},\bm{\Sigma})$ as follows:
%\begin{equation} 
%\mathrm{vec}(\bm{Y}) = \text{diag}( \{\bm{F}\tilde{\bm{g}}(l)\}_{l=1}^L)\bm{\eta} + \mathrm{vec}(\bm{V})
%\end{equation} 
\vspace{-12pt}
\begin{equation} \label{eq:p1_proof_1}
\underbrace{\begin{pmatrix}
	\bm{Y}(1)\\ 
	%\bm{Y}(2)\\ 
	\vdots \\ 
	\bm{Y}(L)
	\end{pmatrix}}_{\vec{\bm{\mathscr{Y}}} } = 
\underbrace{\begin{pmatrix}
	\bm{F} \text{diag}(  \{ \bm{g}_{j}(1) \}_{j=1}^{N})\\ 
	%\bm{F} \text{diag}(  \{ \bm{g}_{j}(2) \}_{j=1}^{N})\\ 
	\vdots \\ 
	\bm{F} \text{diag}(  \{ \bm{g}_{j}(L) \}_{j=1}^{N})
	\end{pmatrix}}_{\bm{\mathscr{X}}^{(\text{P-I})}} \bm{\eta} + 
\underbrace{\begin{pmatrix}
	\bm{v}(1)\\ 
	%\bm{v}(2)\\ 
	\vdots \\ 
	\bm{v}(L)
	\end{pmatrix}}_{\vec{\bm{\mathscr{V}}} } 
\end{equation} 
%Let $\vec{\bm{\mathscr{Y}}}$,  $\bm{\mathscr{X}}^{(\text{P-I})}$, and $\vec{\bm{\mathscr{V}}}$ be defined in (\ref{eq:p1_proof_1}), we re-write (\ref{eq:p1_proof_1}) as, 
%\begin{equation} 
%\vec{\bm{\mathscr{Y}}} = \bm{\mathscr{X}}^{(\text{P-I})}\bm{\eta} + \vec{\bm{\mathscr{V}}}, \quad\quad  \vec{\bm{\mathscr{V}}} \sim N(\bm{0},\bm{\Sigma})
%\end{equation} 

\vspace{-12pt}
\noindent and consider a regularized inverse problem:
\vspace{-12pt}
\begin{equation} \label{eq:PI}
\text{Problem P-I:}\quad\quad
\text{min} \quad \frac{1}{2} (\vec{\bm{\mathscr{Y}}} - \bm{\mathscr{X}}^{(\text{P-I})}\bm{\eta})^T \bm{\Sigma}^{-1} (\vec{\bm{\mathscr{Y}}} - \bm{\mathscr{X}}^{(\text{P-I})}\bm{\eta}) + \mathcal{R}(\bm{\eta}).
\end{equation}

\vspace{-12pt}
Here, the choice of the regularization $\mathcal{R}(\bm{\eta})$ is motivated by two considerations: (\textbf{i}) For smooth or band-limited initial condition (see (\ref{eq:source_Fouier})), the high-frequency modes decay rapidly, i.e., the energy is concentrated in the low-frequency region. This motivates us to impose an $L_1$-regularization to $\bm{\eta}$ (i.e., the \textit{sparsity} of $\bm{\eta}$). (\textbf{ii}) For smooth initial conditions, it is expected that the components in $\bm{\eta}$ gradually decay (not necessarily monotone). This motivates us to impose some level of \textit{smoothness} among the \textit{adjacent} components in $\bm{\eta}$ corresponding to adjacent frequencies to prevent sudden spikes of the estimated special coefficients. %$\bullet$ The \textit{sparsity} of $\bm{\eta}$. For smooth or band-limited initial condition (see (\ref{eq:source_Fouier})), the high-frequency modes decay rapidly, i.e., the energy is concentrated in the low-frequency region. This motivates us to impose the $L_1$-regularization to $\bm{\eta}$. 
%$\bullet$ The \textit{smoothness} of the differences among the components in $\bm{\eta}$. For smooth initial conditions, it is expected that the components in $\bm{\eta}$ gradually decay (not necessarily monotone). This motivates us to impose some level of smoothness among the components in $\bm{\eta}$ corresponding to \textit{adjacent} frequencies to prevent sudden spikes of the estimated special coefficients. 
In particular, an $L_2$-regularization is imposed on the difference between the connected components of $\bm{\eta}$ in both directions (horizontal and vertical) such that
\vspace{-12pt}
\begin{equation} \label{eq:fused}
\left \|  \bm{J}_1 \bm{\eta}   \right\|_2^2 = \sum_{i,j \in \mathcal{K}^{(h)}}  (\eta(\bm{k}_i), \eta(\bm{k}_j))^2, \quad \left \|  \bm{J}_2 \bm{\eta}   \right\|_2^2 = \sum_{i,j \in \mathcal{K}^{(v)}}  (\eta(\bm{k}_i), \eta(\bm{k}_j))^2
\end{equation}

\vspace{-8pt}
\noindent where $\bm{J}_1$ and $\bm{J}_2$ are the matrix difference operators in the horizontal and vertical directions, and the set $\mathcal{K}^{(h)}$ and $\mathcal{K}^{(v)}$ consist of all frequencies $\bm{k} \in \mathcal{K}$ which are connected in the horizontal and vertical directions, i.e., $\mathcal{K}^{(h)} = \{ \bm{k}_i, \bm{k}_j; k_{j,1}-k_{i,1}=1, i<j \}$ and $\mathcal{K}^{(v)} = \{ \bm{k}_i, \bm{k}_j; k_{j,2}-k_{i,2}=1, i<j \}$. 
%\vspace{-12pt} \noindent where $\bm{J}_2$ is a matrix difference operator in the vertical direction, and the set $\mathcal{K}^{(v)}$ consists of all frequencies $\bm{k} \in \mathcal{K}$ which are connected in the second (vertical) direction, i.e., $\mathcal{K}^{(h)} = \{ \bm{k}_i, \bm{k}_j; k_{j,2}-k_{i,2}=1, i<j \}$.  
Note that, (\ref{eq:fused})  modifies the idea of Fussed Lasso \citep{Tibshirani2005}. The difference is that, Fussed Lasso involves an $L_1$-regularization to the differences among the coefficients that leads to a sparse and piecewise constant solution, while it is  appropriate for us to consider an $L_2$-regularization such that the components in $\bm{\eta}$ can smoothly change between the high-frequency and low-frequency regions.

Finally, the regularization $\mathcal{R}(\bm{\eta})$ in (\ref{eq:PI}) is given by
\vspace{-16pt}
\begin{equation} \label{eq:R1}
\mathcal{R}(\bm{\eta}) = \lambda_1 \left \| \bm{\eta} \right\|_1 + \lambda_2 \left \|  \bm{J} \bm{\eta}   \right\|_2^2
\end{equation}

\vspace{-16pt}
\noindent where $\bm{J} = (\bm{J}_1^T, \bm{J}_2^T)^T$ is a 2D difference operator, and $\lambda_1$ and $\lambda_2$ respectively control the sparsity in $\bm{\eta}$ and the smoothness among the adjacent components in $\bm{\eta}$. 
The inverse problem (\ref{eq:PI}) can be solved by the Alternating Direction Method of Multipliers (ADMM) \citep{Zou2005, Ramdas2016}. Details of the ADMM algorithm for our problems are provided in Section \ref{sec:ADMM}.

\vspace{-18pt}
\section{Two Special Cases} \label{sec:special}
\vspace{-6pt}
In this section, we further investigate two special sampling schemes, i.e., non-uniform sampling  and shifted uniform sampling as discussed in Section \ref{sec:problem}, and show that computationally efficient solutions are possible in the spectral domain under the two special schemes.  
%For both sampling schemes, we propose to solve the inverse problems in the \textit{spectrum} domain.

\vspace{-18pt}
\subsection{Non-Uniform Sampling} \label{sec:nu}
\vspace{-8pt}
Consider a rectangular mesh system given by a tensor product of two one-dimensional collocation sets, $\tilde{\mathcal{M}} = \tilde{\mathcal{M}}_1 \otimes \tilde{\mathcal{M}}_2$, 
%\begin{equation} 
%\tilde{\mathcal{M}} = \tilde{\mathcal{M}}_1 \otimes \tilde{\mathcal{M}}_2
%\end{equation} 
where $\tilde{\mathcal{M}}_1 = \{ m_1; m_1=0,1,...,\tilde{M}_1-1 \}$ and $\tilde{\mathcal{M}}_2 = \{ m_2; m_2=0,1,...,\tilde{M}_2-1 \}$ 
are the sets of collocation points.
Here, $\tilde{\mathcal{M}}$ is a mesh system consisting of the candidate locations where sensors can potentially be deployed. 
Let $\mathcal{M}_1  \subseteq \tilde{\mathcal{M}}_1$ and $\mathcal{M}_2 \subseteq  \tilde{\mathcal{M}}_2$, a non-uniform sampling grid is given by a mesh system ${\mathcal{M}} = {\mathcal{M}}_1 \otimes {\mathcal{M}}_2$
%\vspace{-12pt}
%\begin{equation} 
%{\mathcal{M}} = {\mathcal{M}}_1 \otimes {\mathcal{M}}_2
%\end{equation} 
%We consider a problem where sensors are deployed at locations selected from the following set:
%\begin{equation} 
%\mathcal{M} = \left\{ \left(\frac{m_1}{M_1}, \frac{m_2}{M_2}\right)^T; m_1=0,1,...,M_1-1, m_2=0,1,...,M_2-1 \right\}.
%\end{equation} 
%\vspace{-12pt}
%\noindent 
where $|{\mathcal{M}}_1|={M}_1$, $|{\mathcal{M}}_1|={M}_2$, and $|\cdot|$ represents the cardinality of a set. 
%where $\tilde{\mathcal{M}}_1 = \{ \tilde{m}_1; \tilde{m}_1=0,1,...,\tilde{M}_1-1 \}$ and $\mathcal{M}_2 = \{ \tilde{\mathcal{M}}_2; \tilde{m}_2=0,1,...,\tilde{M}_2-1 \}$ 
%be a set of selected locations where sensors are deployed,
Let $y(l, {\bm{m}})$ represent the observation at time $l$ and from location ${\bm{m}} \in {\mathcal{M}}$. 
Then, the non-uniform discrete Fourier transform of type II (NUDFT-II) of $y(l, {\bm{m}})$ is:
\vspace{-12pt}
\begin{equation} \label{eq:beta}
\beta(l, \bm{q}) = \frac{1}{|{\mathcal{M}}|}\sum_{{\bm{m}}\in {\mathcal{M}}}y(l, {\bm{m}}) e^ { -\imath 2 \pi {\bm{m}}^T \bm{q} }, \quad\quad \text{for }\bm{q} \in \mathcal{Q}, 
\end{equation} 

\vspace{-12pt}
\noindent where $\mathcal{Q}=\left\{ (q_1, q_2)^T; q_1 = -\frac{{M}_1}{2}+1, -\frac{{M}_1}{2}+2, \cdots, \frac{{M}_1}{2}, q_2 = -\frac{{M}_2}{2}+1, -\frac{{M}_2}{2}+2, \cdots, \frac{{M}_2}{2} \right\}$. 
%\vspace{-12pt}
%\begin{equation} \label{eq:Q}
%\mathcal{Q}=\left\{ (q_1, q_2)^T; q_1 = -\frac{{M}_1}{2}+1, -\frac{{M}_1}{2}+2, \cdots, \frac{{M}_1}{2}, q_2 = -\frac{{M}_2}{2}+1, -\frac{{M}_2}{2}+2, \cdots, \frac{{M}_2}{2} \right\}
%\end{equation} 

%\vspace{-12pt}
Replacing $y(l, {\bm{m}})$ in (\ref{eq:beta}) by its discrete Fourier transform over the domain $\mathcal{K}$, we have
\vspace{-12pt}
\begin{equation} \label{eq:beta1_nu}
\begin{split}
\beta(l, \bm{q}) 
& = \frac{1}{|{\mathcal{M}}|}\sum_{{\bm{m}}}
\left\{ \sum_{\bm{k} \in \mathcal{K}} \left[\alpha(l, \bm{k})+\varepsilon(l, \bm{k})\right] e^{ \imath 2 \pi {\bm{m}}^T \bm{k} }   \right\} e^{ -\imath 2 \pi {\bm{m}}^T \bm{q} } \\
& = \sum_{i \in \mathcal{I}_{\bm{q}}}\sum_{i \in \mathcal{J}_{\bm{q}}}\left[\alpha(l, \bm{q}+(i{M}_1,j{M}_2)^T)+\varepsilon(l, \bm{q}+(i{M}_1,j{M}_2)^T)\right] 
%& \equiv \sum_{i \in \mathcal{I}_{\bm{q}}}\sum_{i \in \mathcal{J}_{\bm{q}}} \tilde{\alpha}(t, \bm{q}+(iM_1,jM_2)^T).
\end{split}
\end{equation} 

\vspace{-16pt}
\noindent where $\varepsilon$ is due to the observation error, and the sets $\mathcal{I}_{\bm{q}}$ and $\mathcal{J}_{\bm{q}}$ are respectively given by
\vspace{-12pt}
\begin{equation} \label{eq:IJsets}
\begin{split} 
&\mathcal{I}_{\bm{q}} = \left\{ i; -\frac{N_1}{2}+1 \leq (q_1 + i{M}_1) \leq \frac{N_1}{2}, i \in \mathbb{Z} \right\} \\
&\mathcal{J}_{\bm{q}} = \left\{ j; -\frac{N_2}{2}+1 \leq (q_2 + j{M}_2) \leq \frac{N_2}{2}, j \in \mathbb{Z} \right\}.
\end{split}
\end{equation} 

\vspace{-12pt}
Here, the first line of (\ref{eq:beta1_nu}) is obtained by directly inserting the Fourier  transform of $y(l, {\bm{m}})$ into (18). The second line of (\ref{eq:beta1_nu}) is obtained by invoking the well-known orthogonal properties of Fourier bases, i.e., $\sum_{\bm{k} \in \mathcal{K}} \left[\alpha(l, \bm{k})+\varepsilon(l, \bm{k})\right] e^{ \imath 2 \pi {\bm{m}}^T (\bm{k}-\bm{q}) } = 1$ only when $\bm{k} = \bm{q} + (i{M}_1,j{M}_2)^T$ where $i \in \mathcal{I}_{\bm{q}}$ and $j \in \mathcal{J}_{\bm{q}}$; otherwise $\sum_{\bm{k} \in \mathcal{K}} \left[\alpha(l, \bm{k})+\varepsilon(l, \bm{k})\right] e^{ \imath 2 \pi {\bm{m}}^T (\bm{k}-\bm{q}) } = 0$. Hence, given a mesh system ${\mathcal{M}}$ (i.e., the spatial locations where data are collected), (\ref{eq:beta1_nu}) implies that $\beta(l, \bm{q})$ is given by the sum of multiple Fourier coefficients $\alpha(l, \bm{q}+(i{M}_1,j{M}_2)^T)$ where $i \in \mathcal{I}_{\bm{q}}$ and $j \in \mathcal{J}_{\bm{q}}$. In other words, it is not possible to uniquely estimate $\alpha(l, \bm{q}+(i{M}_1,j{M}_2)^T)$ for all $i \in \mathcal{I}_{\bm{q}}$ and $j \in \mathcal{J}_{\bm{q}}$ from $\beta(l, \bm{q})$. This is known as \textit{spectral aliasing}. 

To reveal the spectral aliasing structure clearly, we introduce a set $\mathcal{K}_{\bm{q}}$
\vspace{-16pt}
\begin{equation} 
\mathcal{K}_{\bm{q}}  = \left\{(k_1,k_2)^T; k_1=q_1+i{M}_1, k_2=q_2+j{M}_2, i \in \mathcal{I}_{\bm{q}}, j \in \mathcal{J}_{\bm{q}} \right\} 
\end{equation}

\vspace{-16pt}
\noindent that consists of all wavenumbers in $\mathcal{K}$ corresponding to $\bm{q}\in \mathcal{Q}$. Obviously, the Fourier coefficients corresponding to wavenumbers in $\mathcal{K}_{\bm{q}}$ are all confounded, and cannot be uniquely determined unless the number of temporal observations $L$ is sufficient (see Proposition 2). 
Consider a simple illustrative example where $N_1=N_2=4$ and ${M}_1={M}_2=2$, and define four sets: $\mathcal{K}_{(0,0)^T}  = \{ (0,0)^T, (0,2)^T, (2,0)^T, (2,2)^T\}$, $\mathcal{K}_{(0,1)^T}  = \{ (0,1)^T, (0,-1)^T, (2,1)^T, (2,-1)^T\}$, $\mathcal{K}_{(1,0)^T}  = \{ (1,0)^T, (-1,0)^T, (1,2)^T, (-1,2)^T\}$ and $\mathcal{K}_{(1,1)^T}  = \{ (1,1)^T, (-1,-1)^T, (-1,1)^T, (1,-1)^T \}$
%\begin{equation} 
%\begin{split}
%& \mathcal{K}_{(0,0)^T}  = \{ (0,0)^T, (0,2)^T, (2,0)^T, (2,2)^T\} \\
%& \mathcal{K}_{(0,1)^T}  = \{ (0,1)^T, (0,-1)^T, (2,1)^T, (2,-1)^T\} \\
%& \mathcal{K}_{(1,0)^T}  = \{ (1,0)^T, (-1,0)^T, (1,2)^T, (-1,2)^T\} \\
%& \mathcal{K}_{(1,1)^T}  = \{ (1,1)^T, (-1,-1)^T, (-1,1)^T, (1,-1)^T \} \\
%\end{split} 
%\end{equation}
such that each set consists of the wavenumbers in $\mathcal{K}$ whose corresponding Fourier coefficients are confounded when the number of temporal samples is insufficient. Note that,  $\bigcup \mathcal{K}_{\bm{q}} = \mathcal{K}$ and $\mathcal{K}_{\bm{q}} \bigcap \mathcal{K}_{\bm{q}'} = 0$ for $\bm{q} \neq \bm{q}'$, i.e., $\mathcal{K}_{\bm{q}}$ are mutually exclusive and exhaustive. 

Substituting the temporal dynamics of $\bm{\alpha}$ in (\ref{eq:Y}) into (\ref{eq:beta1_nu}), we obtain
\vspace{-14pt}
\begin{equation} \label{eq:beta2_nu}
\beta(l, \bm{q})
=
\bm{1}^T \mathrm{diag}(\bm{\eta}_{\bm{q}}) \bm{g}_{\bm{q}}(l)
+
\bm{1}^T \bm{\varepsilon}_{\bm{q}}(l), \quad\quad l = 1,2,...,L,
\end{equation} 

\vspace{-14pt}
\noindent where $\bm{1}$ is a column vector of ones, $\bm{g}_{\bm{q}}(l)$, $\bm{{\varepsilon}}_{\bm{q}}(l)$ and $\bm{\eta}_{\bm{q}}$
are respectively the column vectors obtained from $\bm{g}(l)$,  $\bm{{\varepsilon}}(l)$ and $\bm{\eta}$ by keeping only the components corresponding to $\bm{k} \in \mathcal{K}_{\bm{q}}$. 

Combining $\beta(l, \bm{q})$ from all $L$ sampling times, it follows from (\ref{eq:beta2_nu}) that
\vspace{-8pt}
\begin{equation} 
\underbrace{\begin{pmatrix}
	\beta(1, \bm{q})\\ 
	\vdots \\ 
	\beta(L, \bm{q})
	\end{pmatrix}}_{\bm{\beta}_{\bm{q}}} = 
    \underbrace{\begin{pmatrix}
    	 \bm{1}^T &  &  & \\ 
    	&  & \ddots  & \\ 
    	&  &  &  \bm{1}^T
    	\end{pmatrix}}_{\bm{B}_{\bm{q}}} 
    \underbrace{\begin{pmatrix}
    \mathrm{diag}(\bm{g}_{\bm{q}}(1)) \\
    \vdots \\
    \mathrm{diag}(\bm{g}_{\bm{q}}(L)) 
    \end{pmatrix}}_{\bm{G}_{\bm{q}}}\bm{\eta}_{\bm{q}} + 
	\underbrace{\begin{pmatrix}
	\bm{1}^T &  &  & \\ 
	&  & \ddots  & \\ 
	&  &  &  \bm{1}^T
	\end{pmatrix}}_{\bm{B}_{\bm{q}}} 
	\underbrace{\begin{pmatrix}
		\bm{\varepsilon}_{\bm{q}}(1) \\
		\vdots \\
		\bm{\varepsilon}_{\bm{q}}(L)
		\end{pmatrix}}_{\bm{W}_{\bm{q}}}
\end{equation} 

\vspace{-18pt}
\noindent where $\bm{\beta}_{\bm{q}}$ is a $L \times 1$ column vector, $\bm{B}_{\bm{q}}=\mathrm{diag}\{ \bm{1}^T \}$ is a $L \times (L \times d_{\bm{q}})$ block diagonal matrix with $d_{\bm{q}}=|\mathcal{K}_q|$, and $\bm{G}_{\bm{q}}$ is a $(L \times d_{\bm{q}}) \times d_{\bm{q}}$ matrix. Hence, for any $\bm{q} \in \mathcal{Q}$, we obtain from (\ref{eq:beta2_nu}) a linear model

\vspace{-30pt}
\begin{equation} \label{eq:beta3_nu}
\bm{\beta}_{\bm{q}}
=
\bm{B}_{\bm{q}}
\bm{G}_{\bm{q}}
\bm{\eta}_{\bm{q}}
+
\bm{B}_{\bm{q}}
\bm{W}_{\bm{q}}, \quad\quad \bm{B}_{\bm{q}}\bm{W}_{\bm{q}} \sim \mathcal{N}(\bm{0},\sigma^2 d_{\bm{q}} \bm{I}).
\end{equation} 

\vspace{-12pt}
Note that, several factors determine if the components in $\bm{\eta}_{\bm{q}}$ can be uniquely determined from the linear model (\ref{eq:beta3_nu}), including the sensor network layout $\mathcal{M}$, the number of temporal samples $L$, as well as the parameters of the underlying physical process. The following proposition establishes the conditions for the components in $\bm{\eta}_{\bm{q}}$ to be uniquely estimated. 

\begin{proposition3*}
	For $\bm{k}_{j_1},\bm{k}_{j_2} \in \mathcal{K}_q$ and $j_1 \neq j_2$, if (i) $L \geq d_{\bm{q}}$, and (ii) at least one of the conditions,  $\bm{\vec{v}}^T(\bm{k}_{j_1}-\bm{k}_{j_2}) = 0$ and  $\bm{k}_{j_1}^T \bm{D} \bm{k}_{j_1} - \bm{k}_{j_2}^T \bm{D} \bm{k}_{j_2}=0$, does not hold, then, $\bm{B}_{\bm{q}}\bm{G}_{\bm{q}}$ is full column rank, and all spectral coefficients in $\bm{\eta}_{\bm{q}}$ can be uniquely estimated from (\ref{eq:beta3_nu}).
\end{proposition3*}	

\vspace{-12pt}
Although the proposition above suggests that it is possible to estimate $\bm{\eta}_{\bm{q}}$ from a system of linear models for all $\bm{q} \in \mathcal{Q}$, directly solving these individual linear models is rarely appropriate for the following reason: the matrix $\bm{B}_{\bm{q}}\bm{G}_{\bm{q}}$ can be easily ill-conditioned or computationally singular when both $\bm{\vec{v}}^T(\bm{k}_{j_1}-\bm{k}_{j_2})$ and  $\bm{k}_{j_1}^T \bm{D} \bm{k}_{j_1} - \bm{k}_{j_2}^T \bm{D} \bm{k}_{j_2}$ are close to zero. In other words, some columns in $\bm{B}_{\bm{q}}\bm{G}_{\bm{q}}$ can be near identical. 
Hence, we combine the linear models (\ref{eq:beta3_nu}) for all $\bm{q}_1, \bm{q}_2, ..., \bm{q}_{|\mathcal{Q}|}$ and obtain
\vspace{-8pt}
\begin{equation} \label{eq:beta_spectrum}
\underbrace{\begin{pmatrix}
	\bm{\beta}_{\bm{q}_1}\\ 
	\vdots \\ 
	\bm{\beta}_{\bm{q}_{|\mathcal{Q}|}}
	\end{pmatrix}}_{\bm{\vec{\mathscr{Y}}}^{(\text{P-II})}} = 
\underbrace{\begin{pmatrix}
	\bm{B}_{\bm{q}_1}\bm{G}_{\bm{q}_1}  &  &  & \\ 
	&  & \ddots  & \\ 
	&  &  &  \bm{B}_{\bm{q}_{|\mathcal{Q}|}}\bm{G}_{\bm{q}_{|\mathcal{Q}|}} 
	\end{pmatrix}}_{\mathscr{X}^{(\text{P-II})}}
\underbrace{\begin{pmatrix}
	\bm{\eta}_{\bm{q}_1} \\
	\vdots \\
	\bm{\eta}_{\bm{q}_{|\mathcal{Q}|}} 
	\end{pmatrix}}_{\bm{\eta}} + 
\underbrace{\begin{pmatrix}
	\bm{B}_{\bm{q}_1}\bm{W}_{\bm{q}_1}  \\ 
	\vdots   \\ 
\bm{B}_{\bm{q}_{|\mathcal{Q}|}}\bm{W}_{\bm{q}_{|\mathcal{Q}|}} 
	\end{pmatrix}}_{\mathscr{V}^{(\text{P-II})}}
\end{equation} 

\vspace{-12pt}
\noindent where ${\bm{\vec{\mathscr{Y}}}^{(\text{P-II})}}$ is a ($|\mathcal{Q}|\times L) \times 1$ column vector, $\mathscr{X}^{(\text{P-II})}$ is a ($|\mathcal{Q}|\times L) \times |\mathcal{K}|$ matrix, $\bm{\eta}$ is a $|\mathcal{K}| \times 1$ column vector, $\mathscr{V}^{(\text{P-II})} \sim \mathcal{N}(\bm{0}, \bm{\Sigma}^{(\text{P-II})})$ and $\bm{\Sigma}^{(\text{P-II})}=\sigma^2\text{diag}( \{d_{\bm{q}_i} \bm{I}\}_{i=1}^{|\mathcal{Q}|})$.

Similar to Problem P-I (\ref{eq:PI}), we again obtain a regularized inverse problem as follows:
\vspace{-12pt}
%\begin{equation}
%\text{Inverse Problem (P-II):}\quad\quad
%\text{argmin} \quad \frac{1}{2}\left \| \bm{\vec{\mathscr{\beta}}}^{(\text{P-II})} - \bm{\mathscr{X}}^{(\text{P-II})}\bm{\eta} \right \|_2^2 + \lambda_1 \left \| \bm{\eta} \right \|_1 + \lambda_2 \left \| \bm{\eta} \right \|_2^2
%\end{equation}
%where $\lambda_1$ and $\lambda_2$ are respectively the $\l_1$ sparsity and $\l_2$ stability constraints. 
\begin{equation}  \label{eq:PII}
\text{Problem P-II:    } \text{min } \frac{1}{2} (\bm{\vec{\mathscr{Y}}}^{(\text{P-II})} - \bm{\mathscr{X}}^{(\text{P-II})}\bm{\eta})^T (\bm{\Sigma}^{(\text{P-II})})^{-1} (\bm{\vec{\mathscr{Y}}}^{(\text{P-II})} - \bm{\mathscr{X}}^{(\text{P-II})}\bm{\eta}) + \mathcal{R}(\bm{\eta}).
\end{equation}

\vspace{-12pt}
\noindent where  %$\bm{\vec{\mathscr{Y}}}^{(\text{P-II})}=\bm{F}^{(\text{P-II})}\bm{\vec{\mathscr{\beta}}}^{(\text{P-II})}$ represents the non-uniform discrete Fourier transform, $\bm{F}^{(\text{P-II})}$ is the matrix operator for non-uniform Fourier transform, $\bm{\mathscr{X}}^{(\text{P-II})} = \bm{F}^{(\text{P-II})}\bm{X}^{(\text{P-II})}$, $\bm{\Sigma}^{(\text{P-II})} =\sigma^2 \bm{F}^{(\text{P-II})} \text{diag}( \{d_{\bm{q}_i} \bm{I}\}_{i=1}^{|\mathcal{Q}|}) (\bm{F}^{(\text{P-II})})^T$, and 
$\mathcal{R}(\bm{\eta})$ is defined in (\ref{eq:R1}).

\vspace{-12pt}
\subsection{Shifted Uniform Sampling} \label{sec:shift}
Shifted uniform sensor arrays or platforms  consists of two nested rectangular mesh systems (Figure \ref{fig:grid1}c), $\mathcal{M}^{(1)}$ and $\mathcal{M}^{(2)}$, which are respectively defined by the tensor product of two one-dimensional collocation sets:
\vspace{-12pt}
\begin{equation} 
\mathcal{M}^{(1)} = ( \frac{m_1}{M_1}; m_1=0,1,...,M_1-1) \otimes( \frac{m_2}{M_2}; m_2=0,1,...,M_2-1)
\end{equation} 
\vspace{-40pt}
\begin{equation} 
\mathcal{M}^{(2)} = (\frac{m_1}{M_1} + \delta_1; m_1=0,1,...,M_1-1) \otimes (\frac{m_2}{M_2}  + \delta_2; m_2=0,1,...,M_2-1)
\end{equation} 

\vspace{-8pt}
\noindent where $0 < \delta_1 < M_1^{-1}$ and $0 < \delta_2 < M_2^{-1}$, and $\bm{\delta} = (\delta_1, \delta_2)^T$ is the spatial shift between the two sensor platforms. 
%As an illustration, Figure \ref{fig:grid} shows a sampling network with $M_1=M_2=3$. The underlying process is defined on a $20\times20$ mesh system shown in the background, and data are sampled from 18 spatial locations (9 from $\mathcal{M}^{(1)}$ and 9 from $\mathcal{M}^{(2)}$). 
%\begin{figure}[h!]  
%	\begin{center}
%		\includegraphics[width=0.6\textwidth]{../code_inverse/figures/grid.png} 
%		\centering
%		\caption{Illustration of a sampling network that consists of two rectangular grids, $\mathcal{M}^{(1)}$ and $\mathcal{M}^{(2)}$. In this figure, ``1'' and ``2'' respectively represent the grid points from  $\mathcal{M}^{(1)}$ and $\mathcal{M}^{(2)}$. The underlying process is defined on a $20\times20$ mesh system shown in the background.}
%		\label{fig:grid}
%	\end{center}
%\end{figure}
Let $y^{(1)}(l, \bm{m})$ represent the observation at time $l$ and location $\bm{m}$ from $\mathcal{M}^{(1)}$, where $\bm{m} = (\frac{m_1}{M_1}, \frac{m_2}{M_2} )^T$. 
For any $\bm{q}=(q_1, q_2)^T\in\mathcal{Q}$, the Fourier coefficient at $\beta^{(1)}(l, \bm{q})$ based on the observations from the first mesh system $\mathcal{M}^{(1)}$ at time $l$ is:
\vspace{-12pt}
\begin{equation} \label{eq:beta_1}
\begin{split}
\beta^{(1)}(l, \bm{q}) & = \frac{1}{M_1M_2}\sum_{{\bm{m}}}y^{(1)}(l, \bm{m}) e^ { -\imath 2 \pi \bm{m}^T \bm{q} }  = \frac{1}{M_1M_2}\sum_{\bm{m}}
\left\{ \sum_{\bm{k} \in \mathcal{K}} \left[\alpha(l, \bm{k})+\varepsilon(l, \bm{k})\right] e^{ \imath 2 \pi \bm{m}^T \bm{k} }   \right\} e^{ -\imath 2 \pi \bm{m}^T \bm{q} } \\
& = \sum_{i \in \mathcal{I}_{\bm{q}}}\sum_{i \in \mathcal{J}_{\bm{q}}}\left\{\alpha(l, \bm{q}+(iM_1,jM_2)^T)+\varepsilon(l, \bm{q}+(iM_1,jM_2)^T)\right\}
%& \equiv \sum_{i \in \mathcal{I}_{\bm{q}}}\sum_{i \in \mathcal{J}_{\bm{q}}} \tilde{\alpha}(t, \bm{q}+(iM_1,jM_2)^T).
\end{split}
\end{equation} 

\vspace{-20pt}
\noindent where the sets $\mathcal{I}_{\bm{q}}$ and $\mathcal{J}_{\bm{q}}$ are defined in (\ref{eq:IJsets}). %Here, (\ref{eq:beta_1}) is obtained in a similar way of how (\ref{eq:beta1_nu}) is obtained. 
%\begin{equation} 
%\begin{split}
%&\mathcal{I}_{\bm{q}} = \left\{ i; -\frac{N_1}{2}+1 \leq |q_1 + iM_1| \leq \frac{N_1}{2}, i \in \mathbb{Z} \right\} \\
%&\mathcal{J}_{\bm{q}} = \left\{ j; -\frac{N_2}{2}+1 \leq |q_2 + jM_2| \leq \frac{N_2}{2}, j \in \mathbb{Z} \right\}
%\end{split}
%\end{equation} 

Because the second mesh system $\mathcal{M}^{(2)}$ is obtained from the first mesh system $\mathcal{M}^{(2)}$ given a shift $\bm{\delta}$ in space, the Fourier coefficient, $\beta^{(2)}(l, \bm{q})$, obtained from the data collected from $\mathcal{M}^{(2)}$, can be immediately obtained:
\vspace{-12pt}
\begin{equation} 
\beta^{(2)}(l, \bm{q}) = \sum_{i \in \mathcal{I}_{\bm{q}}}\sum_{i \in J_{\bm{q}}} \left\{\alpha(l, \bm{q}+(iM_1,jM_2)^T)+\varepsilon(l, \bm{q}+(iM_1,jM_2)^T)\right\} e^{\imath 2\pi \bm{\delta}^T (\bm{q}+(iM_1,jM_2)^T)}
\end{equation} 

\vspace{-12pt}
\noindent where the last term $e^{\imath 2\pi \bm{\delta}^T (\bm{q}+(iM_1,jM_2)^T)}$ is due to the spatial shift from $\mathcal{M}^{(1)}$ to $\mathcal{M}^{(2)}$.   

Hence, for $\bm{q}\in\mathcal{Q}$, one may define a set 
\vspace{-16pt}
\begin{equation} 
\mathcal{K}_{\bm{q}}  = \left\{(k_1,k_2); k_1=q_1+iM_1, k_2=q_2+jM_2, i \in \mathcal{I}_{\bm{q}}, j \in \mathcal{J}_{\bm{q}} \right\}, 
\end{equation}
%and note that $\bigcup \mathcal{K}_{\bm{q}} = \mathcal{K}$ and $\mathcal{K}_{\bm{q}} \bigcap \mathcal{K}_{\bm{q}'} = 0$ for $\bm{q} \neq \bm{q}'$, i.e., $\mathcal{K}_{\bm{q}}$ are mutually exclusive and exhaustive. 

\vspace{-18pt}
\noindent and for $\bm{k} \in \mathcal{K}_{\bm{q}}$, we have
\vspace{-12pt}
\begin{equation} 
\begin{pmatrix}
\beta^{(1)}(l, \bm{q}) \\
\beta^{(2)}(l, \bm{q}) 
\end{pmatrix}
=
\begin{pmatrix}
\bm{1}^T\\
\bm{b}^T_{\bm{\delta}, \bm{q}}
\end{pmatrix} \mathrm{diag}(\bm{\eta}_{\bm{q}}) \bm{g}_{\bm{q}}(l)
+
\begin{pmatrix}
\bm{1}^T\\
\bm{b}^T_{\bm{\delta}, \bm{q}}
\end{pmatrix} \bm{\varepsilon}_{\bm{q}}(l)
\end{equation} 

\vspace{-12pt}
\noindent for $l = 1,2,...,L$. Here, $\bm{1}$ and $\bm{b}_{\bm{\delta}, \bm{q}} =\{e^{\imath 2\pi \bm{\delta}^T (\bm{q}+(iM_1,jM_2)^T)}\}_{i\in \mathcal{I}_{\bm{q}},j\in \mathcal{J}_{\bm{q}}}$ are column vectors of length $d_{\bm{q}}=|\mathcal{I}_{\bm{q}}\otimes\mathcal{J}_{\bm{q}}|$, and $\bm{g}_{\bm{q}}(l)$, $\bm{\varepsilon}_{\bm{q}}(l)$, and $\bm{\eta}_{\bm{q}}$
are respectively the column vectors obtained from $\bm{g}(l)$, $\bm{{\varepsilon}}(l)$ and $\bm{\eta}$ by keeping only the components corresponding to $\bm{k} \in \mathcal{K}_{\bm{q}}$. Similar to the discussions in Section \ref{sec:nu}, $ \mathcal{K}_{\bm{q}}$ consists of all wavenumbers in $\mathcal{K}$ corresponding to $\bm{q}$ which are aliased and cannot be uniquely determined unless the number of temporal observations $L$ is sufficiently large. 
% and $\bm{\alpha}_0(\bm{k})=\{\alpha_0(\bm{k})\}_{{\bm{\kappa}} \in \mathcal{K}_{\bm{q}}}^T$ and $\bm{\tilde{\varepsilon}}_l(\bm{\kappa})=\{\tilde{\varepsilon}_l(\bm{\kappa})\}_{{\bm{\kappa}} \in \mathcal{K}_{\bm{q}}}^T$.
Combining $\beta^{(1)}(l, \bm{q})$ and $\beta^{(2)}(l, \bm{q})$ from all sampling times, we have
\vspace{-12pt}
\begin{equation} 
\underbrace{\begin{pmatrix}
	\beta^{(1)}(1, \bm{q})\\ 
	\beta^{(2)}(1, \bm{q})\\ 
	\vdots \\ 
	\beta^{(1)}(L, \bm{q})\\ 
	\beta^{(2)}(L, \bm{q})
	\end{pmatrix}}_{\bm{\beta}_{\bm{q}}^{(\text{P-III})}} = 
\underbrace{\begin{pmatrix}
	\bm{1}^T &  &  & \\
	\bm{b}_{\bm{\delta}, \bm{q}}^T &  &  & \\ 
	&  & \ddots  & \\ 
	&  &  &  \bm{1}^T \\
	&  &  &  \bm{b}_{\bm{\delta}, \bm{q}}^T
	\end{pmatrix}}_{\bm{B}_{\bm{\delta}, \bm{q}}} 
\underbrace{\begin{pmatrix}
	\mathrm{diag}(\bm{g}_{\bm{q}}(1)) \\
	\vdots \\
	\mathrm{diag}(\bm{g}_{\bm{q}}(L)) 
	\end{pmatrix}}_{\bm{G}_{\bm{q}}}\bm{\eta}_{\bm{q}} + 
\underbrace{\begin{pmatrix}
	\bm{1}^T &  &  & \\
	\bm{b}_{\bm{\delta}, \bm{q}}^T &  &  & \\ 
	&  & \ddots  & \\ 
	&  &  &  \bm{1}^T \\
	&  &  &  \bm{b}_{\bm{\delta}, \bm{q}}^T
	\end{pmatrix}}_{\bm{B}_{\bm{\delta}, \bm{q}}}  
\underbrace{\begin{pmatrix}
	\bm{\varepsilon}_{\bm{q}}(1) \\
	\vdots \\
	\bm{\varepsilon}_{\bm{q}}(L)
	\end{pmatrix}}_{\bm{W}_{\bm{q}}}
\end{equation} 

\vspace{-14pt}
\noindent where $\bm{\beta}_{\bm{q}}^{(\text{P-III})}$ is a $2L \times 1$ column vector, $\bm{B}_{\bm{\delta}, \bm{q}}=\mathrm{diag}\{  (\bm{1}, \bm{b}_{\bm{\delta}, \bm{q}})^T \}$ is a $2L \times (L \times d_{\bm{q}})$ block diagonal matrix with $d_{\bm{q}}=|\mathcal{K}_q|$, and $\bm{G}_{\bm{q}}$ is a $(L \times d_{\bm{q}}) \times d_{\bm{q}}$ matrix. For any $\bm{q} \in \mathcal{Q}$, we have
\vspace{-14pt}
\begin{equation}  \label{eq:beta1_shift}
\bm{\beta}_{\bm{q}}^{(\text{P-III})}
=
\bm{B}_{\bm{\delta}, \bm{q}}
\bm{G}_{\bm{q}}
\bm{\eta}_{\bm{q}}
+
\bm{B}_{\bm{\delta}, \bm{q}}
\bm{W}_{\bm{q}}
\end{equation}

\vspace{-14pt} 
\noindent where $\bm{B}_{\bm{\delta}, \bm{q}}\bm{G}_{\bm{q}}$ is  a $2L\times d_{\bm{q}}$ matrix. 
Because $0 < \delta_1 < M_1^{-1}$ and $0 < \delta_2 < M_2^{-1}$, the elements in $\bm{b}_{\bm{\delta}, \bm{q}}$ are identical, which immediately makes $2L \geq d_{\bm{q}}$ the sufficient condition for $\bm{B}_{\bm{\delta}, \bm{q}}\bm{G}_{\bm{q}}$ to be full column rank, i.e.,
\vspace{-12pt}
\begin{equation} 
L \geq \frac{d_{\bm{q}}}{2} \geq \frac{1}{2} \left \lfloor \frac{N_1-1}{M_1}+1 \right \rfloor \left \lfloor \frac{N_2-1}{M_2}+1 \right \rfloor.
\end{equation} 

\vspace{-12pt}
As discussed in Section \ref{sec:nu}, directly solving the individual linear models in (\ref{eq:beta1_shift}) is not an appropriate choice for ill-conditioned problems. Hence, combining the linear models (\ref{eq:beta1_shift}) for all $\bm{q}_1, \bm{q}_2, ..., \bm{q}_{|\mathcal{Q}|}$, we have
\vspace{-8pt} 
\begin{equation}  \label{eq:beta_spectrum_36}
\underbrace{\begin{pmatrix}
	\bm{\beta}_{\bm{q}_1}\\ 
	\vdots \\ 
	\bm{\beta}_{\bm{q}_{|\mathcal{Q}|}}
	\end{pmatrix}}_{\bm{\vec{\mathscr{Y}}}^{(\text{P-III})}} = 
\underbrace{\begin{pmatrix}
	\bm{B}_{\bm{\delta}, \bm{q}_1}\bm{G}_{\bm{q}_1}  &  &  & \\ 
	&  & \ddots  & \\ 
	&  &  &  \bm{B}_{\bm{\bm{\delta}, q}_{|\mathcal{Q}|}}\bm{G}_{\bm{q}_{|\mathcal{Q}|}} 
	\end{pmatrix}}_{\mathscr{X}^{(\text{P-III})}}
\underbrace{\begin{pmatrix}
	\bm{\eta}_{\bm{q}_1} \\
	\vdots \\
	\bm{\eta}_{\bm{q}_{|\mathcal{Q}|}} 
	\end{pmatrix}}_{\bm{\eta}} + 
\underbrace{\begin{pmatrix}
	\bm{B}_{\bm{\delta}, \bm{q}_1}\bm{W}_{\bm{q}_1}  \\ 
	\vdots   \\ 
	\bm{B}_{\bm{\delta}, \bm{q}_{|\mathcal{Q}|}}\bm{W}_{\bm{q}_{|\mathcal{Q}|}} 
	\end{pmatrix}}_{\mathscr{V}^{(\text{P-III})}}
\end{equation} 

\vspace{-12pt}
\noindent where $\bm{\vec{\mathscr{Y}}}^{(\text{P-III})}$ is a ($2|\mathcal{Q}|\times L) \times 1$ column vector, $\mathscr{X}^{(\text{P-III})}$ is a ($2|\mathcal{Q}|\times L) \times |\mathcal{K}|$ matrix, $\bm{\eta}$ is a $|\mathcal{K}| \times 1$ column vector, $\mathscr{V}^{(\text{P-III})} \sim \mathcal{N}(\bm{0}, \bm{\Sigma}^{(\text{P-III})})$, $\bm{\Sigma}^{(\text{P-III})}=\sigma^2 \text{diag} ( 
\{\bm{S}_i\}_{i=1}^{|\mathcal{Q}|}))$ and $\bm{S}_i  = \text{diag}(d_{\bm{q}_i}\bm{I},  ||\bm{b}_{\bm{\delta},\bm{q}_i}||_2^2\bm{I})$ for $i=1,2,...,|\mathcal{Q}|$.
%with $\bm{\Sigma}^{(\text{P-III})}$ being a block diagonal matrtix
%\begin{equation} 
%\bm{\Sigma}^{(\text{P-III})}  = \sigma^2 \text{diag} ( 
%\{\bm{S}_i\}_{i=1}^{|\mathcal{Q}|})
%\end{equation} 
%and 
%\begin{equation} 
%\bm{S}_i  = \text{diag}(d_{\bm{q}_i}\bm{I},  ||\bm{b}_{\bm{\delta},\bm{q}_i}||_2^2\bm{I})
%\end{equation} 
%for $i=1,2,...,|\mathcal{Q}|$.

Similar to Problem P-II (\ref{eq:PII}), %instead of directly solving the inverse problem in the spectral domain, we again convert the spectral coefficients back to the space-time domain, and 
we obtain a regularized inverse problem:
%\begin{equation}
%\text{Inverse Problem (P-II):}\quad\quad
%\text{argmin} \quad \frac{1}{2}\left \| \bm{\vec{\mathscr{\beta}}}^{(\text{P-II})} - \bm{\mathscr{X}}^{(\text{P-II})}\bm{\eta} \right \|_2^2 + \lambda_1 \left \| \bm{\eta} \right \|_1 + \lambda_2 \left \| \bm{\eta} \right \|_2^2
%\end{equation}
%where $\lambda_1$ and $\lambda_2$ are respectively the $\l_1$ sparsity and $\l_2$ stability constraints. 
\vspace{-12pt}
\begin{equation}  \label{eq:PIII}
\text{Problem P-III:  } \text{min} \frac{1}{2} (\bm{\vec{\mathscr{Y}}}^{(\text{P-III})} - \bm{\mathscr{X}}^{(\text{P-III})}\bm{\eta})^T (\bm{\Sigma}^{(\text{P-III})})^{-1} (\bm{\vec{\mathscr{Y}}}^{(\text{P-IIII})} - \bm{\mathscr{X}}^{(\text{P-II})}\bm{\eta}) + \mathcal{R}(\bm{\eta})
\end{equation}

\vspace{-12pt}
\noindent where  %$\bm{\vec{\mathscr{Y}}}^{(\text{P-III})}=\bm{F}^{(\text{P-III})}\bm{\vec{\mathscr{\beta}}}^{(\text{P-III})}$ represents the discrete Fourier transform, $\bm{F}^{(\text{P-III})}$ is the matrix operator, $\bm{\mathscr{X}}^{(\text{P-III})} = \bm{F}^{(\text{P-III})}\bm{X}^{(\text{P-III})}$, $\bm{\Sigma}^{(\text{P-III})} =\sigma^2 \bm{F}^{(\text{P-III})} \text{diag} ( \{\bm{S}_i\}_{i=1}^{|\mathcal{Q}|}) (\bm{F}^{(\text{P-III})})^T$, and 
$\mathcal{R}(\bm{\eta})$ is defined in (\ref{eq:R1}). 

\textbf{Remarks}. Problem P-I estimates the special coefficients $\bm{\eta}$ by minimizing the squared distance between forward model outputs and observations in the space-time domain, while Problems P-II and P-III estimate the spectral coefficients $\bm{\eta}$ by minimizing the squared distance between the spectral coefficients of forward model outputs and that of observations in the spectral domain. For this reason, when we convert the the optimal solution obtained by P-II and P-III from the spectral domain back to the space-time domain through the inverse Fourier transform, the solution is no longer optimal in the least squares sense (i.e., the squared distance between forward model outputs and observations is not minimized in the space-time domain). This is due to the fact that the least squares estimator is not invariant under transformation. If computational cost is not the primary concern, we  recommend one to solve the inverse problem in the space-time domain using Problem P-I (\ref{eq:PI}).
\color{black}

%Note that, as is often the case in the literature, the inverse problem can be directly solved in the spectral domain by estimating $\bm{\vec{\mathscr{\beta}}}^{(\text{P-II})}$ from the linear model (\ref{eq:beta_spectrum}) using the Least Squares (LS). However, this paper proposes to solve the inverse problem (\ref{eq:PII}) after converting the spectral coefficients back to the space-time domain, and minimizes the distance between observations and forward predictions. The main consideration behind this strategy is due to the fact that the LS estimator is not invariant under transformation (i.e., Fourier transformation in our case). Hence, there is no guarantee that the optimal solutions obtained in the spectral domain is still near-optimal in the space-time domain. In fact, our numerical investigations have demonstrated poorer source detection performance if the inverse model is solved in the spetral domain. 

\vspace{-12pt}
\section{Solving the Problems P-I, P-II and P-III using ADMM}\label{sec:ADMM}

This section provides the algorithm required to solve the inverse problems P-I, P-II and P-III (throughout this section, the superscripts, $\cdot^{\text{P-I}}$, $\cdot^{\text{P-II}}$ and $\cdot^{\text{P-III}}$, are dropped without causing ambiguity). Note that, the dimension of $\bm{\eta}$ in these three inverse problems is given by $|\mathcal{K}|=N_1 \times N_2$. Hence, even for moderate size of $N_1$ and $N_2$, the dimension of $\bm{\eta}$ can be large. The Alternating Direction Method of Multipliers (ADMM) for large-scale optimization problems becomes a sensible choice. 

We first convert an unconstrained problem of the general form
\vspace{-12pt}
\begin{equation} \label{eq:inverse_general}
\text{min}_{\bm{\eta}} \quad \frac{1}{2} (\vec{\bm{\mathscr{Y}}} - \bm{\mathscr{X}}\bm{\eta})^T \bm{\Sigma}^{-1} (\vec{\bm{\mathscr{Y}}} - \bm{\mathscr{X}}\bm{\eta})+ \mathcal{R}(\bm{\eta})
\end{equation}

\vspace{-16pt}
\noindent to a constrained problem:
\vspace{-16pt}
\begin{equation} \label{eq:PI_constrained}
\text{min}_{\bm{\eta}, \bm{\psi}} \quad f(\bm{\eta}) + \mathcal{R}(\bm{\psi}), \quad\quad \text{s.t. } \bm{\eta}=\bm{\psi}
\end{equation}

\vspace{-14pt}
\noindent where $f(\bm{\eta})=\frac{1}{2} (\vec{\bm{\mathscr{Y}}} - \bm{\mathscr{X}}\bm{\eta})^T \bm{\Sigma}^{-1} (\vec{\bm{\mathscr{Y}}} - \bm{\mathscr{X}}\bm{\eta})$.
For $\rho>0$, the scaled form of the augmented Lagrangian is written as:
\vspace{-16pt}
\begin{equation} \label{eq:PI_Lagrangian}
f(\bm{\eta}) + \mathcal{R}(\bm{\psi}) + \frac{\rho}{2} \left\| \bm{\eta}-\bm{\psi} + \bm{u}  \right\|_2^2 +  \frac{\rho}{2} \left\| \bm{u} \right\|_2^2.
\end{equation}

\vspace{-16pt}
Then, the ADMM solves the constrained problem (\ref{eq:PI_constrained}) by repeating the following iterations \citep{Zou2005, Ramdas2016}: 
\vspace{-16pt}
\begin{subequations} \label{eq:ADMM}
\begin{align}
 \bm{\eta}^{(i)} = \text{argmin}_{\bm{\eta}} f(\bm{\eta}) + \frac{\rho}{2}\left\| \bm{\eta} - \bm{\psi}^{(i-1)} + \bm{u}^{(i-1)} \right\|_2^2\\
 \bm{\psi}^{(i)} = \text{argmin}_{\bm{\psi}} \mathcal{R}(\bm{\psi}) + \frac{\rho}{2}\left\| \bm{\eta}^{(i)} - \bm{\psi} + \bm{u}^{(i-1)} \right\|_2^2\\
 \bm{u}^{(i)} = \bm{u}^{(i-1)} + \bm{\eta}^{(i)} - \bm{\psi}^{(i)}
\end{align}
\end{subequations}

\vspace{-18pt}
\noindent for $i=1,2,\cdots$. The iterations satisfy: residual convergence (i.e., $\bm{\eta}^{(i)} - \bm{\psi}^{(i)} \rightarrow 0$ as $i\rightarrow \infty$), objective convergence (i.e., $f(\bm{\eta}^{(i)}) + \mathcal{R}(\bm{\psi}^{(i)}) \rightarrow f^*+\mathcal{R}^*$ where $f^*$ and $\mathcal{R}^*$ are the primal optimal values), and dual convergence (i.e., $\bm{u}^{(i)} \rightarrow \bm{u}^{*}$ where $\bm{u}^{*}$ is the dual solution). 
Algorithm 1 summarizes the ADMM algorithm developed for solving (\ref{eq:inverse_general}). In the Supplemental Materials, we provide technical details of how each step in Algorithm 1 is obtained.

\textbf{Remarks}. Although Problems P-I, P-II and P-III can be solved by the ADMM algorithm, it is noted that Problem P-I is formulated in the space-time domain, while Problems P-II and P-III are constructed in the spectral domain. As a result, the design matrix $\bm{\mathscr{X}}^{(\text{P-I})}$ in (\ref{eq:p1_proof_1}) is a dense matrix, while the design matrices $\bm{\mathscr{X}}^{(\text{P-II})}$ and $\bm{\mathscr{X}}^{(\text{P-III})}$ in (\ref{eq:beta_spectrum}) and (\ref{eq:beta_spectrum_36}) are sparse (block diagonal), making the computation of $\bm{\mathscr{X}}^T \bm{\Sigma}^{-1}\bm{\mathscr{X}}$, $\bm{\mathscr{X}}^T \bm{\Sigma}^{-1}$ and $\bm{\Sigma}^{-1}\bm{\mathscr{X}}$ faster in the ADMM algorithm. In addition, Problems P-II and P-III enable one to truncate the high-frequency components because each block of $\bm{\mathscr{X}}$,  in both (\ref{eq:beta_spectrum}) and (\ref{eq:beta_spectrum_36}), corresponds to a frequency level. This further helps to reduce the computational time and details are provided in the Supplemental Materials. 

In many applications, a non-negativity constraint can be added to the output of the inverse model (e.g., the detected emissions or initial conditions need to be non-negative). When a non-negativity constraint is added,  the inverse modeling problem (\ref{eq:inverse_general}) becomes:

\vspace{-18pt}
\begin{equation} \label{eq:inverse_general_constraint}
%\begin{split}
\text{min}_{\bm{\eta}} \text{ }  \frac{1}{2} (\vec{\bm{\mathscr{Y}}} - \bm{\mathscr{X}}\bm{\eta})^T \bm{\Sigma}^{-1} (\vec{\bm{\mathscr{Y}}} - \bm{\mathscr{X}}\bm{\eta})+ \mathcal{R}(\bm{\eta}), \text{      s.t. }  \bm{\mathscr{X}}\bm{\eta} \geq 0 
%\end{split}
\end{equation}

\vspace{-8pt}
In the Supplemental Materials, we show that the constrained problem (\ref{eq:inverse_general_constraint}) can be efficiently solved by modifying the ADMM algorithm described in Section \ref{sec:ADMM}, which expands the applicability of the proposed model for a wider range of problems. 

\vspace{-8pt}
\singlespacing
\begin{algorithm}[H]
	
	\KwData{ $\vec{\bm{\mathscr{Y}}}$,  $\bm{\mathscr{X}}$, $\vec{\bm{\mathscr{V}}}$, $\lambda_1$, $\lambda_2$}
	
	$\bm{\eta}^{(0)}$, $\bm{\psi}^{(0)}$, $\bm{u}^{(0)}$, $\rho>0$, $\omega>0$ //initialize
	
	$i \leftarrow 1$
	
	\vspace{0.2cm}
	\textbf{(the outer loop)}
	
	\While{convergence criterion is not met}{
		\vspace{0.25cm}
		$\bm{\eta}^{(i)} \leftarrow \left(\bm{\mathscr{X}}^T\bm{\Sigma}^{-1}\bm{\mathscr{X}}+\frac{1}{2}\rho\bm{I}\right)^{-1} \left\{  \bm{\mathscr{X}}^T\bm{\Sigma}^{-1}\vec{\bm{\mathscr{Y}}}  + \rho (\bm{\psi^{(i-1)}}+\bm{u}^{(i-1)})  \right\} $

		\vspace{0.25cm}
		\textbf{(the inner loop)}
		
		$\tilde{\bm{\psi}}^{(0)}$ //initialization for the inner loop
		
		$j \leftarrow 1$ 
		
		%\vspace{0.2cm}
		\While{convergence criterion is not met}{
			\vspace{0.25cm}
			$\tilde{\bm{\psi}}^{(j)} \leftarrow  \frac{\rho}{2\lambda_2}\left(\bm{J}^T\bm{J}+\frac{1}{2}(\rho + \omega)\bm{I} \right)^{-1} (\bm{\eta}^{(i)}+\bm{u}^{(i-1)} + \omega \bm{\theta}^{(j-1)} - \omega \bm{v}^{(j-1)})$
			
			$\bm{\theta}^{(j)} \leftarrow S_{\lambda_1/\omega}(\tilde{\bm{\psi}}^{(j)} + \bm{v}^{(j-1)})$,
			$\bm{v}^{(j)} \leftarrow \bm{v}^{(j-1)} + \tilde{\bm{\psi}}^{(j)} - \bm{\theta}^{(j)}$ 
			
			$j \leftarrow j + 1$
		}
		
		\vspace{0.25cm}
		$\bm{\psi}^{(i)} \leftarrow \tilde{\bm{\psi}}^{(j)}$,
		$\bm{u}^{(i)} \leftarrow \bm{u}^{(i-1)} + \bm{\eta}^{(i)} - \bm{\psi}^{(i)}$
		
		$i \leftarrow i + 1$
	}
	\vspace{0.2cm}
	%\KwResult{Simulated $n$ particles and their weights}

	\caption{ADMM for solving the Inverse Problems}
\end{algorithm}
\doublespacing

\vspace{-16pt}
\section{Numerical Examples} \label{sec:numerical}

\vspace{-6pt}
This section presents two numerical examples to illustrate the application of the proposed inverse models and generate some useful insights of the approach. 

\vspace{-18pt}
\subsection{Example I}
\vspace{-6pt}
%The proposed approach is firstly investigated using a simulated dataset. 
We first simulate an advection-diffusion process from the PDE (\ref{eq:SPDE1}) on a $40 \times 40$ rectangular grid. The parameters of the advection-diffusion operator $\mathcal{A}$ are chosen as: $\bm{\vec{v}}=(0.5\times 10^{-2},0.5\times 10^{-2})^T$, $\bm{D}=\text{diag}\{0.25\times 10^{-3}\}$ and $\zeta=0$. The initial condition contains three spatially-sparse instantaneous sources given by $\Phi(t, \bm{s}) = \delta(t-0) \sum_{j=1}^{3} \phi_j(\bm{s})$. Here, $\phi_j(\bm{s}) = \phi_0 \exp\{-\frac{||\bm{s}-\bm{s}_0^{(j)}||_2}{0.09}\}$ where $\phi_0=300$, $\bm{s}_0^{(1)}=(0.4,0.2)^T$, $\bm{s}_0^{(2)}=(0.2,0.4)^T$ and $\bm{s}_0^{(3)}=(0.5,0.5)^T$.

\vspace{-10pt}
\begin{figure}[h!]  
	\begin{center}
		\includegraphics[width=0.80\textwidth]{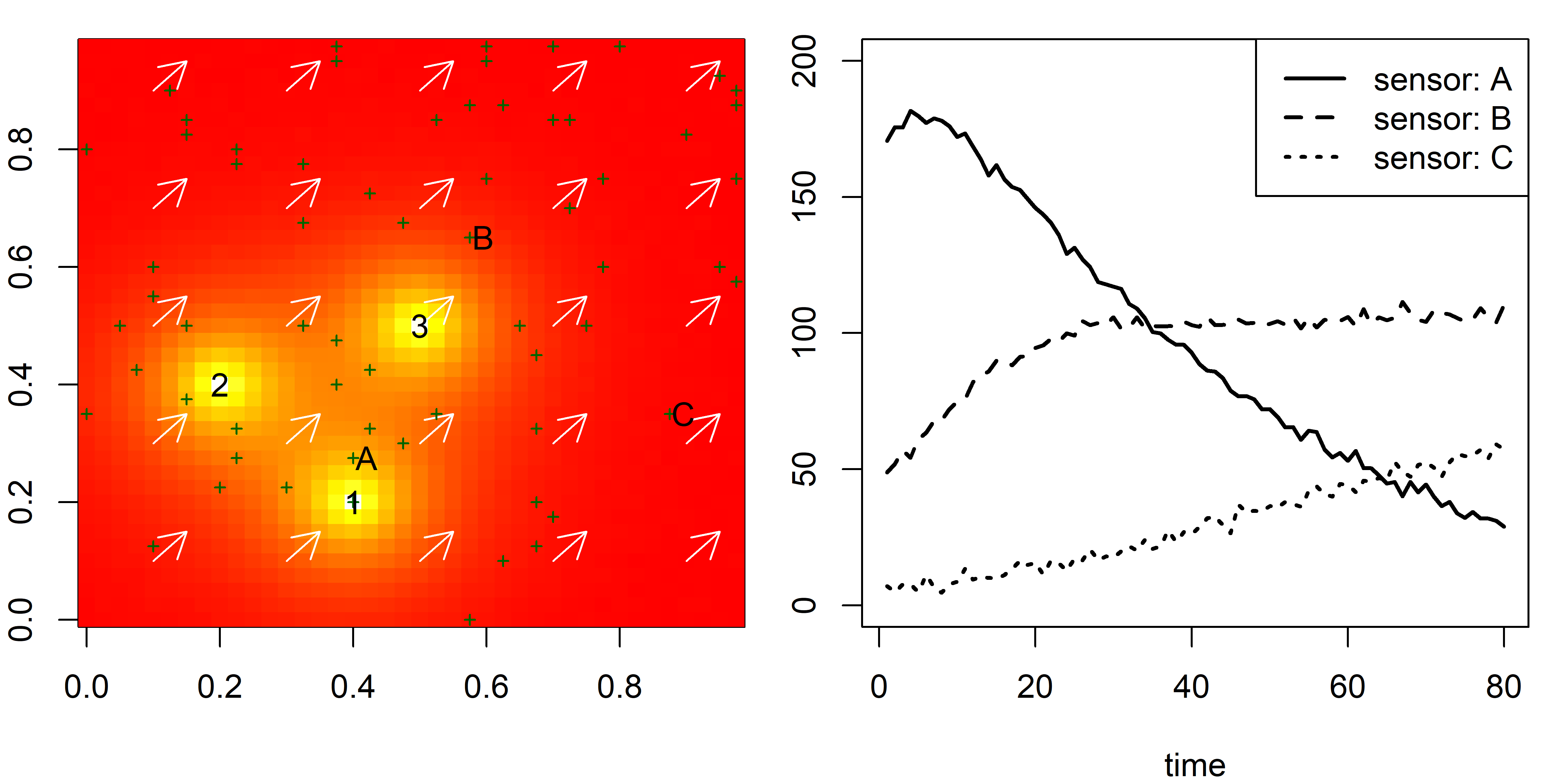}
		\centering
		\vspace{-12pt}
		\caption{(left panel) the initial condition with three instantaneous sources, velocity field (indicated by white arrows), and locations of 64 randomly distributed sensors; (right panel) noisy sensor measurements over time from three selected sensors, ``A'', ``B'' and ``C''.}
		\label{fig:irregular_initial}
	\end{center}
\end{figure}

\vspace{-26pt}
Figure \ref{fig:irregular_initial} (left panel) shows the initial condition, velocity field (indicated by arrows), and the locations of 64 randomly distributed sensors (indicated by small crosses).  
%It is assumed that a sensor can only observe the spatio-temporal process at the location where the sensor is located. 
Figure \ref{fig:irregular_initial} also highlights the locations of three selected sensors, ``A'', ``B'' and ``C'', and the measurements over time are shown in the right panel of this figure. The measurement errors are \textit{i.i.d.} samples from a Normal distribution with mean zero and standard deviation two. The strength of the signal from sensor ``A'' firstly increases when the process (primarily from source 1) quickly reaches location ``A''. After that, the signal decreases as the process propagates away and diffuses. Sensor ``B'' gradually picks up the signal (firstly from source 3, and then, from the other two sources), while sensor ``C'' slowly picks up relatively weak signal because this sensor is far from all three sources. The goal is to estimate the initial condition in the absence of the ``complete picture'' of the spatio-temporal process over the entire spatial domain. 

\vspace{-10pt}
\begin{figure}[h!]  
	\begin{center}
		\includegraphics[width=0.90\textwidth]{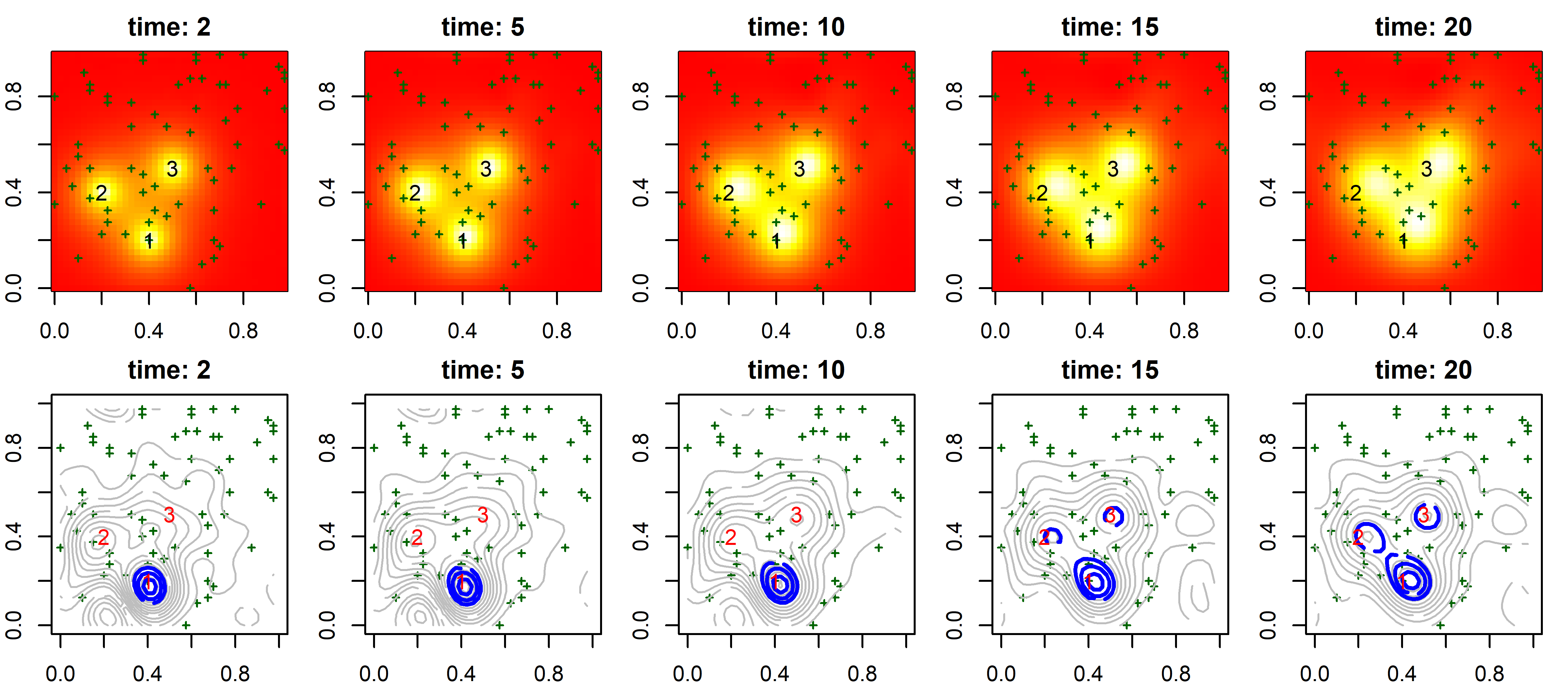}
		\centering
		\vspace{-4pt}
		\caption{(top row) snapshots of the advection-diffusion process at times 2, 5, 10, 15 and 20; (bottom row) detected spatial sources based on the streaming data from a network of 64 sensors up to times 2, 5, 10, 15 and 20.}
		\label{fig:irregular_64}
	\end{center}
\end{figure}
\vspace{-32pt}
\begin{figure}[h!]  
	\begin{center}
		\includegraphics[width=0.90\textwidth]{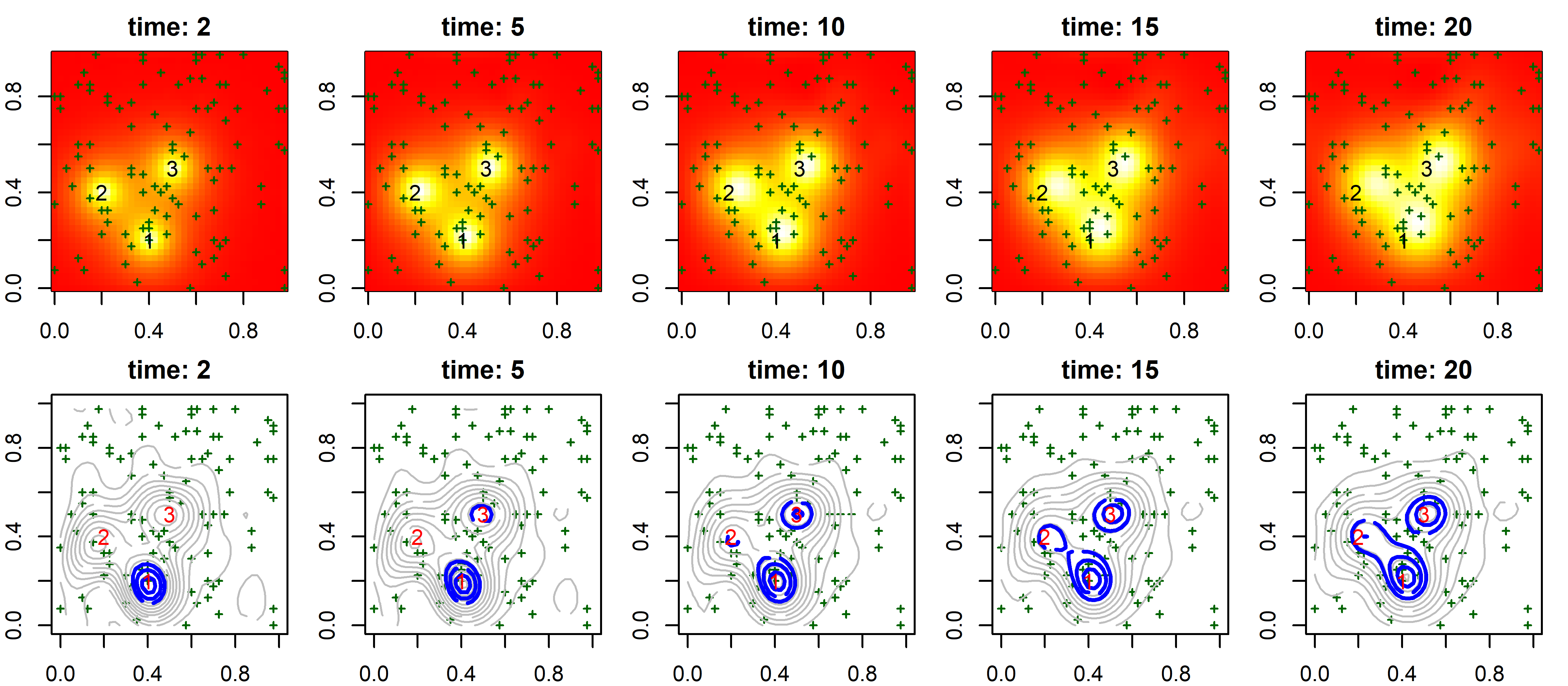}
		\centering
		\vspace{-4pt}
		\caption{(top row) snapshots of the advection-diffusion process at times 2, 5, 10, 15 and 20; (bottom row) detected spatial sources based on the streaming data from a network of 100 sensors up to times 2, 5, 10, 15 and 20.}
		\label{fig:irregular_100}
	\end{center}
\end{figure}
%\vspace{-30pt}

\vspace{-28pt}
The first row of Figure \ref{fig:irregular_64} shows the snapshots of the process at times 2, 5, 10, 15 and 20. Solving the Inverse Problem P-I (\ref{eq:PI}) for irregular sampling grid using Algorithm 1, the second row of Figure \ref{fig:irregular_64} shows the contour plots of the estimated initial condition using the sensor observations up to times 2, 5, 10, 15 and 20. The thick blue level sets are respectively the 75th, 85th and 95th percentiles of the output generated by the inverse model. It is seen that, source 1 is quickly detected at time 2. This is only because there happens to be a sensor located near source 1. It is seen that source 2 might also has been detected (circled by a contour line). However, the estimated strength of source 2 is weaker than that of source 1. Source 3 cannot be detected at all at time 2 because most of the sensors have not yet picked up any signal from this source. At time 15, both sources 2 and 3 are clearly detected as the downstream sensors have picked up the signal from these two sources.

Expanding the size of the sensor network is expected to reduce the detection latency. We randomly add another 36 sensors to the existing sensor network (note that, only those sensors added to the downstream areas of the sources may help to reduce the detection latency). Figure \ref{fig:irregular_100} presents the updated results: with the additional 36 sensors, all three sources can be detected using the sensor data up to time 10. Figures \ref{fig:irregular_64} and \ref{fig:irregular_100} well demonstrate the dynamic nature of the inverse problem based on spatially-distributed sensor data streams. 

\vspace{-8pt}
\begin{figure}[h!]  
	\begin{center}
		\includegraphics[width=0.9\textwidth]{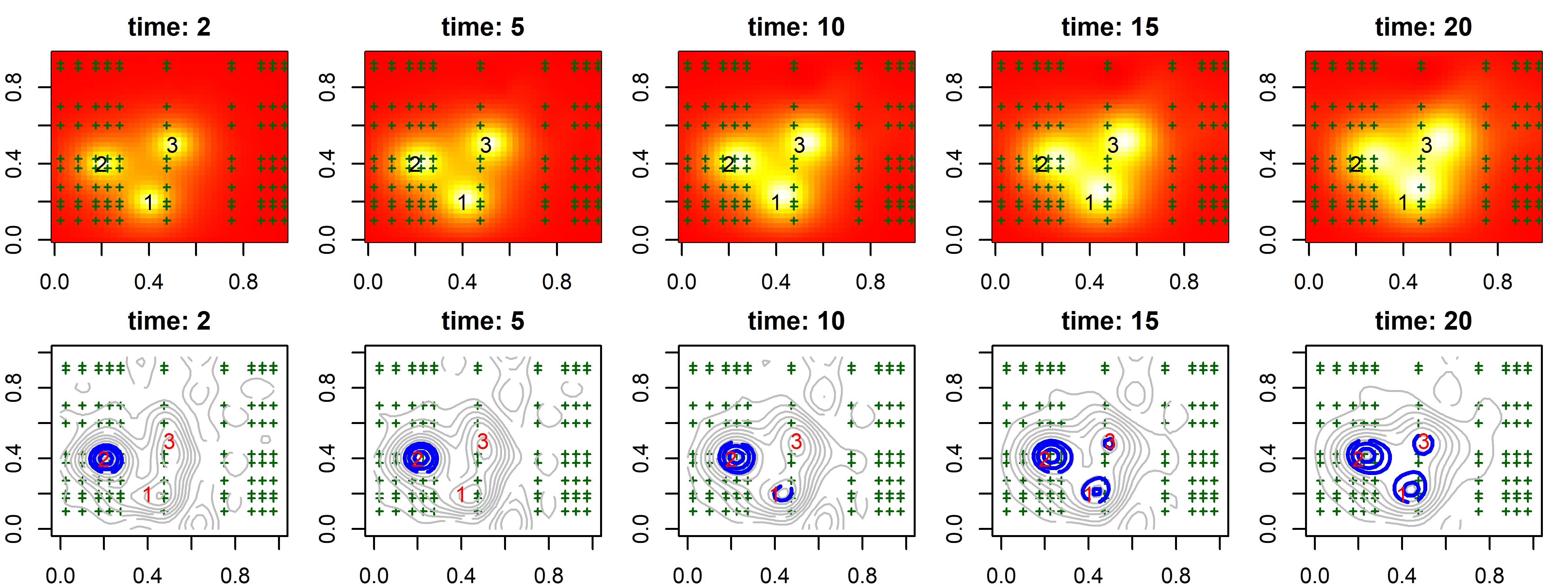}
		\centering
		\vspace{-4pt}
		\caption{(top row) snapshots of the simulated advection-diffusion process at times 2, 5, 10, 15 and 20; (bottom row) detected spatial sources based on the streaming data from a non-uniform sampling grid up to times 2, 5, 10, 15 and 20.}
		\label{fig:nonuniform_10}
	\end{center}
\end{figure}

\vspace{-28pt}
Next, we investigate the Inverse Problems P-II (\ref{eq:PII}) and P-III (\ref{eq:PIII}) for non-uniform sampling grid and shifted sampling grids. Figure \ref{fig:nonuniform_10} shows the output of the Problem P-II based on the sensor data streams from a non-uniform sampling grid, which is given by a $10 \times 10$ mesh system generated from a $40 \times 40$ uniform mesh system as described in Section \ref{sec:nu}. Similar to Figure \ref{fig:irregular_64}, the first row of  Figure \ref{fig:nonuniform_10} shows the snapshots of the process at times 2, 5, 10, 15 and 20, while the second row shows the contour plots of the estimated initial condition using the streaming observations up to times 2, 5, 10, 15 and 20. The thick blue level sets are respectively the 75th, 85th and 95th percentiles of the output generated by the inverse model. 
We see that, source 2 is almost immediately detected because of its proximity to nearby sensors. Sources 1 and 3 are detected later at times 10 and 15 only when sensors in the downstream areas have picked up the signal originated from these two sources.  

\vspace{-8pt}
\begin{figure}[h!]  
	\begin{center}
		\includegraphics[width=0.9\textwidth]{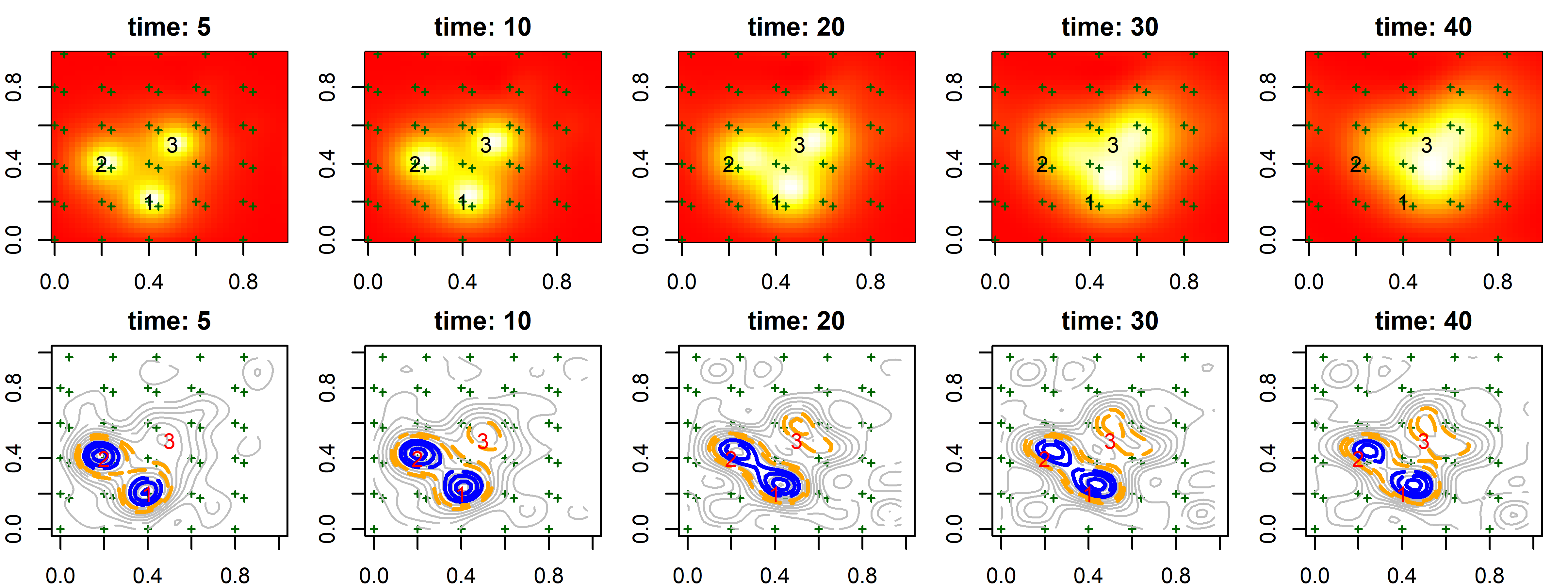}
		\centering
		\vspace{-4pt}
		\caption{(top row) snapshots of the simulated advection-diffusion process at times 5, 10, 20, 30 and 40; (bottom row) detected spatial sources based on the streaming data from two shifted uniform sampling grids up to times 5, 10, 20, 30 and 40.}
		\label{fig:shifted_5}
	\end{center}
\end{figure}

\vspace{-26pt}
Figure \ref{fig:shifted_5} shows the dynamic output from the Inverse Problem P-III based on the sensor data streams from two shifted sampling grids. The first sampling grid is a $5 \times 5$ mesh system, while the second grid is obtained by shifting the first grid by $\bm{\delta}=(0.04, 0.175)$. The first row of Figure \ref{fig:shifted_5} shows the process at times 5, 10, 20, 30 and 40, while the second row shows the model output up to times 5, 10, 20, 30 and 40. Similarly, the solid blue thick level sets are respectively the 75th, 85th and 95th percentiles of the output generated by the model, while the dashed thick level sets are the 50th and 60th percentiles of the output generated by the inverse model. We see that, sources 1 and 2 are quickly detected at time 5 because of their proximity to nearby sensors. Source 3 is detected later when sensors in the downstream areas of source 3 have picked up the signal.  

\vspace{-2pt}
\begin{figure}[h!]  
	\begin{center}
		\includegraphics[width=1\textwidth]{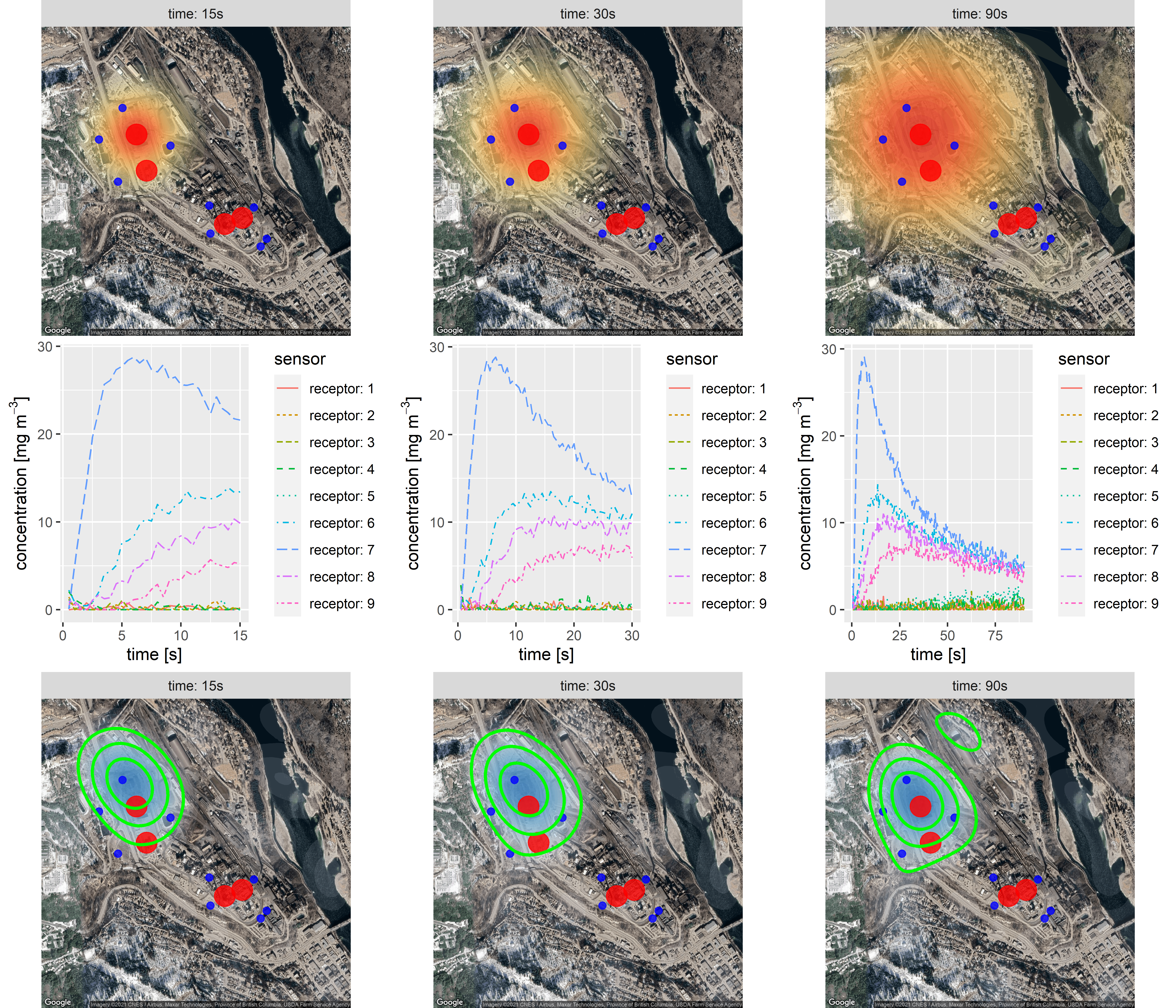}
		\centering
		\caption{Inverse modeling for a pure diffusion process: (a) concentration of the pollutant 15, 30 and 90 seconds after its release; (b) ground measurements by receptors; (c) contour plots of the output from the inverse model}
		\label{fig:diffusion}
	\end{center}
\end{figure}

\vspace{-20pt}
Example I generates useful insights on the proposed inverse models, and successfully reveals the dynamic nature of the inverse problem using spatially-distributed data streams. A source can be detected when sensors in the downstream areas (if there are any) pick up the signal originated from that source. In the Supplemental Materials,  we compare the bias and Mean-Squared-Error (MSE) of the estimated initial condition $\hat{\xi}(0, \bm{s})$ for different choices of regularizations, including the proposed regularization, generalized Lasso, Elastic Net, $L_1$ and $L_2$ regularizations.

\vspace{-2pt}
\begin{figure}[h!]  
	\begin{center}
		\includegraphics[width=1\textwidth]{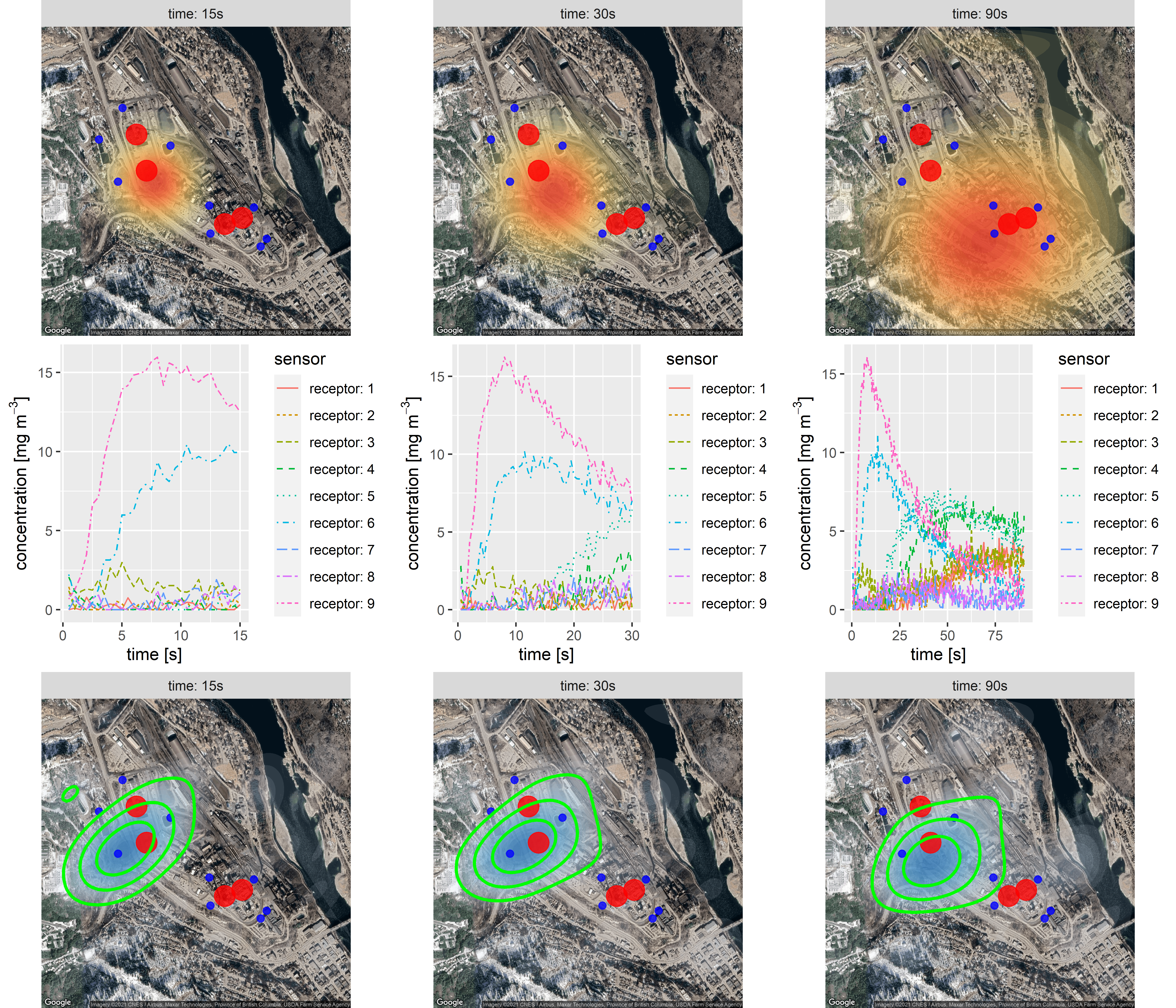}
		\centering
		\vspace{-8pt}
		\caption{Inverse modeling for an advection-diffusion process: (a) concentration of the pollutant 15, 30 and 90 seconds after its release; (b) ground measurements by receptors; (c) contour plots of the output from the inverse model}
		\label{fig:advection}
	\end{center}
\end{figure}

%\section{A Case Study} \label{sec:case}
\vspace{-26pt}
\subsection{Example II} \label{sec:example2}
%In Example II, we further investigate the proposed inverse modeling approach in the context of a real application. The application is modified from recent studies on emissions from point contaminant sources at a large lead-zinc smelter located in Trail, British Columbia, Canada. As shown in Figure \ref{fig:map}, the four large red circles indicate the potential emission sources of Zinc Sulphate (ZnSO$_\text{4}$), while the small blue circles indicate the locations of nine receptors (i.e., sensors) deployed to obtain the ground-level measurements. The spatial area shown in this figure is about $2\times 2$ $\text{km}^2$, and more details of this application can be found in \cite{Hosseini2016}. 
%\begin{figure}[h!]  
%	\begin{center}
%		\includegraphics[width=0.9\textwidth]{figures/map.png}
%		\centering
%		\caption{The spatial domain of interest in Example II (big red circles: potential ZnSO$_\text{4}$ emission sources; small blue circles: receptors for ground-level measurements)}
%	\end{center}
%\end{figure}
Example II revisits the motivating example described in Section \ref{sec:motivation}. This example is concerned with the estimation of emission locations of accidental ZnSO$_\text{4}$ releases using sensor monitoring data; see  Section \ref{sec:motivation} and \cite{Hosseini2016} for more details. 

The release and transport of a single contaminant in the atmosphere can be well described by an advection-diffusion equation, $c(t, \bm{s}) + \bm{\vec{v}}^T\triangledown c(t, \bm{s}) - \triangledown \cdot [\bm{D} \triangledown  c(t, \bm{s})] = \Phi(t, \bm{s})$, 
%\begin{equation} \label{eq:SPDE1_case}
%c(t, \bm{s}) + \bm{\vec{v}}^T\triangledown c(t, \bm{s}) - \triangledown \cdot [\bm{D} \triangledown  c(t, \bm{s})] = \Phi(t, \bm{s})
%\end{equation}	
where $c(t, \bm{s})$ is the contamination concentration [mg $\text{m}^{-3}$], $\bm{\vec{v}}$ is the wind vector [m $\text{s}^{-1}$], $\bm{D}$ is the diffusivity [$\text{m}^{2}$ $\text{s}^{-1}$], and $\Phi(t, \bm{s}) = \delta(t-0) \sum_{j=1}^{4} \phi_j(\bm{s})$ is the instantaneous emission source [mg $\text{m}^{-3}$ $\text{s}^{-1}$]. 
%\begin{equation} \label{eq:source_case}
%\Phi(t, \bm{s}) = \delta(t-0) \sum_{j=1}^{4} \phi_j(\bm{s}), \quad\quad J\geq 0
%\end{equation}	
This equation is a special case of the general form (\ref{eq:SPDE1}) in Section \ref{sec:general}.

$\bullet$ Scenario 1: pure diffusion. We first consider the scenario where the propagation of ZnSO$_\text{4}$, after its release, is driven by a pure diffusion process (i.e., the case when there is no wind). The first row of Figure \ref{fig:diffusion} shows the concentration of the pollutant at times 15, 30 and 90 seconds after its release. As shown by this figure, the pollutant is released from the first source in the north, and propagates to all directions following a pure diffusion process. The second row of Figure \ref{fig:diffusion} shows the noisy measurements over time. The third row of Figure \ref{fig:diffusion} presents the contour plots of the output generated by the inverse model (\ref{eq:PI}) using the streaming observations up to 15, 30 and 90 seconds. As seen in the first and second rows of Figure \ref{fig:diffusion}, the sensor located to the north of the source firstly picks up signal. Hence, at times 15 and 30 seconds, the peaks indicated by the contour plot are somewhere between that sensor and the actual source. As more data become available from the other three sensors located near the source (the second row of Figure \ref{fig:diffusion}), the peak moves closer to the actual source at time 90 and the emission source is successfully identified. Such observations rationalize the dynamic nature of the proposed inverse problems based on sensor data streams. 

$\bullet$ Scenario 2: advection and diffusion. We now consider a more common scenario where the propagation of ZnSO$_\text{4}$, after its release, is driven by both advection and diffusion. The pollutant is released from the second source from the top of the spatial domain, and propagates to the southeast direction due to wind. The first row of Figure \ref{fig:advection} shows the pollutant concentration  at times 15, 30, 90 seconds after its release.  The second row shows the noisy ground-level measurements. The third row of presents the contour plots of the output generated by the inverse model (\ref{eq:PI}) based on the data up to 15, 30 and 90 seconds. As seen in the first and second rows of Figure \ref{fig:advection}, the sensor located to the west of the source firstly picks up signal. Hence, at times 15 and 30 seconds, the peaks indicated by the contour plot are somewhere between that sensor and the actual source. As more data become available from the other sensors located near the source (shown in the second row of Figure \ref{fig:advection}), the peak moves closer to the actual source at time 90, accurately pinpointing the source of emission.

$\bullet$ Sensitivity analysis. The accuracy of the input velocity, i.e., $\vec{\bm{v}}$ in the advection-diffusion operator $\mathcal{A}$, significantly affects the performance of the inverse model. %For many applications, information on the velocity field (e.g., wind) is obtained through sensor measurements or computer models. 
Imagine that, in the second scenario above, if the specified wind direction is far from the true direction, we no longer expect the model to yield accurate results. 
%Hence, it is necessary to further investigate the robustness of the model by introducing random error to the intput wind vector. 
Hence, sensitivity analysis is performed to investigate the robustness of the model against the mis-specification of the input velocity and the tuning parameters in the regularization, $\mathcal{R}$. 
Let $\vec{\bm{v}}=(v_x, v_y)^T=(10,-10)^T$ be the actual wind vector $[\text{m} \text{s}^{-1}]$, and let $\vec{\bm{v}}^{\text{input}}=(v_{x}+\epsilon^{\text{wind}}_x, v_{y}+\epsilon^{\text{wind}}_y)^T$ be the input wind vector to the inverse model with random errors, $\epsilon^{\text{wind}}_x \sim N(0, |v_{x}|\tau)$ and $\epsilon^{\text{wind}}_y \sim N(0, |v_{y}|\tau)$, i.e., the variance of the error is proportional to the magnitude of the wind vector given a factor $\tau$. 

\vspace{-6pt}
\begin{figure}[h!]  
	\begin{center}
		\includegraphics[width=0.70\textwidth]{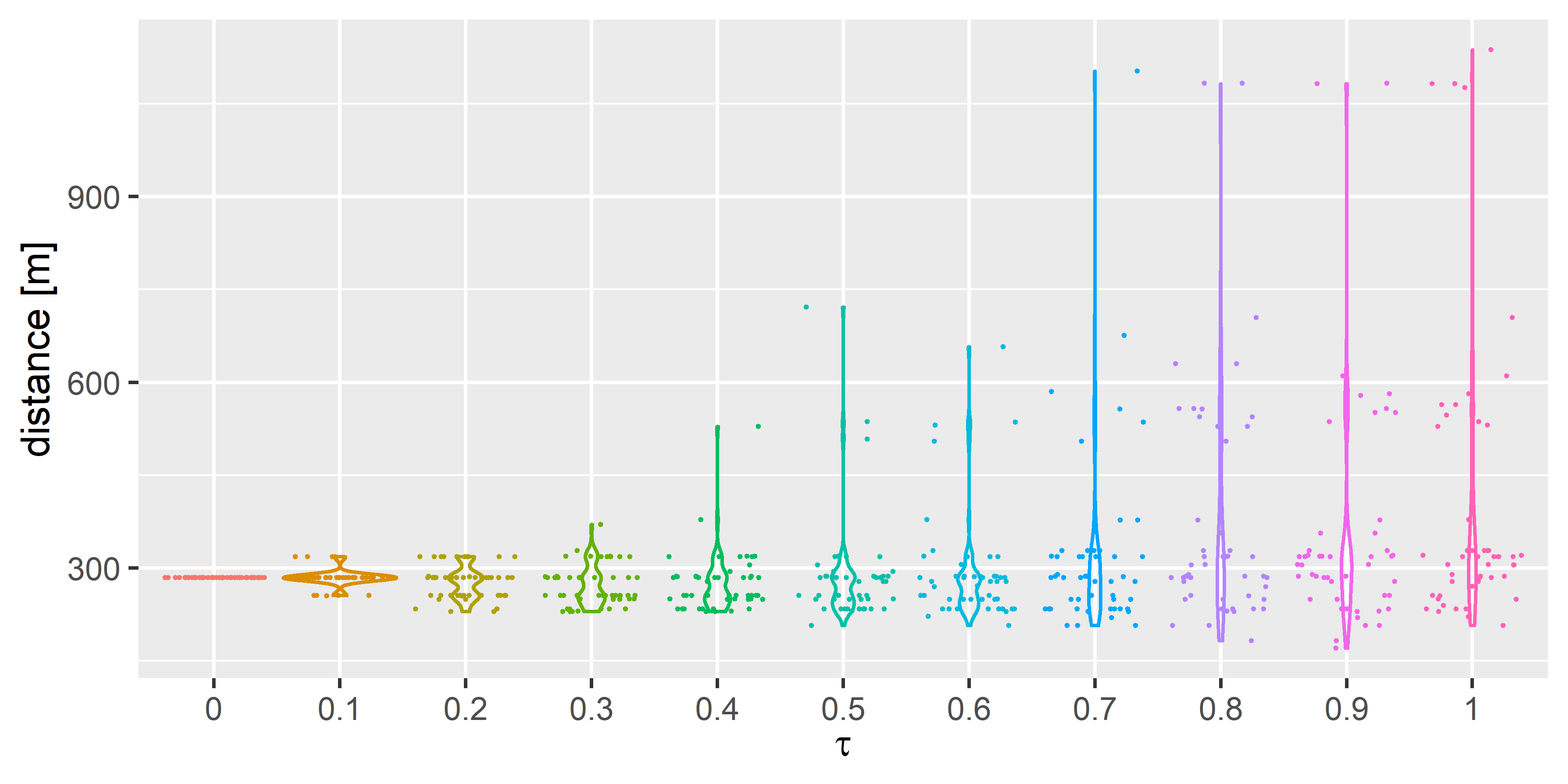}
		\centering
		\vspace{-14pt}
		\caption{Sensitivity analysis against the mis-specification of input velocity vector.}
		\label{fig:sensitivity}
	\end{center}
\end{figure}

\vspace{-16pt}
Consider the second scenario above, Figure \ref{fig:sensitivity} shows the violin plot of the distance (in meters) between the detected source and the actual source for different values of $\tau$ ranging from 0.1 to 1. For each value of $\tau$, the experiment is repeated by 50 times, and a new input wind vector is simulated for each run. It is seen that, the distance between the detected source and the actual source remains within the range from 225m to 380m when $\tau \leq 0.3$. In other words, the model performance appears to be robust when the input wind vector does not significantly deviate away from the actual wind vector. In the context of this problem, since the horizontal and vertical components of the actual wind vector are both 10m $\text{s}^{-1}$, the standard deviation of the random error associated with input wind vector ranges from 1m to 3m in both the horizontal and vertical directions when $0.1 \leq \tau \leq 0.3$. Such a margin of specification error is reasonable and can be achieved in many applications where both wind speed and directions are observable \citep{Ding2021}. When $\tau \geq 0.4$, we note that the variance of the detection error dramatically increases, indicating a rapid performance deterioration of the inverse model, as expected.

\vspace{-8pt}
\begin{figure}[h!]  
	\begin{center}
		\includegraphics[width=0.85\textwidth]{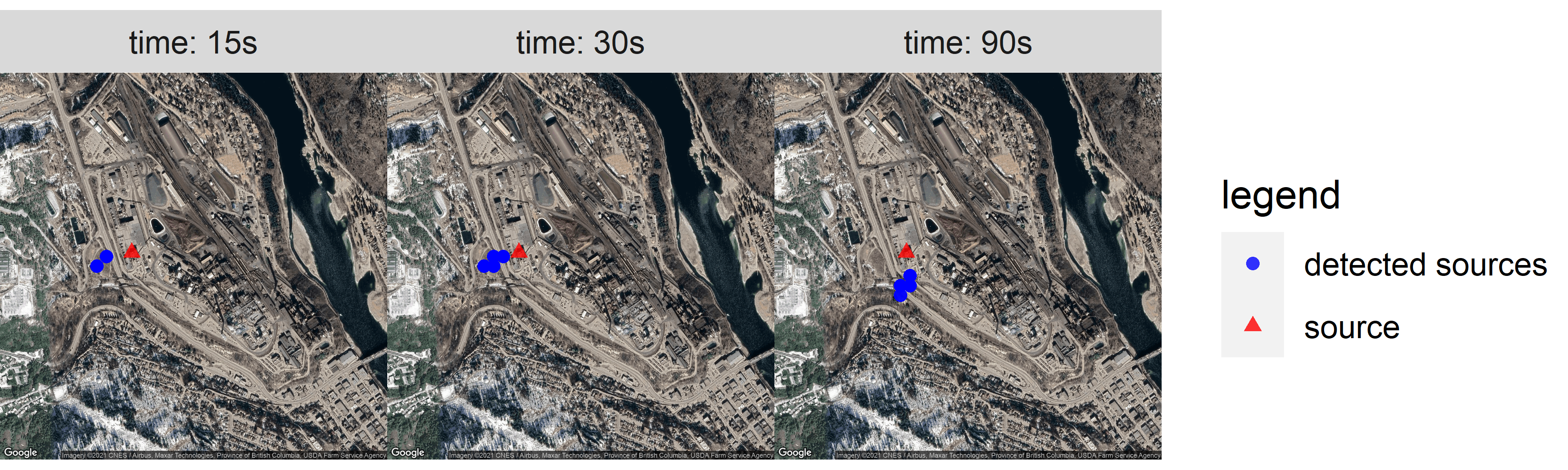}
		\centering
		\vspace{-8pt}
		\caption{Sensitivity analysis of the detected sources, at times 15, 30 and 90 seconds, for combinations of $\lambda_1$ and $\lambda_2$ taken from a mesh grid $\{5,10,15,\cdots,50\}\otimes\{5,10,15,\cdots,50\}$.}
		\label{fig:robustness}
	\end{center}
\end{figure}
%Finally, Figure \ref{fig:robustness2} shows the distance between the detected source and the actual source for diffenrent values of $\lambda_1$ and $\lambda_2$. This figure provides some insights how the appropriate choices of the two tuning parameters. Of course, for general cases where the actual source locations are not known, it is impossible to generate figures like Figure \ref{fig:robustness2}. 
%\begin{figure}[h!]  
%	\begin{center}
%		\includegraphics[width=1\textwidth]{figures/sensitivity3.png}
%		\centering
%		\vspace{-8pt}
%		\caption{Distance between the detected source and the actual source for different values of $\lambda_1$ and $\lambda_2$.}
%		\label{fig:robustness2}
%	\end{center}
%\end{figure}

\vspace{-26pt}
The performance of the model also depends on the choice of the tuning parameters, $\lambda_1$ and $\lambda_2$, in the regularization $\mathcal{R}$. Note that, for many forward prediction problems, such as Lasso regression, Ridge regression, Elastic Net, etc., the tuning parameters can be chosen through cross-validation. However, the idea of cross-validation no longer applies to inverse problems because the true source is never known and it is impossible to establish the link between the tuning parameters and detection accuracy. Hence, it is more meaningful to investigate the robustness (sensitivity) of the proposed inverse model against $\lambda_1$ and $\lambda_2$. Figure \ref{fig:robustness} shows the detected sources, at 15, 30 and 90 seconds, for combinations of $\lambda_1$ and $\lambda_2$ taken from a mesh grid $\{5,10,15,\cdots,50\}\otimes\{5,10,15,\cdots,50\}$. The figure shows that the detect sources are robust enough against the choices of the tuning parameters parameters. In other words, the true source can be correctly identified for different combinations of $\lambda_1$ and $\lambda_2$ chosen from a relatively wide range, which is certainly desirable in practice. 

Finally, it is worth noting that f the initial conditions are strictly modeled by delta functions (e.g., ``point'' sources), the Fourier series of the space-time process does not converge and the proposed approach may generate an oscillatory solution known as the Gibbs phenomenon \citep{Willard1898}. Hence, the proposed approach works well if the initial condition is a smooth function, as seen in the numerical examples above. Even if the initial conditions consist of point sources, the model seeks a solution for a little bit later than the time of release when the point sources become smoother functions due to diffusion. 
%It is also noted that, in the atmospheric inverse problems, the majority of the existing approaches are within the Bayesian framework and relies on more complex CFD solver to account for unsteady wind conditions. On the other hand, the proposed model, which is based on a much simpler advection-diffusion PDE can find multiple source locations without a-priori information (most of the inverse models assume that modeler knows the number of sources). In addition, once the source location is found by the proposed inverse model, it is capable of predicting how the concentration field will evolve in the future by solving the forward model. It can be a good example of a rapid response model that when hazardous materials are released by a terrorist attack or an accident (as in Example II in Section \ref{sec:example2}). 

\vspace{-20pt}
\section{Conclusions}
Based on a PDE-based statistical model for spatio-temporal data, this paper proposed an inverse modeling approach for advection-diffusion processes utilizing data streams generated by three spatial sampling schemes. The paper obtained both necessary and sufficient conditions under which the Fourier coefficients of the initial condition of the advection-diffusion process can be uniquely estimated. Detailed iteration steps of the ADMM have been obtained, which solves the inverse problems in a computational efficient manner. The algorithm has also been extended for handling a linear inequality constraint on the model output. Numerical examples have been presented to demonstrate the robustness of the proposed inverse models against input model parameters, and reveal the dynamic nature of the inverse problem based on sensor data streams. Note that, the paper considers data arising from a deterministic PDE. One critical future direction is to extend the proposed models for stochastic PDEs, which aim at minimizing the distance between the forward distribution and that of the observations. Computer code is available at \url{https://github.com/dnncode/inverse-model}.

\vspace{-16pt}
\section*{Acknowledgments}
We are grateful to two reviewers, the Associate Editor and the Editor for their constructive comments, which improved the quality of the paper. 

\vspace{-16pt}
\section*{Funding}
This material is based upon work supported by the National Science Foundation under Grant No. 2143695. 

\vspace{-16pt}
\section*{Supplementary Materials}
The Supplementary Materials provide (i) the proof of Propositions 1 and 2, (ii) derivation of the ADMM Algorithm 1, (iii) ADMM with non-negativity constraint, (iv) discussions on the computational time of Problems P-I, P-II and P-III, and (v) Numerical comparison on different choices of regularizations.

\singlespacing
\bibliographystyle{asa}
\bibliography{references}
\doublespacing
%\end{thebibliography}

\clearpage
\appendix

\begin{center}
\textbf{\Large{Inverse Models for Estimating the Initial Condition of Spatio-Temporal Advection-Diffusion Processes}}
\end{center}

\begin{center}
Xiao Liu\\
Department of Industrial Engineering, University of Arkansas\\

Kyongmin Yeo\\
IBM T. J. Watson Research Center
\end{center}

\vspace{10pt}
\color{black}\textbf{\Large{Supplemental Materials}}\color{black}

\vspace{-16pt}
\section*{1. Proof of Proposition 1}
\textbf{\textit{Proof}.} To show Proposition 1, note that%we re-write (\ref{eq:Y}) as
%\begin{equation} 
%\mathrm{vec}(\bm{Y}) = \text{diag}( \{\bm{F}\tilde{\bm{g}}(l)\}_{l=1}^L)\bm{\eta} + \mathrm{vec}(\bm{V})
%\end{equation} 
%\begin{equation} \label{eq:p1_proof_1}
%\underbrace{\begin{pmatrix}
%	\bm{Y}(1)\\ 
%	\bm{Y}(2)\\ 
%	\vdots \\ 
%	\bm{Y}(L)
%	\end{pmatrix}}_{\vec{\bm{\mathscr{Y}}} } = 
%\underbrace{\begin{pmatrix}
%	\bm{F} \text{diag}(  \{ \bm{g}_{j}(1) \}_{j=1}^{N})\\ 
%	\bm{F} \text{diag}(  \{ \bm{g}_{j}(2) \}_{j=1}^{N})\\ 
%	\vdots \\ 
%	\bm{F} \text{diag}(  \{ \bm{g}_{j}(L) \}_{j=1}^{N})
%	\end{pmatrix}}_{\bm{\mathscr{X}}^{(\text{P-I})}} \bm{\eta} + 
%\underbrace{\begin{pmatrix}
%	\bm{v}(1)\\ 
%	\bm{v}(2)\\ 
%	\vdots \\ 
%	\bm{v}(L)
%	\end{pmatrix}}_{\vec{\bm{\mathscr{V}}} } 
%\end{equation} 

$\bullet$ For any $\bm{k}_{j_1}$ and $\bm{k}_{j_2}$ ($\bm{k}_{j_1},\bm{k}_{j_2} \in \mathcal{K}$ and $j_1 \neq j_2$) such that $\bm{\vec{v}}^T(\bm{k}_{j_1}-\bm{k}_{j_2}) = 0$ and  $\bm{k}_{j_1}^T \bm{D} \bm{k}_{j_1} = \bm{k}_{j_2}^T \bm{D} \bm{k}_{j_2}$, it is immediately implied by $\gamma_{j}= -\bm{k}^T_j \bm{D}\bm{k}_j - \zeta - \imath  \bm{\vec{v}}^T \bm{k}_j$ that the two vectors $(g_{j_1}(1), ...,g_{j_1}(L))$ and $(g_{j_2}(1), ...,g_{j_2}(L))$ become identical; 

$\bullet$ For a given $\bm{s}$, the complex exponential $e^{\imath 2\pi \bm{s}^T \bm{k}_{j_1}}$ and $e^{\imath 2\pi \bm{s}^T \bm{k}_{j_2}}$ are not linearly independent in the complex domain $\mathbb{C}$. For example, there exist $u,v\in\mathbb{R}$ (at least one of them is non-zero) such that $(u+v\imath)e^{\imath 2\pi \bm{s}^T \bm{k}_{j_1}} = (u-v\imath)e^{\imath 2\pi \bm{s}^T \bm{k}_{j_2}}$;

$\bullet$ For a pair of sampling locations $\bm{s}$ and $\bm{s}'$, if both $2\bm{k}_{j_1}(\bm{s}-\bm{s}')$ and $2\bm{k}_{j_2}(\bm{s}-\bm{s}')$ return even numbers, then, $e^{\imath 2\pi \bm{s}^T \bm{k}} = e^{\imath 2\pi \bm{s}'^T \bm{k}}$. Similarly, if both $2\bm{k}_{j_1}(\bm{s}-\bm{s}')$ and $2\bm{k}_{j_2}(\bm{s}-\bm{s}')$ return odd numbers, then, $e^{\imath 2\pi \bm{s}^T \bm{k}} = - e^{\imath 2\pi \bm{s}'^T \bm{k}}$.

Hence, the matrix $\bm{\mathscr{X}}^{(\text{P-I})}$ in (\ref{eq:p1_proof_1}) has independent columns if one of the conditions A and B in Proposition 1 holds. When $ML \geq N$, $\bm{\mathscr{X}}^{(\text{P-I})}$ has a full column rank of $N$, and the condition in Proposition 1 is also sufficient for components in $\bm{\eta}$ to be uniquely determined.$\blacksquare$

\vspace{-16pt}
\section*{2. proof of Proposition 2}
\textbf{\textit{Proof}.} 
We first construct a new matrix $\bm{\tilde{G}}$ by eliminating the redundant rows (if there is any) in $\bm{G}$ as follows: 
Let $\bm{g}_{i,\cdot}$ be the $i$th row $\bm{G}$. For $i=2,...N$, if there exists $i'=1,...,i-1$ such that $\bm{g}_{i,\cdot}=\bm{g}_{i',\cdot}$, the row $\bm{g}_{i,\cdot}$ is eliminated from $\bm{G}$. 
By eliminating the repeated rows from $\bm{G}$, the matrix $\bm{\tilde{G}}$ is full row rank when $L \geq \tilde{N}$.

Let $\bm{H}=(\bm{h}_{\cdot,1},\bm{h}_{\cdot,2},...,\bm{h}_{\cdot,\tilde{N}})$ be a $N\times\tilde{N}$ matrix with $\bm{h}_{\cdot,j}$ representing the $j$th column vector of $\bm{H}$. For any column vector $\bm{h}_{\cdot,j}$, its $i$th element $h_{i,j}=1$ if $i \in \Psi_j$; otherwise, $h_{i,j}=0$.
Then, (\ref{eq:Y}) can be re-written as
\vspace{-6pt}
\begin{equation} \label{eq:proof1}
\bm{Y} = \bm{F} (\bm{E}\bm{H})\tilde{\bm{G}}
+ \bm{V}.
\end{equation}
 
Since $\bm{\tilde{G}}$ is full row rank of $\tilde{N}$ (when $L \geq \tilde{N}$), we re-write (\ref{eq:proof1}) as follows:
\vspace{-14pt}
\begin{equation} \label{eq:proof2}
\bm{Y}\tilde{\bm{G}}^{-1}_R = \bm{F} (\bm{E}\bm{H})
+ \bm{V}\tilde{\bm{G}}^{-1}_R 
\end{equation} 

\vspace{-16pt}
\noindent where $\tilde{\bm{G}}^{-1}_R$ is the right inverse of $\bm{\tilde{G}}$.

Let $\bm{Y}\tilde{\bm{G}}^{-1}_R \equiv \tilde{\bm{Y}} = (\tilde{\bm{Y}}_1, \tilde{\bm{Y}}_2, ..., \tilde{\bm{Y}}_{\tilde{N}})$, 
$\tilde{\bm{F}}_i = \{f_{m,j}\}_{m=1,...,M, j\in\Psi_i}$ be a $M\times|\Psi_i|$ matrix, and $\bm{V}\tilde{\bm{G}}^{-1}_R \equiv \bm{\Omega} = (\bm{\Omega}_1, \bm{\Omega}_2, ..., \bm{\Omega}_{\tilde{N}})$. Then, (\ref{eq:proof2}) defines a system of $\tilde{N}$ linear models
\vspace{-12pt}
\begin{equation} \label{eq:proof3}
\tilde{\bm{Y}}_i = \tilde{\bm{F}}_i\bm{\eta}_i  + \bm{\Omega}_i, \quad\quad \forall i=1,2,...,\tilde{N}
\end{equation} 

\vspace{-12pt}
\noindent where $\bm{\eta}_i$ is a column vector $\{\eta(\bm{k}_j)\}_{j\in\Psi_i}$.
Hence, the sufficient condition for all components in $\bm{\eta}$ to be uniquely determined is $\mathrm{rank}(\tilde{\bm{F}}_i) = |\Psi_i|$ for all $i=1,...,\tilde{N}$ $\blacksquare$.

\vspace{-12pt}
\section*{3. derivation of the ADMM Algorithm 1}

The ADMM solves the constrained problem (\ref{eq:PI_constrained}) by repeating the following iterations \citep{Zou2005, Ramdas2016}: 
\vspace{-16pt}
\begin{subequations} \label{eq:ADMM}
\begin{align}
 \bm{\eta}^{(i)} = \text{argmin}_{\bm{\eta}} f(\bm{\eta}) + \frac{\rho}{2}\left\| \bm{\eta} - \bm{\psi}^{(i-1)} + \bm{u}^{(i-1)} \right\|_2^2\\
 \bm{\psi}^{(i)} = \text{argmin}_{\bm{\psi}} \mathcal{R}(\bm{\psi}) + \frac{\rho}{2}\left\| \bm{\eta}^{(i)} - \bm{\psi} + \bm{u}^{(i-1)} \right\|_2^2\\
 \bm{u}^{(i)} = \bm{u}^{(i-1)} + \bm{\eta}^{(i)} - \bm{\psi}^{(i)}
\end{align}
\end{subequations}

\vspace{-16pt}
\noindent for $i=1,2,\cdots$. Note that, the two minimization problems in (\ref{eq:ADMM}a) and (\ref{eq:ADMM}b) can be efficiently solved as follows.  

\textbf{Solving (\ref{eq:ADMM}a)}. For (\ref{eq:ADMM}a), it is possible to show that:
\vspace{-12pt}
\begin{equation} \label{eq:ADMM_1}
\begin{split}
\bm{\eta}^{(i)} & =  \text{argmin}_{\bm{\eta}} \frac{1}{2}\langle (\vec{\bm{\mathscr{Y}}} - \bm{\mathscr{X}}\bm{\eta}), \bm{\Sigma}^{-1}(\vec{\bm{\mathscr{Y}}} - \bm{\mathscr{X}}\bm{\eta})\rangle + \frac{\rho}{2} \langle \bm{\eta} - \bm{\psi}^{(i-1)} + \bm{u}^{(i-1)}, \bm{\eta} - \bm{\psi}^{(i-1)} + \bm{u}^{(i-1)} \rangle\\
& = \text{argmin}_{\bm{\eta}} -\frac{1}{2} \langle \bm{\eta}, \bm{\mathscr{X}}^T \bm{\Sigma}^{-1}\vec{\bm{\mathscr{Y}}}\rangle -\frac{1}{2} \langle \bm{\eta}, (\bm{\Sigma}^{-1}\bm{\mathscr{X}})^{T} \vec{\bm{\mathscr{Y}}}\rangle + \frac{1}{2}\langle \bm{\eta},  \bm{\mathscr{X}}^T \bm{\Sigma}^{-1}\bm{\mathscr{X}}\bm{\eta}\rangle \\ & \quad\quad\quad\quad\quad\quad+ \frac{\rho}{2}\langle \bm{\eta}, \bm{\eta}\rangle - \frac{1}{2}\langle \bm{\eta}, \bm{\psi}^{(i-1)}\rangle + \frac{1}{2}\langle \bm{\eta}, \bm{u}^{(i-1)}\rangle \\
& = \frac{1}{2}\langle \bm{\eta}, (\bm{\mathscr{X}}^T \bm{\Sigma}^{-1}\bm{\mathscr{X}}+\frac{\rho}{2}\bm{I})\bm{\eta} \rangle + \langle \bm{\eta}, -\frac{1}{2} (\bm{\mathscr{X}}^T \bm{\Sigma}^{-1} + (\bm{\Sigma}^{-1}\bm{\mathscr{X}})^T)\vec{\bm{\mathscr{Y}}}-\rho(\bm{\psi}^{(i-1)}-\bm{u}^{(i-1)})\rangle
\end{split}
\end{equation}
where $\langle \cdot, \cdot \rangle$ represents the inner product in a vector space. 

Because the gradient vector of the right hand side of (\ref{eq:ADMM_1}) can be obtained as
\vspace{-12pt}
\begin{equation}
(\bm{\mathscr{X}}^T \bm{\Sigma}^{-1}\bm{\mathscr{X}}+\frac{\rho}{2}\bm{I})\bm{\eta}  -\frac{1}{2} (\bm{\mathscr{X}}^T \bm{\Sigma}^{-1} + (\bm{\Sigma}^{-1}\bm{\mathscr{X}})^T)\vec{\bm{\mathscr{Y}}}-\rho(\bm{\psi}^{(i-1)}-\bm{u}^{(i-1)}),
\end{equation}

\vspace{-12pt}
\noindent we obtain the closed-form solution of (\ref{eq:ADMM}a) by setting the gradient vector above to zero:
\vspace{-12pt}
\begin{equation}
\bm{\eta}^{(i)} = (\bm{\mathscr{X}}^T \bm{\Sigma}^{-1}\bm{\mathscr{X}}+\frac{\rho}{2}\bm{I})^{-1} \left\{\frac{1}{2} (\bm{\mathscr{X}}^T \bm{\Sigma}^{-1} + (\bm{\Sigma}^{-1}\bm{\mathscr{X}})^T)\vec{\bm{\mathscr{Y}}}-\rho(\bm{\psi}^{(i-1)}-\bm{u}^{(i-1)}) \right\}.
\end{equation}

\textbf{Solving (\ref{eq:ADMM}b)}. We re-write (\ref{eq:ADMM}b) as
\vspace{-12pt}
\begin{equation}  \label{eq:ADMM_b}
	\begin{split}
\bm{\psi}^{(i)} & = \text{argmin}_{\bm{\psi}} \mathcal{R}(\bm{\psi}) + \frac{\rho}{2}\left\| \bm{\eta}^{(i)} - \bm{\psi} + \bm{u}^{(i-1)} \right\|_2^2 \\
& = \text{argmin}_{\bm{\psi}}  \lambda_1 \left \| \bm{\psi} \right\|_1 + \lambda_2 \left \|  \bm{J} \bm{\psi}   \right\|_2^2 + \frac{\rho}{2}\left\| \bm{\eta}^{(i)} - \bm{\psi} + \bm{u}^{(i-1)} \right\|_2^2 \\
& = \text{argmin}_{\bm{\psi}} \lambda_1 \left \| \bm{\psi} \right\|_1  + \lambda_2 \langle\bm{\psi}, (\bm{J}^T\bm{J}+ \frac{\rho}{2}\bm{I}) \bm{\psi}  \rangle - \rho \langle \bm{\psi}, \bm{u}^{(i-1)} + \bm{\eta}^{(i)} \rangle
	\end{split}.
\end{equation}

\vspace{-12pt}
The optimization problem (\ref{eq:ADMM_b}) can again by solved numerically using ADMM. Converting (\ref{eq:ADMM_b}) to a constrained problem yields:
\vspace{-12pt}
\begin{equation}  \label{eq:ADMM_4}
\text{min}_{\bm{\tilde{\psi}}, \bm{\theta}} \quad  \lambda_2 \langle \bm{\tilde{\psi}}, (\bm{J}^T\bm{J}+ \frac{\rho}{2}\bm{I})\bm{\tilde{\psi}}  \rangle - \rho \langle \bm{\tilde{\psi}}, \bm{u}^{(i-1)} + \bm{\eta}^{(i)}\rangle+ \lambda_1 \left \| \bm{\theta} \right\|_1, \quad\quad \text{s.t. } \bm{\psi} = \bm{\theta}. 
\end{equation}

\vspace{-12pt}
Then, for a given $\omega>0$, the constrained problem is solved by repeating the following steps (for $j=1,2,...$):
\vspace{-12pt}
\begin{equation}  
	\begin{split}
\bm{\tilde{\psi}}^{(j)} & = \text{argmin}_{\bm{\tilde{\psi}}}   \lambda_2 \langle \bm{\tilde{\psi}}, (\bm{J}^T\bm{J}+ \frac{\rho}{2}\bm{I})\bm{\tilde{\psi}}  \rangle - \rho \langle \bm{\tilde{\psi}}, \bm{\tilde{\psi}} + \bm{\eta}^{(i)} \rangle + \frac{\omega}{2} \left\| \bm{\tilde{\psi}} -\bm{\theta}^{(j-1)}  + \bm{v}^{(j-1)} \right\|_2^2 \\
&  = \text{argmin}_{\bm{\tilde{\psi}}}  \frac{1}{2} \langle \bm{\tilde{\psi}}, 2\lambda_2(\bm{J}^T\bm{J}+ \frac{\rho+\omega}{2}\bm{I}) \bm{\tilde{\psi}} \rangle - \rho \langle \bm{\tilde{\psi}}, \bm{u}^{(i-1)} + \bm{\eta}^{(i)}+\omega\bm{\theta}^{(j-1)} - \omega\bm{v}^{(j-1)} \rangle \\
& = \frac{\rho}{2\lambda_2}(\bm{J}^T\bm{J}+ \frac{\rho+\omega}{2}\bm{I})^{-1}(\bm{u}^{(i-1)} + \bm{\eta}^{(i)}+\omega\bm{\theta}^{(j-1)} - \omega\bm{v}^{(j-1)}), 
\end{split}
\end{equation}
\begin{equation}  
\begin{split}
\bm{\theta}^{(j)} & = \text{argmin}_{\bm{\theta}} \quad \lambda_1 \left \| \bm{\theta} \right\|_1 +  \frac{\omega}{2} \left\| \bm{\tilde{\psi}}^{(j)} -\bm{\theta}  + \bm{v}^{(j-1)} \right\|_2^2 \\
& = S_{\lambda_1/\omega}(\bm{\tilde{\psi}}^{(j)}+\bm{v}^{(j-1)})
\end{split}
\end{equation}

\vspace{-12pt}
\noindent with $S_{\lambda_1/\omega}(\cdot)$ being a soft-thresholding operator 
\vspace{-12pt}
\begin{equation} \label{eq:cases}
S_{\lambda_1/\omega}(x) = \begin{cases}
x-\lambda_1/\omega,& \text{if } x > \lambda_1/\omega\\
x+\lambda_1/\omega,& \text{if } x < -\lambda_1/\omega\\
0,              & \text{otherwise} 
\end{cases}
\end{equation}

\vspace{-12pt}
\noindent and $\bm{v}^{(j)} = \bm{v}^{(j-1)} + \bm{\tilde{\psi}}^{(j)} - \bm{\theta}^{(j)}$.

%\begin{equation}
%\bm{v}^{(j)} = \bm{v}^{(j-1)} + \bm{\tilde{\psi}}^{(j)} - \bm{\theta}^{(j)}
%\end{equation}

\vspace{-12pt}
\section*{4. ADMM with Non-negativity Constraint}
%When a non-negativity constraint is added,  the inverse modeling problem (\ref{eq:inverse_general}) becomes (\ref{eq:inverse_general_constraint}). 
%\begin{equation} \label{eq:inverse_general_constraint}
%\begin{split}
%\text{min}_{\bm{\eta}} \quad & \frac{1}{2} (\vec{\bm{\mathscr{Y}}} - \bm{\mathscr{X}}\bm{\eta})^T \bm{\Sigma}^{-1} (\vec{\bm{\mathscr{Y}}} - \bm{\mathscr{X}}\bm{\eta})+ \mathcal{R}(\bm{\eta}) \\
%& \text{s.t. }  \bm{\mathscr{X}}\bm{\eta} \geq 0 
%\end{split}
%\end{equation}
%Here, the linear constraint forces the estimated initial condition to be non-negative. 
We show that the constrained problem (\ref{eq:inverse_general_constraint}) can be efficiently solved by modifying the ADMM algorithm described in Section \ref{sec:ADMM}. 
Firstly, we replace the inequality constraint in (\ref{eq:inverse_general_constraint}) with an equality constraint by introducing a penalty function and variable substitution:
\vspace{-12pt}
\begin{equation} \label{eq:inverse_general_constraint_2}
\text{min}_{\bm{\eta}} \quad  \frac{1}{2} (\vec{\bm{\mathscr{Y}}} - \bm{\mathscr{X}}\bm{\eta})^T \bm{\Sigma}^{-1} (\vec{\bm{\mathscr{Y}}} - \bm{\mathscr{X}}\bm{\eta})+ \mathcal{R}(\bm{\eta}) + \mathcal{I}(\tilde{\bm{\eta}}) \quad\quad
 \text{s.t. }  \bm{\mathscr{X}}\bm{\eta} - \tilde{\bm{\eta}} = 0 
\end{equation}

\vspace{-2pt}
\noindent where $\mathcal{I}(\tilde{\bm{\eta}})=0$ if $\tilde{\bm{\eta}}\geq 0$, and $\mathcal{I}(\tilde{\bm{\eta}})=\infty$ otherwise. 
Then, for $\rho>0$, the scaled form of the augmented Lagrangian of (\ref{eq:inverse_general_constraint_2}) is written as:
\vspace{-12pt}
\begin{equation} \label{eq:inverse_general_constraint_Lagrangian}
f(\bm{\eta}) + \mathcal{R}(\bm{\eta}) + \mathcal{I}(\tilde{\bm{\eta}}) + \frac{\rho}{2} \left\| \bm{\mathscr{X}}\bm{\eta}-\tilde{\bm{\eta}}+ \bm{u}  \right\|_2^2 +  \frac{\rho}{2} \left\| \bm{u} \right\|_2^2.
\end{equation}

\vspace{-12pt}
\noindent where $f(\bm{\eta})=\frac{1}{2} (\vec{\bm{\mathscr{Y}}} - \bm{\mathscr{X}}\bm{\eta})^T \bm{\Sigma}^{-1} (\vec{\bm{\mathscr{Y}}} - \bm{\mathscr{X}}\bm{\eta})$.

Similar to (\ref{eq:ADMM}), solving (\ref{eq:inverse_general_constraint_Lagrangian}) using ADMM requires repeating the following iterations:
\vspace{-12pt}
\begin{subequations} \label{eq:inverse_general_constraint_ADMM}
\begin{align}
 \bm{\eta}^{(i)} = \text{argmin}_{\bm{\eta}} f(\bm{\eta}) + \mathcal{R}(\bm{\eta}) + \frac{\rho}{2}\left\| \bm{\mathscr{X}}\bm{\eta} - \tilde{\bm{\eta}}^{(i-1)} + \bm{u}^{(i-1)} \right\|_2^2\\
\tilde{\bm{\eta}}^{(i)} = \text{max}(0, \bm{\mathscr{X}}\bm{\eta}^{(i)}+ \bm{u}^{(i-1)})\\
 \bm{u}^{(i)} = \bm{u}^{(i-1)} + \bm{\mathscr{X}}\bm{\eta}^{(i)} - \tilde{\bm{\eta}}^{(i)}
\end{align}
\end{subequations}

\vspace{-12pt}
The evaluation of (56b) and (56c) are trivial, and the computational challenge lies in the minimization problem (68a). Fortunately, ADMM can again be used to solve (56a) by converting (56a) to a constrained problem:
\vspace{-12pt}
\begin{equation} \label{eq:inverse_general_constraint_ADMM_2}
\text{min}_{\bm{\eta}} f(\bm{\eta}) + \mathcal{R}(\bm{\psi}) + \frac{\rho}{2}\left\| \bm{\mathscr{X}}\bm{\psi} - \bm{c} \right\|_2^2, \quad \text{s.t. } \bm{\eta} = \bm{\psi}
\end{equation}

\vspace{-12pt}
\noindent where $\bm{c}=\tilde{\bm{\eta}}^{(i-1)} - \bm{u}^{(i-1)}$. Then, the ADMM for solving (\ref{eq:inverse_general_constraint_ADMM_2}) involves the iterations:
\vspace{-12pt}
\begin{subequations} \label{eq:inverse_general_constraint_ADMM_3}
\begin{align}
 \bm{\eta}^{(i)} = \text{argmin}_{\bm{\eta}} f(\bm{\eta})  + \frac{\rho}{2}\left\| \bm{\eta} -  \bm{\psi}^{(i-1)} + \bm{v}^{(i-1)}\right\|_2^2\\
\bm{\psi}^{(i)} = \text{argmin}_{\bm{\psi}} \mathcal{R}(\bm{\psi}) + \frac{\rho}{2}\left\| \bm{\mathscr{X}}\bm{\psi}-\bm{c} \right\|_2^2+ \frac{\rho}{2}\left\| \bm{\eta}^{(i)} - \bm{\psi} + \bm{v}^{(i-1)} \right\|_2^2\\
 \bm{v}^{(i)} = \bm{v}^{(i-1)} + \bm{\eta}^{(i)} - \bm{\psi}^{(i)}
\end{align}
\end{subequations}

\vspace{-12pt}
A closer examination of (\ref{eq:inverse_general_constraint_ADMM_3}) yields the following critical observations:

$\bullet$ (59a) takes exactly the same form of (46a) and the closed-form solution of (59a) is already given by (\ref{eq:ADMM_1}). 

$\bullet$ (59b) can be re-written as
\vspace{-12pt}
\begin{equation} 
	\begin{split}
\bm{\psi}^{(i)} & = \text{argmin}_{\bm{\psi}} \mathcal{R}(\bm{\psi}) + \frac{\rho}{2}\left\| \bm{\mathscr{X}}\bm{\psi}-\bm{c}  \right\|_2^2 + \frac{\rho}{2}\left\| \bm{\eta}^{(i)} - \bm{\psi} + \bm{v}^{(i-1)} \right\|_2^2 \\
& = \text{argmin}_{\bm{\psi}}  \lambda_1 \left \| \bm{\psi} \right\|_1 + \lambda_2 \left \|  \bm{J} \bm{\psi}   \right\|_2^2 + \frac{\rho}{2}\left\| \bm{\mathscr{X}}\bm{\psi}-\bm{c} \right\|_2^2 + \frac{\rho}{2}\left\| \bm{\eta}^{(i)} - \bm{\psi} + \bm{v}^{(i-1)} \right\|_2^2 \\
& = \text{argmin}_{\bm{\psi}} \lambda_1 \left \| \bm{\psi} \right\|_1  + \lambda_2 \langle\bm{\psi}, (\bm{J}^T\bm{J}+ \frac{\rho}{2}\bm{\mathscr{X}}^T\bm{\mathscr{X}} + \frac{\rho}{2}\bm{I}) \bm{\psi}  \rangle - \rho \langle \bm{\psi}, \mathscr{X}\bm{c} + \bm{u}^{(i-1)} + \bm{\eta}^{(i)} \rangle
	\end{split}
\end{equation}
which takes exactly the same form of (\ref{eq:ADMM_b}), and can be solved following the same steps described from (\ref{eq:ADMM_4}) to (\ref{eq:cases}).

The discussions above show that the proposed inverse model can be extended to handle a linear inequality constraint (on the model output), which expands the applicability of the proposed model for a wider range of problems. 

%\section{Appendix E: Additional remarks on ``point'' sources}
%Theoretically speaking, if the initial conditions are strictly modeled by delta functions (e.g., ``point'' sources), the Fourier series of the spectral decomposition of the space-time process does not converge and the proposed approach may generate an oscillatory solution known as the Gibbs phenomenon \citep{Willard1898}. Hence, the proposed approach works well if the initial condition is a smooth function, as seen in the numerical examples above. Even if the initial conditions consist of point sources, the model seeks a solution for a little bit later than the time of release when the point sources become smoother functions due to diffusion. It is also noted that, in the atmospheric inverse problems, the majority of the existing approaches are within the Bayesian framework and relies on more complex CFD solver to account for unsteady wind conditions. On the other hand, the proposed model, which is based on a much simpler advection-diffusion PDE can find multiple source locations without a-priori information (most of the inverse models assume that modeler knows the number of sources). In addition, once the source location is found by the proposed inverse model, it is capable of predicting how the concentration field will evolve in the future by solving the forward model. It can be a good example of a rapid response model that when hazardous materials are released by a terrorist attack or an accident (as in Example II in Section \ref{sec:example2}). 

\vspace{-12pt}
\section*{5. About the Computational Time of Problems P-I, P-II and P-III.}

\begin{figure}[h!] 
	\begin{center}
		\includegraphics[width=0.75\textwidth]{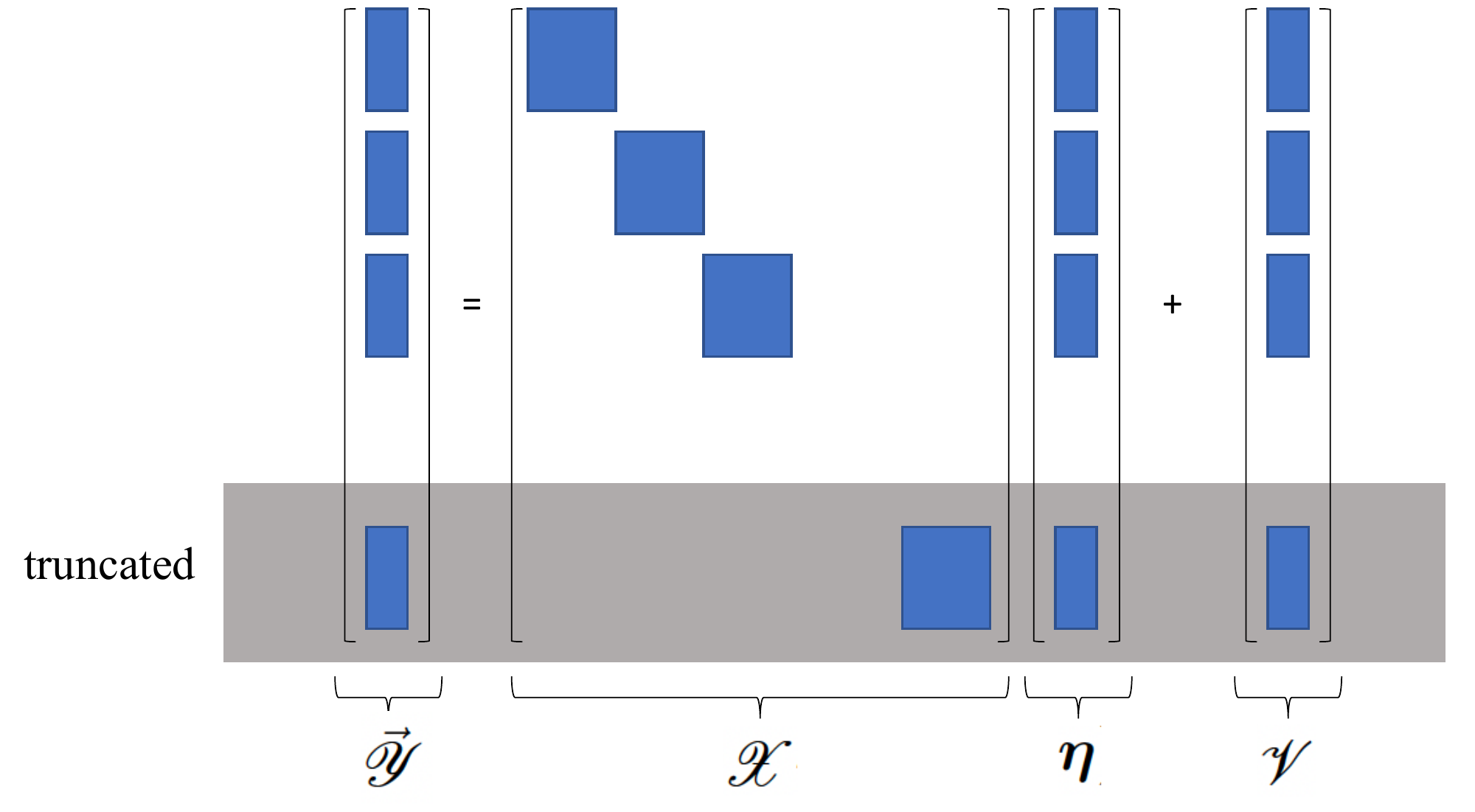}
		\centering
		\vspace{-12pt}
		\caption{In problems P-II and P-III, $\bm{\mathscr{X}}$ is a block diagonal matrix with each block corresponding to a frequency, and it is possible to drop the blocks corresponding to high frequencies to further reduce the computational time.}
		\label{fig:computation}
	\end{center}  
\end{figure}

Although Problems P-I, P-II and P-III can be solved by the ADMM algorithm, it is noted that Problem P-I is formulated in the space-time domain, while Problems P-II and P-III are constructed in the spectral domain. As a result, the design matrix $\bm{\mathscr{X}}^{(\text{P-I})}$ in (\ref{eq:p1_proof_1}) is a dense matrix, while the design matrices $\bm{\mathscr{X}}^{(\text{P-II})}$ and $\bm{\mathscr{X}}^{(\text{P-III})}$ in (\ref{eq:beta_spectrum}) and (\ref{eq:beta_spectrum_36}) are sparse (block diagonal), making the computation of $\bm{\mathscr{X}}^T \bm{\Sigma}^{-1}\bm{\mathscr{X}}$, $\bm{\mathscr{X}}^T \bm{\Sigma}^{-1}$ and $\bm{\Sigma}^{-1}\bm{\mathscr{X}}$ faster in the ADMM algorithm. 
In addition, like many spectral methods, Problems P-II and P-III enable one to truncate the high-frequency components because each block of $\bm{\mathscr{X}}$,  in both (\ref{eq:beta_spectrum}) and (\ref{eq:beta_spectrum_36}), corresponds to a frequency level. This may help further reduce the computational time and is illustrated in Figure \ref{fig:computation}.

\vspace{-12pt}
\section*{6. Numerical Comparison on Different Choices of Regularizations}

We compare the bias and MSE of the estimated initial condition $\hat{\xi}(0, \bm{s})$ for different choices of regularizations, including

Approach 1). The proposed regularization

Approach 2). Generalized Lasso

Approach 3). Elastic Net

Approach 4). $L_1$ regularization only

Approach 5). $L_2$ regularization only

The comparison is based on Example-I presented in the paper. We repeated the ``data simulation---statistical inference'' process for 100 times, and compute the bias and MSE of the estimated initial condition $\hat{\xi}(0, \bm{s})$. \\

\begin{figure}[h!]  
	\begin{center}
		\includegraphics[width=1\textwidth]{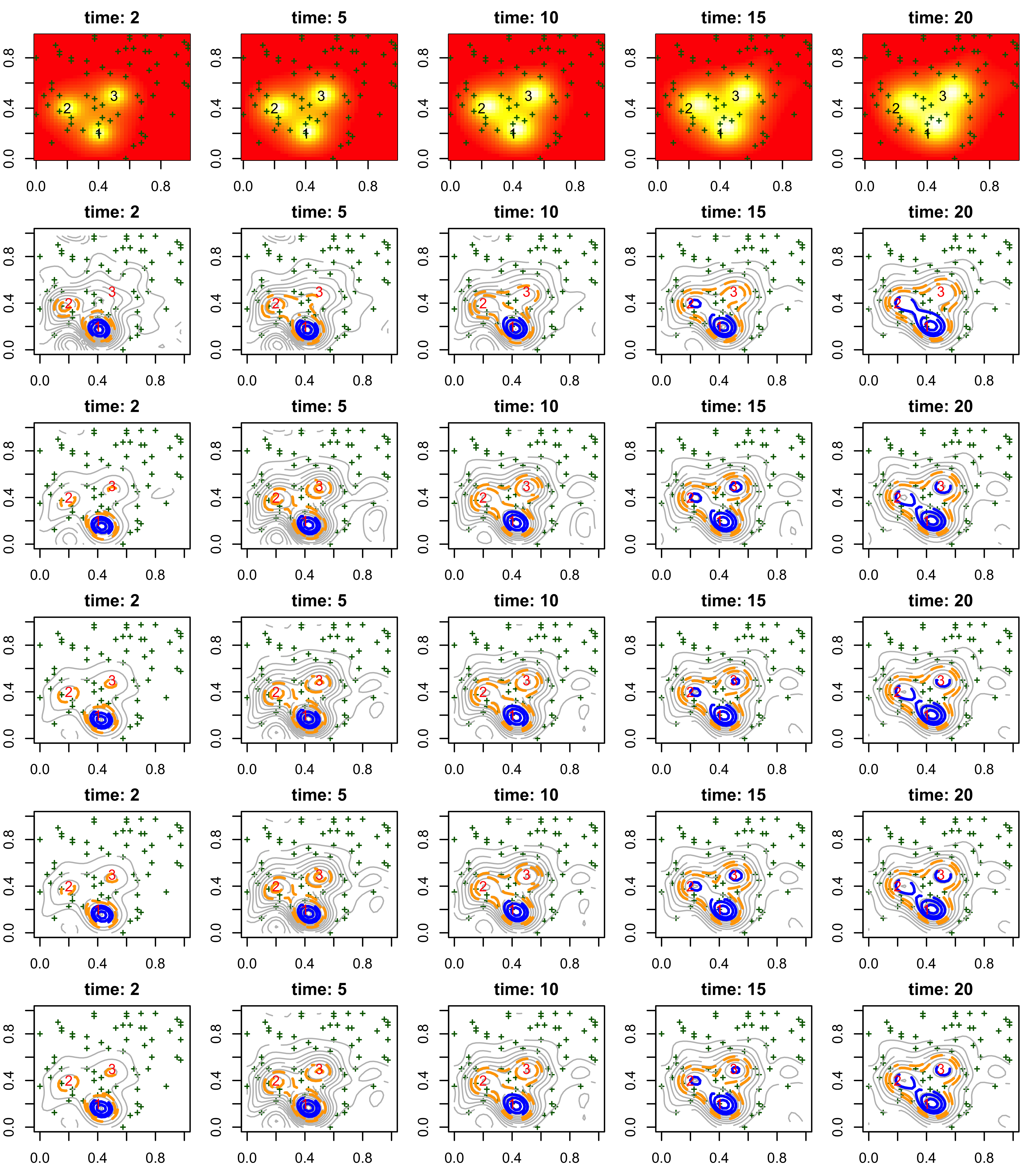}
		\centering
		\vspace{-6pt}
		\caption{(1st row) snapshots of the advection-diffusion process at times 2, 5, 10, 15 and
20; (2nd-6th rows) estimated initial conditions at times 2, 5, 10, 15 and 20 using the proposed regularization (2nd row), Generalized Lasso (3rd row), Elastic Net (4th row),  $L_1$ regularization (5th row) and  $L_2$ regularization (6th row).}
		\label{fig:comparison}
	\end{center}
\end{figure}

Figure (\ref{fig:comparison}) on the next page firstly shows the estimated initial condition using different regularizations. In particular, the 1st row of this figure shows the snapshots of the process at times 2, 5, 10, 15 and 20. The 2nd-6th rows of this figure show the contour plots of the estimated initial condition using the sensor observations up to times 2, 5, 10, 15 and 20. 
The thick blue level sets are respectively the 75th, 85th and 95th percentiles of the output generated by the inverse model, while the dashed thick level sets are the 50th and 60th percentiles. 

\begin{figure}[h!]  
	\begin{center}
		\includegraphics[width=1\textwidth]{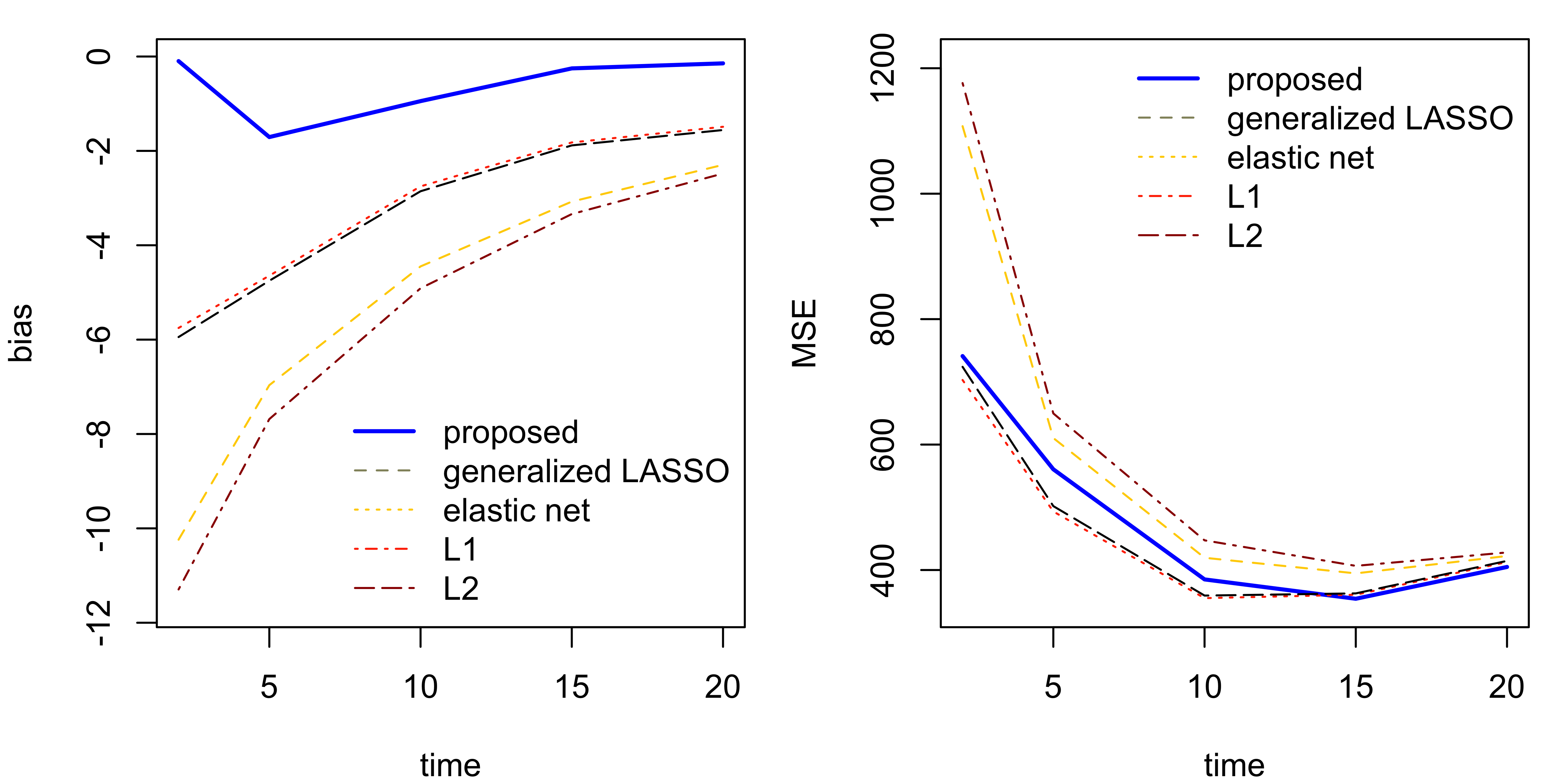}
		\centering
		\vspace{-20pt}
		\caption{Bias and the Mean Squared Error (MSE) of the estimated initial conditions obtained based on different choices of regularizations.}
		\label{fig:comparison2}
	\end{center}
\end{figure}

\clearpage
Next, Figure \ref{fig:comparison2} shows the bias and MSE of the estimated initial condition for all choices of regularizations. 

\vspace{-12pt}

\vspace{-20pt}
It is  seen that:

$\diamond$ The proposed regularization yields the lowest bias. The $L_2$ regularization and the Elastic Net yield the highest bias, while the $L_1$ regularization and the generalized Lasso yield similar performance. 

$\diamond$ The proposed regularization yields a slightly higher MSE than the $L_1$ regularization and generalized Lasso, while the $L_2$ regularization and the Elastic Net yield higher MSE. 

$\diamond$  During our investigation, we spent tremendous amount of time on experimenting on different choices and combinations of how regularizations can be added. One challenge that we found was that the estimated spectral coefficients in $\bm{\eta}$ may dramatically vary from one frequency to another (for example, we may see a sudden spike of the coefficient corresponding to a higher frequency). However, because this paper focuses on the estimation of smooth initial condition, ideally the spectral coefficients should gradually decay as the frequency increases (not necessarily monotone). Adding an $L_1$ or $L_2$ regularization alone does not automatically overcome this difficulty, although it promotes sparsity. Hence,  we modified the Fussed Lass and added a second regularization which helps to impose the ``smoothness among the components in corresponding to adjacent frequencies'' (to prevent the neighboring spectral coefficients from varying dramatically). As a result, we may see the gradual decay of the estimated special coefficients that give rise to a smooth initial conditions. 

\end{document}